\documentclass{article}

\usepackage{set_arxiv}

\usepackage[utf8]{inputenc} 
\usepackage[T1]{fontenc}    
\usepackage{hyperref}       
\usepackage{url}            
\usepackage{caption}
\usepackage{subcaption}
\usepackage{booktabs}       
\usepackage{amsmath}
\usepackage{mathtools}
\usepackage{amssymb}
\usepackage{bbm}
\usepackage{nicefrac}       
\usepackage{microtype}      
\usepackage{lipsum}		
\usepackage{graphicx}
\usepackage{doi}
\usepackage{bookmark}
\usepackage{algorithm}
\usepackage{algpseudocodex}
\usepackage[authoryear,round]{natbib}
\bibpunct{(}{)}{;}{a}{,}{,}
\renewcommand{\cite}{\citet}

\title{Waveform-Based Probabilistic Seismic Hazard Analysis Using Ground-Motion Generative Models\thanks{Submitted to Bull. Seismol. Soc. Am.}}

\author{\href{https://orcid.org/0000-0002-8929-9453}{\includegraphics[scale=0.06]{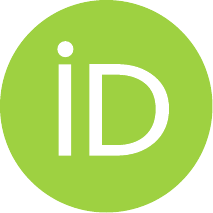}\hspace{1mm}Yuma Matsumoto} \\
    National Research Institute for \\
    Earth Science and Disaster Resilience \\
    Ibaraki, Japan \\
	\And
	\href{https://orcid.org/0000-0001-9589-4630}{\includegraphics[scale=0.06]{orcid.pdf}\hspace{1mm}Taro Yaoyama} \\
    Resilience Engineering Research Center \\
    Graduate School of Engineering \\
	  The University of Tokyo\\
	  Tokyo, Japan \\
    \And
    \href{https://orcid.org/0009-0001-3083-9953}{\includegraphics[scale=0.06]{orcid.pdf}\hspace{1mm}Sangwon Lee} \\
    Department of Architecture \\
    Graduate School of Engineering \\
	The University of Tokyo\\
	Tokyo, Japan \\
    \And
	\href{https://orcid.org/0000-0002-0210-5514}{\includegraphics[scale=0.06]{orcid.pdf}\hspace{1mm}Asako Iwaki} \\
    National Research Institute for \\
    Earth Science and Disaster Resilience \\
    Ibaraki, Japan \\
    \And
    \href{https://orcid.org/0000-0003-3522-7101}{\includegraphics[scale=0.06]{orcid.pdf}\hspace{1mm}Tatsuya Itoi} \\
    Department of Architecture \\
    Graduate School of Engineering \\
	The University of Tokyo\\
	Tokyo, Japan \\
}

\begin{document}
\maketitle

\begin{abstract}
In probabilistic seismic hazard analysis (PSHA),
the exceedance probability of a ground-motion intensity measure (IM) is typically evaluated.
However, in recent years, dynamic response analyses using ground-motion time histories as input
have been increasingly common in seismic design and risk assessment, and thus there is a growing demand for representing seismic hazard in terms of ground-motion waveforms.
In this study, we propose a novel PSHA framework, referred to as waveform-based PSHA,
that enables the direct evaluation of the probability distribution of ground-motion waveforms by introducing ground-motion models (GMMs) based on deep generative models (ground-motion generative models; GMGMs) into the PSHA framework.
In waveform-based PSHA, seismic hazard is represented, in a Monte Carlo sense, as a set of ground-motion waveforms.
We propose the formulation of such a PSHA framework as well as an algorithm for performing the required Monte Carlo simulations. Three different GMGMs based on generative adversarial networks (GANs) are constructed.
After verifying the performance of each GMGM, hazard evaluations using the proposed method are conducted for two numerical examples: one assuming a hypothetical area source and the other assuming an actual site and source faults in Japan.
We demonstrate that seismic hazard can be represented as a set of ground-motion waveforms,
and that the IM-based hazard obtained from these waveforms is consistent with the results of conventional PSHA using GMMs.
Finally, nonlinear dynamic response analyses of a building model are performed using the evaluated seismic hazard as input, and it is shown that exceedance probabilities of engineering demand parameters (EDPs) as well as hazard disaggregation with respect to EDPs can be carried out in a straightforward manner within the proposed framework.
\end{abstract}

\section{Introduction}

Probabilistic seismic hazard analysis (PSHA) is an essential methodology for seismic design and seismic risk assessment,
and it has been widely applied in the field of earthquake engineering.
In conventional PSHA,
seismic hazard is evaluated in terms of the exceedance probability that a ground-motion intensity measure (IM) exceeds a specified intensity level.
The distribution of IM is typically modeled as a lognormal distribution based on ground-motion models (GMMs).

  Within this framework,
numerous studies have been conducted on each of its major components,
including seismic source characterization (SSC; e.g., \cite{Field2014, Morikawa2016, Papadopoulos2021}),
ground-motion characterization (GMC; e.g., \cite{Morikawa2013}; \cite{Bozorgnia2014NGA}; \cite{Abrahamson2018}),
and aleatory variability and epistemic uncertainty (e.g., \cite{SSHAC1997}; \cite{Kiureghian2009105}; \cite{Liou2025}).
Further studies have also been conducted on regional variability (e.g., \cite{Bora2019}; \cite{kotha2020regionally}; \cite{Contreras2025}) as well as on the ergodic assumption and the development of non-ergodic GMMs (e.g., \cite{Anderson1999}; \cite{Milner2021}; \cite{IAEA2025TECDOC}).
As a result of these studies, PSHA has evolved to enable detailed seismic hazard evaluations in various regions (\cite{Gerstenberger2020}; \cite{Fujiwara2023};
\cite{Petersen2024}).

However, in recent years, it has become increasingly common in seismic design and detailed seismic risk assessment to conduct nonlinear dynamic response analyses of structures
using ground-motion time histories, which are difficult to evaluate directly in conventional PSHA.
For example, dynamic response analysis is required by building codes in some regions for the design of specific buildings, such as important structures and high-rise buildings.
In performance-based earthquake engineering (PBEE) represented by the PEER-PBEE framework (\cite{Moehle2014}; \cite{FEMAp58}),
the engineering demand parameters (EDPs) and damage states are evaluated based on dynamic response analyses when detailed risk assessments are conducted.
In the PSHA framework, the input ground motions are generally obtained through methods that select waveforms matching a target response spectrum
(e.g., \cite{Baker2006}).
These methods are widely used in earthquake engineering practice; however, they may not properly capture the phase characteristics of ground motion,
which are important in dynamic response analysis (\cite{Ohsaki1979}; \cite{Sakai2012}).
Extending the PSHA framework to directly evaluate the probabilistic properties of ground-motion waveforms is an essential research direction,
with the potential to advance the engineering applications of PSHA results while strengthening the integration between engineering seismology and earthquake engineering \citep{sewell2009}.

Recently, several approaches have been proposed for directly generating ground-motion waveforms using deep generative models,
such as the variational autoencoder \citep{Ning2024}, generative adversarial networks (GANs; e.g., \cite{gatti2020towards}; \cite{wang2021seismogen}; \cite{matinfar2023deep}; \cite{matsumoto2023fundamental};
\cite{xu2024high}),
and the denoising diffusion model \citep{Huang2025109274}.
In particular, GANs \citep{Goodfellow2014} have achieved considerable success in the modeling of ground-motion waveforms
conditioned on source, path, and site characteristics, similar to the GMMs.
\cite{Florez2022} developed a model using conditional GANs (cGAN; \cite{mirza2014conditional}) that can generate ground-motion waveforms conditioned on magnitude, distance, and the average shear-wave velocity in the top 30 m ($V_{\mathrm{S}30}$).
\cite{esfahani2023tfcgan} developed a model named TFCGAN, which learns the time-frequency domain amplitudes of ground
motions conditioned on magnitude, distance, and $V_{\mathrm{S}30}$, and demonstrated that ground-motion waveforms can be retrieved
from the time-frequency amplitudes.
\cite{Shi2024} employed an extension of GANs, known as the generative adversarial neural operator \citep{rahman2022generative},
to construct a model capable of generating ground-motion waveforms conditioned on magnitude, distance, $V_{\mathrm{S}30}$,
and the style of faulting.
\cite{Yamaguchi2024} developed a model capable of generating site-specific ground-motion waveforms by combining a cGAN
with the generalized spectral inversion technique \citep{Shible2022}.
\cite{matsumoto2024ssgmgm} examined the use of a model known as StyleGAN2 \citep{karras2020analyzing} and developed a model
to generate ground-motion waveforms conditioned on magnitude, distance, and five proxies for site conditions.
\cite{Huang2024} proposed a cGAN model based on two-dimensional convolution, enabling the generation of ground-motion waveforms
conditioned on magnitude, distance, $V_{\mathrm{S}30}$, and the style of faulting.
\cite{Chen2025} developed a cGAN model that generates ground-motion waveforms conditioned on
magnitude, distance, and $V_{\mathrm{S}30}$, and investigated its applicability to regions where only a small number of observed records are available.
\cite{LIN2025107740} constructed a cGAN model that can generate near-fault ground motions.

One of the most important features of such deep generative models is their ability to capture the inherent probability distribution
of observed records and generate new data samples that are statistically similar to the observations \citep{Ruthotto2021}.
In this context, the set of strong-motion observed records is regarded as a collection of samples drawn from an underlying probability distribution,
and the deep generative model is trained to learn that distribution from the data.
Accordingly, just as conventional GMMs represent not only the mean but also the variability of IMs,
a deep generative model trained on ground-motion waveforms can learn not only to accurately generate the waveforms themselves but also
to capture their variability in a non-parametric manner.
We refer to this type of deep generative model-based GMM as a ground-motion generative model (GMGM).
Our previous studies (\cite{matsumoto2023fundamental}; \cite{matsumoto2024ssgmgm}) showed that GMGMs successfully captured the
distributions of indices representing the temporal and frequency characteristics of ground-motion waveforms.
By incorporating GMGM into the framework of PSHA, it may be possible to establish a waveform-based PSHA,
in which the seismic hazard is evaluated directly using ground-motion waveforms.

In this study, toward establishing the waveform-based PSHA framework,
we first focus on the ground-motion modeling component of PSHA and propose a novel PSHA framework that incorporates GMGMs.
We present a mathematical formulation of the waveform-based PSHA,
and introduce computational methods for performing the proposed PSHA integration.
As examples of GMGMs used for numerical experiments, we train three different models.
The first is the StyleGAN-based GMGM (S-GMGM) proposed by \cite{matsumoto2024ssgmgm}.
To simplify the PSHA problem settings, we adopt the deep neural network (DNN) architecture of \cite{matsumoto2024ssgmgm},
while modifying the model to target three conditional labels: magnitude, distance, and $V_{\mathrm{S}30}$.
The second model is that proposed by \cite{Florez2022}.
\cite{Florez2022} developed a cGAN model based on the Wasserstein GAN (WGAN) framework \citep{Gulrajani2017}, which we hereafter refer to as the conditional Wasserstein GAN-based GMGM (CW-GMGM).
As the third GMGM, we develop a model by modifying the S-GMGM into a cGAN framework,
utilizing a DNN architecture similar to that of \cite{Florez2022}.
This model is referred to as the conditional StyleGAN-based GMGM (CS-GMGM).
These three GMGMs are trained on the same dataset based on the strong-motion observed records in Japan compiled in our previous study \citep{matsumoto2024ssgmgm},
and their performance is evaluated by checking the waveforms as well as the distributions of generated data.
We also propose a method for objectively determining the optimal number of training epochs based on quantitative evaluation metrics.
Then, numerical experiments of PSHA using the trained three GMGMs are conducted.
We demonstrate that the proposed methods can represent the seismic hazard as a set of ground-motion waveforms,
and the hazard curves of IMs calculated from the evaluated waveforms are consistent with those of the PSHA results based on conventional GMMs.
Furthermore, to demonstrate the engineering applicability of the proposed PSHA framework,
we perform nonlinear dynamic response analyses of a building inputting ground-motion waveforms evaluated through the proposed method.
In addition to demonstrating that the exceedance probabilities of EDPs can be evaluated in a straightforward manner,
by applying hazard disaggregation \citep{Bazzurro1999} to the hazard curves of the EDPs,
we show that the relationship between the EDPs and the seismic sources can be clearly analyzed.

\section{Methods}

In this section, we first present the proposed waveform-based PSHA formulation for evaluating the distribution of ground-motion waveforms.
We then describe the computational methods used to perform the integration in waveform-based PSHA.
Furthermore, we demonstrate a method for calculating the exceedance probabilities of IMs from the hazard analysis results.

\subsection{Proposed waveform-based PSHA formulation}

In conventional PSHA, seismic hazard is expressed in terms of exceedance probability.
However, defining the exceedance probability of ground-motion waveforms is challenging, as the target level for exceedance cannot be clearly specified.
We propose to evaluate the seismic hazard in the form of probability distribution of ground-motion waveforms as follows:
\begin{equation}
  p(\mathbf{g}\mid\mathbf{s}) = \iint p(\mathbf{g}\mid\mathbf{m}, \mathbf{r}, \mathbf{s})p(\mathbf{m}\mid\mathbf{s})p(\mathbf{r}\mid\mathbf{m}, \mathbf{s})\mathrm{d}\mathbf{m}\mathrm{d}\mathbf{r}
  \label{eq:psha_gmgm}
\end{equation}
in which $\mathbf{g}\in\mathbb{R}^{M\times L}$ is an $M$-component ground-motion waveform with $L$ sampled time steps.
$\mathbf{m}$, $\mathbf{r}$, and $\mathbf{s}$ are vector-valued random variables representing information regarding source, path, and site, respectively.
In conventional PSHA, magnitude and distance are used as the source and path characteristics, respectively.
Here, to provide a more general formulation, we define $\mathbf{m}$ and $\mathbf{r}$ as vectors of random variables that include
multiple attributes to be considered, for example, $\mathbf{m} = [\text{magnitude}, \text{hypocenter coordinates}, \text{focal mechanism}]$.
Although the feasibility of actually modeling such probability distributions is uncertain, this issue is not considered here,
as the focus is on presenting a generalized formulation.

In equation \ref{eq:psha_gmgm}, the probability distribution $p(\mathbf{g}\mid\mathbf{s})$ represents the distribution of ground-motion waveforms at a site of interest. The vector of random variables $\mathbf{s}$ is assumed to include not only site characteristics such as $V_{\mathrm{S}30}$, but also deterministic information such as the geographic location of the site.
Accordingly, $p(\mathbf{m}\mid\mathbf{s})$ represents the seismicity around the site, and $p(\mathbf{r}\mid\mathbf{m}, \mathbf{s})$ represents information related to attenuation properties and geographic relationships between the sources and the site.
The term $p(\mathbf{g}\mid\mathbf{m}, \mathbf{r}, \mathbf{s})$ represents the distribution of ground-motion waveforms conditioned on
the source, path, and site characteristics.
This can be modeled by constructing a GMGM whose conditional labels include $\mathbf{m}$, $\mathbf{r}$, and $\mathbf{s}$.
It should be noted that a GMGM is not the only possible approach; for example, simulation-based models capable of representing ground-motion variability could also be employed, but such considerations are left for future work.
Based on this setting, a set of ground-motion waveforms following the distribution $p(\mathbf{g}\mid\mathbf{s})$ can be
interpreted as representing the seismic hazard at the site of interest.

\begin{algorithm}[t]
  \caption{Sampling ground-motion waveforms based on equation (\ref{eq:psha_gmgm})}
  \label{alg:mcs_algo}
  \begin{algorithmic}[1]
    \Require Time period $t$; number of seismic sources $N_s$; target site condition $\mathbf{s}_\ast$; target number of waveforms $N_{w}$; generator $G$ (joint or conditional); distance metric $d(\cdot,\cdot)$; tolerance $\epsilon$;
    and earthquake occurrence models for each source.
    \Ensure Set of ground-motion waveforms $\mathcal{G}_{wav}$; total number of Monte Carlo cycles $c_{sim}$;
    cumulative event counts $(c_i)_{i=1}^{N_s}$, in which $c_i$ is the total number of earthquakes from source $i$ across all Monte Carlo cycles.
    \Ensure Ground-motion waveform $\mathbf{g}_{i, j, k}\in\mathcal{G}_{wav}$, denoting the $k$-th waveform from source $i$ in Monte Carlo cycle $j$; at generation time $j = c_{sim}$ and $k = c_i$.

    \State $c \gets 0$;\quad $c_{sim} \gets 0$;\quad $\mathcal{G}_{wav} \gets \varnothing$

    \For{$i \gets 1$ to $N_s$}
      \State $c_i \gets 0$
    \EndFor

    \While{$c < N_{w}$}
      \State $c_{sim} \gets c_{sim} + 1$
      \For{$i \gets 1$ to $N_s$}
        \State Sample the event counter $s_i$ for source $i$ from its occurrence model
        \If{$s_i = 0$}
          \State \textbf{continue}
        \EndIf
        \For{$\ell \gets 1$ to $s_i$}
          \State $c_i \gets c_i + 1$
          \State Sample $\mathbf{m}_\ast \sim p(\mathbf{m}\mid\mathbf{s}_\ast)$
          \State Sample $\mathbf{r}_\ast \sim p(\mathbf{r}\mid\mathbf{m}_\ast, \mathbf{s}_\ast)$
          \State Sample $\mathbf{z} \sim p_{\mathcal{Z}}(\mathbf{z})$
          \State $\mathbf{y}_\ast \gets (\mathbf{m}_\ast, \mathbf{r}_\ast, \mathbf{s}_\ast)$
          \If{$G$ is a joint model}
            \While{true}
            \State Generate $(\mathbf{g}_{i, j, k}, \mathbf{y}) \gets G(\mathbf{z})$
            \If{$d(\mathbf{y}, \mathbf{y}_\ast) \le \epsilon$}
            \State $\mathcal{G}_{wav} \gets \mathcal{G}_{wav} \cup \{\mathbf{g}_{i, j, k}\}$
              \State $c \gets c + 1$
              \State \textbf{break}
            \Else
              \State Sample $\mathbf{z} \sim p_{\mathcal{Z}}(\mathbf{z})$
            \EndIf
            \EndWhile
          \Else
            \State Generate $\mathbf{g}_{i, j, k} \gets G(\mathbf{z}, \mathbf{y}_\ast)$
            \State $\mathcal{G}_{wav} \gets \mathcal{G}_{wav} \cup \{\mathbf{g}_{i, j, k}\}$
            \State $c \gets c + 1$
          \EndIf
        \EndFor
      \EndFor
    \EndWhile
  \end{algorithmic}
\end{algorithm}

\subsection{Computational methods for integration}

When using GMGMs to evaluate the distribution $p(\mathbf{g}\mid\mathbf{m}, \mathbf{r}, \mathbf{s})$,
it is generally possible to obtain samples that follow this distribution, but not to evaluate the probability density $p(\mathbf{g}\mid\mathbf{m}, \mathbf{r}, \mathbf{s})$
itself. Therefore, the analytical evaluation of equation (\ref{eq:psha_gmgm}) is intractable, and
equation (\ref{eq:psha_gmgm}) is evaluated using Monte Carlo integration \citep{Assatourians2013}.

Algorithm \ref{alg:mcs_algo} presents the details of the Monte Carlo integration of the proposed method.
The overall flow of the Monte Carlo simulation (MCS) is presented here,
whereas more specific formulations for practical applications
are described in the \nameref{sec:ne} section for each problem setting.
The MCS is generally performed in three steps.
The first step is the generation of an earthquake catalog for given source geometries and seismicity parameters.
In this step, to provide a more general formulation,
not only the stationary Poisson distribution but also time-dependent earthquake occurrence models such as the Brownian passage time (BPT) distribution \citep{Matthews2002}
are considered.
Therefore, seismic hazard is evaluated directly over a time period of $t$ years, rather than as an annualized hazard.
For each seismic source $E_i$ $(i = 1, \cdots, N_s)$, the number of earthquake occurrences $s_i$ within the $t$-year period
is simulated based on the specified earthquake occurrence model.
Subsequently, $s_i$ samples are drawn from the distribution $p(\mathbf{m}\mid\mathbf{s}_\ast)$ for the given site condition $\mathbf{s}_\ast$, and an earthquake catalog is constructed.

The second step is the sampling of $\mathbf{r}$.
Based on the sampled source characteristics $\mathbf{m}_\ast$ and given site condition $\mathbf{s}_\ast$,
a sample following the distribution $p(\mathbf{r}\mid\mathbf{m}, \mathbf{s})$ is obtained.
Here, we assumed that samples can be drawn from the distributions $p(\mathbf{m}\mid\mathbf{s})$ and $p(\mathbf{r}\mid\mathbf{m}, \mathbf{s})$.
Such an assumption is common in probabilistic modeling and, as shown in the \nameref{sec:ne} section, also holds in practical implementations of PSHA.

The final step is the sampling of ground-motion waveforms based on the sampled $\mathbf{m}_\ast$, $\mathbf{r}_\ast$, and the given site condition $\mathbf{s}_\ast$.
In the case of the GMGMs based on GANs, sampling is performed through the generator of the GAN. Let the generators of the S-GMGM, CW-GMGM, and CS-GMGM be denoted as $G_s$, $G_{cw}$, and $G_{cs}$, respectively.
For the S-GMGM, which models the joint distribution, a noise vector $\mathbf{z}\sim p_{\mathcal{Z}}(\mathbf{z})$ is used as the input to generate data as follows:
\begin{equation}
  \mathbf{g}, \mathbf{y} = G_{s}(\mathbf{z})
  \label{eq:sgmgm_gen}
\end{equation}
in which $\mathbf{y}$ is the conditional labels and is a vector containing $\mathbf{m}$, $\mathbf{r}$, and $\mathbf{s}$.
Although the generator in a GAN is a deterministic function, it is trained as a mapping function that transforms the prior distribution $p_{\mathcal{Z}}(\mathbf{z})$ into the data distribution.
Therefore, the distribution of the generated samples follows a distribution approximates the distribution of the observed records.
The conditional distribution is evaluated using rejection sampling \citep{Tavare1997}, as described in \cite{matsumoto2024ssgmgm}:
\begin{equation}
  p(\mathbf{g}\mid\mathbf{y}_\ast) \simeq
  \int_{\mathcal{Z}}\int_{\mathcal{Y}}
  p(\mathbf{g}, \mathbf{y}\mid\mathbf{z})
  p_{\mathcal{Z}}(\mathbf{z})
  \mathbbm{1}\left[d(\mathbf{y}, \mathbf{y}_\ast) \le \epsilon\right]
  \mathrm{d}\mathbf{y}\mathrm{d}\mathbf{z}
\end{equation}
in which $p(\mathbf{g}, \mathbf{y}\mid\mathbf{z})$ is modeled by the S-GMGM, $\mathbbm{1}[\cdots]$ is the indicator function, $d(\cdot, \cdot)$ is a distance metric, and $\epsilon > 0$ is a tolerance.
In practice, a sample that follows the conditional distribution is obtained by repeatedly generating ground-motion waveforms using equation (\ref{eq:sgmgm_gen}) until the condition $d(\mathbf{y}, \mathbf{y}_\ast) \le \epsilon$ is satisfied.

For the CW-GMGM and CS-GMGM, the ground-motion waveforms are generated using the noise vector $\mathbf{z}$ and conditional label $\mathbf{y}$ as inputs:
\begin{equation}
  \mathbf{g} = G_{cw}(\mathbf{z}, \mathbf{y}), \qquad \mathbf{g} = G_{cs}(\mathbf{z}, \mathbf{y})
  \label{eq:cwgmgm_gen}
\end{equation}
As in the case of the S-GMGM, the distribution of the noise vector is transformed into the data distribution through the generator, and the ground-motion waveforms generated by equation \ref{eq:cwgmgm_gen} follow the distribution $p(\mathbf{g}\mid\mathbf{m}, \mathbf{r}, \mathbf{s})$.
Accordingly, in the CW-GMGM and CS-GMGM, a ground-motion waveform following the conditional distribution 
$p(\mathbf{g}\mid\mathbf{m}, \mathbf{r}, \mathbf{s})$ can be obtained simply by sampling one noise vector and inputting it into the generator together with $\mathbf{m}_\ast$, $\mathbf{r}_\ast$, and $\mathbf{s}_\ast$.

The above three steps are repeated until the target number of ground-motion waveforms, $N_w$, is obtained.
To maintain consistency among the sampled conditions, the MCS is not terminated immediately when $c = N_w$, but rather after completing the entire MCS cycle in which $c=N_w$ is reached.  Consequently, the total number of sampled ground-motion waveforms does not necessarily equal $N_w$, and it holds that $|{\mathcal{G}_{wav}}| \ge N_w$.
The resulting set of ground-motion waveforms:
\begin{equation}
  \mathcal{G}_{wav} = \{\mathbf{g}_{i, j, k}\mid i=1, \cdots, N_s, j = 1, \cdots, c_i, k= 1, \cdots, c_{sim}\}
\end{equation}
represents an ensemble of ground-motion waveforms following the distribution $p(\mathbf{g}\mid\mathbf{s})$ in equation \ref{eq:psha_gmgm}.
The ground-motion waveforms with the same subscript $i$ represent the waveforms associated with source $i$.
The subscript $k$ indicates the MCS cycle in which each waveform was obtained.
The number of waveforms $\mathbf{g}_{i, j, k}$ with the same subscripts $i$ and $k$ corresponds to the number of earthquakes that occurred from the source $i$ during the $k$-th MCS cycle.
By keeping track of the number of MCS cycles $c_{sim}$, weighting information that accounts for the earthquake occurrence probability associated with each waveform is also retained.
For example, the earthquake occurrence rate $\nu_i$ of the source $i$ over the $t$-year period can be evaluated as:
\begin{equation}
  \nu_i = \frac{c_i}{c_{sim}}
\end{equation}
in which $c_i$ is the total number of earthquakes from the source $i$.
These three quantities, $\mathcal{G}_{wav}$, $c_{sim}$, and $c_i$, together provide the basis for the waveform-based PSHA proposed in this study.

\subsection{Calculation of IM exceedance probability}\label{subsec:IM_exceed}

Hazard curves for IMs can be computed from the sampled waveforms $\mathcal{G}_{wav}$.
Let $IM_{i, j, k}$ be the IM calculated from the ground-motion waveform $\mathbf{g}_{i, j, k}\in\mathcal{G}_{wav}$,
and let $P_i(IM > x; t)$ be the $t$-year exceedance probability
that the IM exceeds an intensity level $x$ for source $i$
We assume that $\{IM_{i, j, k}\}_{k=1}^{c_{sim}}$ are independent and identically distributed (i.i.d.) across $k$.
The estimator $\hat{P}_i(IM > x; t)$ for $P_i(IM > x; t)$ can be obtained as:
\begin{equation}
  P_i(IM > x; t) \simeq \hat{P}_i(IM > x; t) = \frac{1}{c_{sim}}\sum_{k=1}^{c_{sim}}\left(1 - \prod_{j\in J_i(k)}\mathbbm{1}\left[
      IM_{i, j, k} \le x
    \right]\right)
\end{equation}
in which $J_i(k)$ is the set of indices $j$ corresponding to waveforms with the same subscripts $i$ and $k$.
Since we consider both the Poisson and time-dependent earthquake occurrence models here, the $t$-year exceedance probability $P(IM > x; t)$ for all target sources can be computed using the complementary event formulation as:
\begin{equation}
  P(IM > x; t) \simeq \hat{P}(IM > x; t) = 1 - \prod_{i=1}^{N_s}\left\{1 - \hat{P}_i(IM > x; t)\right\}
\end{equation}
in which $\hat{P}(IM > x; t)$ is the estimator for $P(IM > x; t)$.

The coefficient of variation (COV) is a widely used metric for evaluating the uncertainty of exceedance probability estimates obtained from MCS (e.g., \cite{Au2003subset}; \cite{Houng2025}).
It is well known that MCS estimator $\hat{P}(IM > x; t)$ is unbiased, and its expectation is given by:
\begin{equation}
\mathbb{E}\left[\hat{P}(IM > x; t)\right] = P(IM > x; t)
\end{equation}
while the variance of $\hat{P}(IM > x; t)$ is
\begin{equation}
\mathrm{var}\left[\hat{P}(IM > x; t)\right] = \frac{P(IM > x; t)(1 - P(IM > x; t))}{c_{sim}}
\end{equation}
Therefore, the COV is calculated as:
\begin{equation}
\mathrm{CV}\left(\hat{P}(IM > x; t)\mid P(IM > x; t), c_{sim}\right) = \sqrt{
  \frac{1 - P(IM > x; t)}{c_{sim}P(IM > x; t)}
} \label{eq:cv_ff}
\end{equation}
In the numerical experiments conducted in this study,
the true value of $P(IM > x; t)$ cannot be evaluated.
Therefore, equation (\ref{eq:cv_ff}) is evaluated by using the value of $\hat{P}(IM > x; t)$ instead.

 \begin{figure}[t]
   \centering
   \includegraphics[width=0.45\columnwidth]{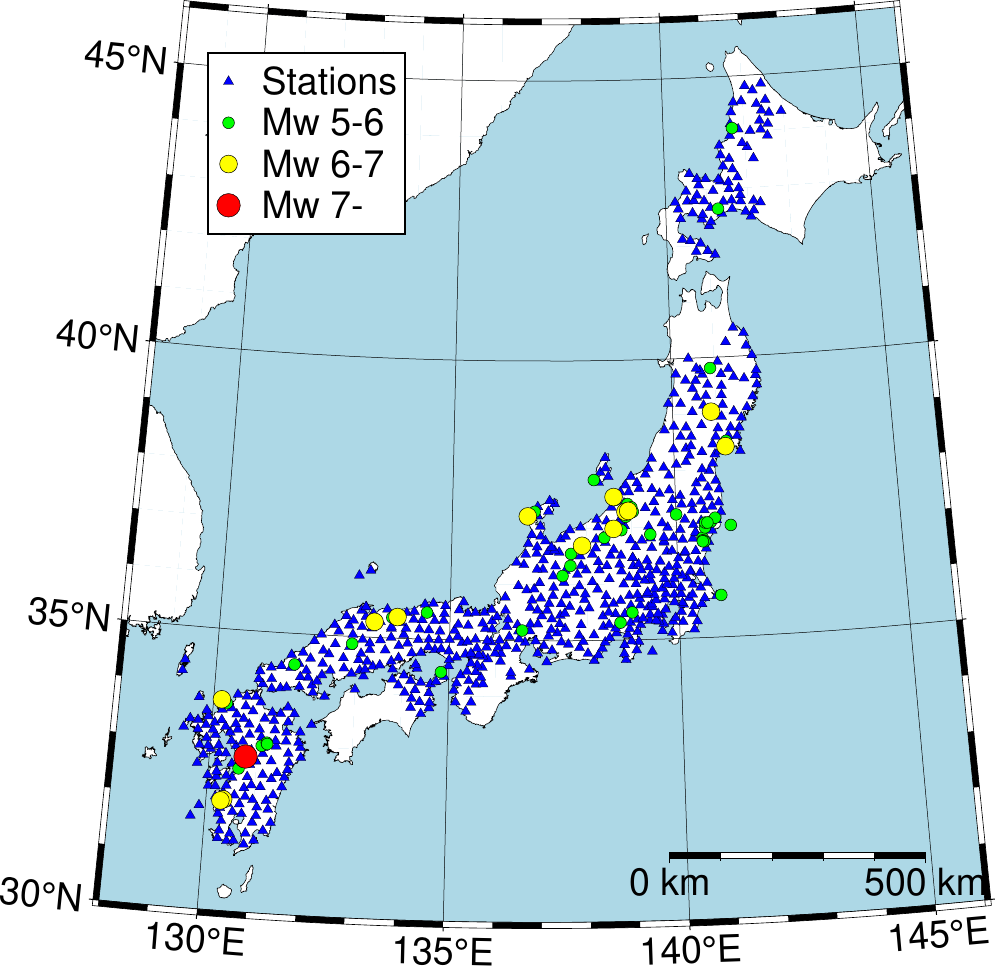}
   \caption{Locations of the earthquake epicenters (circles) and stations (triangles).}
   \label{fig:map_hyp_station}
 \end{figure}%

 \begin{figure}[t]
   \centering
   \includegraphics[width=0.45\columnwidth]{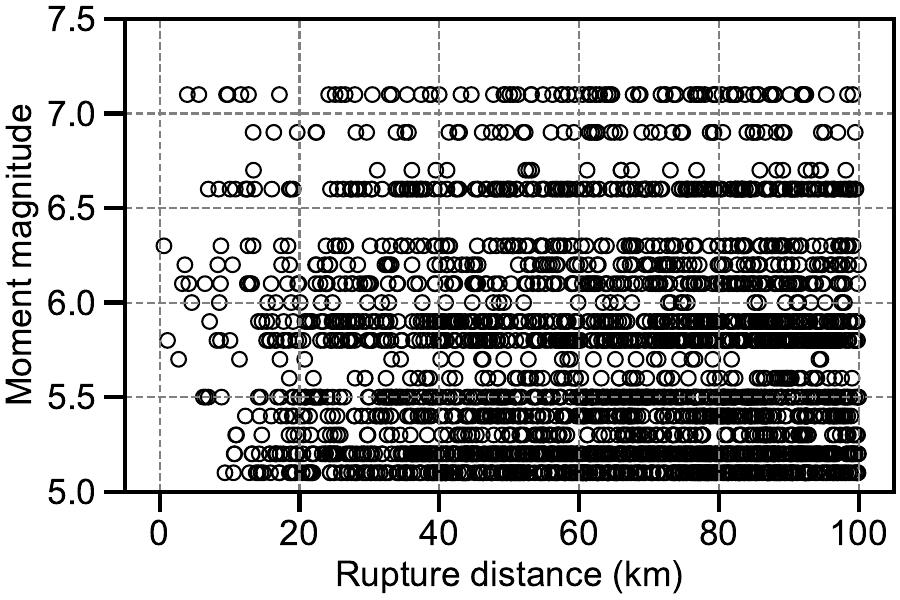}
   \caption{Magnitude-distance distribution of the training dataset.}
   \label{fig:mw_dist_scat}
 \end{figure}%

\section{GMGMs}
\subsection{Training dataset}

For training the three GMGMs (S-GMGM, CW-GMGM, and CS-GMGM), we use the dataset of crustal earthquakes in Japan compiled in our previous studies (\cite{matsumoto2023fundamental}; \cite{matsumoto2024ssgmgm}).
Strong-motion observed records were collected from K-NET stations \citep{KNET} for earthquakes that occurred within the Eurasian Plate between 1997 and 2016, with $M_W > 5$ and the rupture distance $R_{\mathrm{RUP}} \le 100$ km.
$M_W$ was determined using the moment tensor solutions from the F-net (Full Range Seismograph Network of Japan) database \citep{Fnet}.
$R_{\mathrm{RUP}}$ was calculated as the shortest distance from the rupture area to the station.
When $M_W$ was sufficiently small that the earthquake could be considered a point source,
$R_{\mathrm{RUP}}$ was calculated as the hypocentral distance.
The two horizontal components of ground motions were treated independently.
To increase the amount of training data, the ground-motion time histories were rotated in the horizontal plane at
45-degrees intervals, and the separated components were used after removing duplicates.
As a result, a total of 21,696 one-component horizontal ground-motion waveforms from 62 earthquakes were obtained.
As a proxy for site characteristics, $V_{\mathrm{S}20}$ was calculated from the P-S logging results provided in the K-NET database,
and $V_{S\mathrm{30}}$ was then estimated using the empirical formula proposed by \cite{Kanno2006}:
\begin{equation}
  V_{\mathrm{S}30} = 1.13V_{\mathrm{S}20} + 19.5
\end{equation}
The locations of the earthquake epicenters and stations are shown in Figure \ref{fig:map_hyp_station},
and magnitude-distance distribution is shown in Figure \ref{fig:mw_dist_scat}.
Further details on the training dataset are provided in \cite{matsumoto2023fundamental} and \cite{matsumoto2024ssgmgm}.

Several preprocessing steps were applied to the ground-motion waveforms.
Each record was aligned with respect to the P-wave arrival time, and signals caused by aftershocks were removed by visually inspecting the waveforms.
The S-GMGM and CS-GMGM \citep{matsumoto2024ssgmgm} were modeled using ground-motion data with a length of $L = 8192$ time steps,
whereas the CW-GMGM \citep{Florez2022} used data with a length of $L=4000$.
For each case, the end of the record was truncated, a cosine taper was applied to the final 100 steps,
and 100 zeros were appended to both the beginning and the end of the record to match the required data length.
The sampling frequency was kept at 100 Hz, and no bandpass filter was applied.
Each waveform amplitude was normalized by its PGA, and the PGA value was included as one of the conditional label elements.
Accordingly, the conditional label is defined as the four-dimensional vector $\mathbf{y} = [M_W, R_{\mathrm{RUP}}, V_{\mathrm{S}30}, \mathrm{PGA}]^{\mathrm{T}}$,
in which the natural logarithm is applied to $R_{\mathrm{RUP}}$, and the base-10 logarithm to $V_{\mathrm{S}30}$ and PGA.
Finally, each component of $\mathbf{y}$ was normalized so that the mean and standard deviation across the entire dataset were 0 and 0.1, respectively.

\subsection{DNN architectures}

\begin{figure}[t]
  \centering
  \includegraphics[width=\columnwidth]{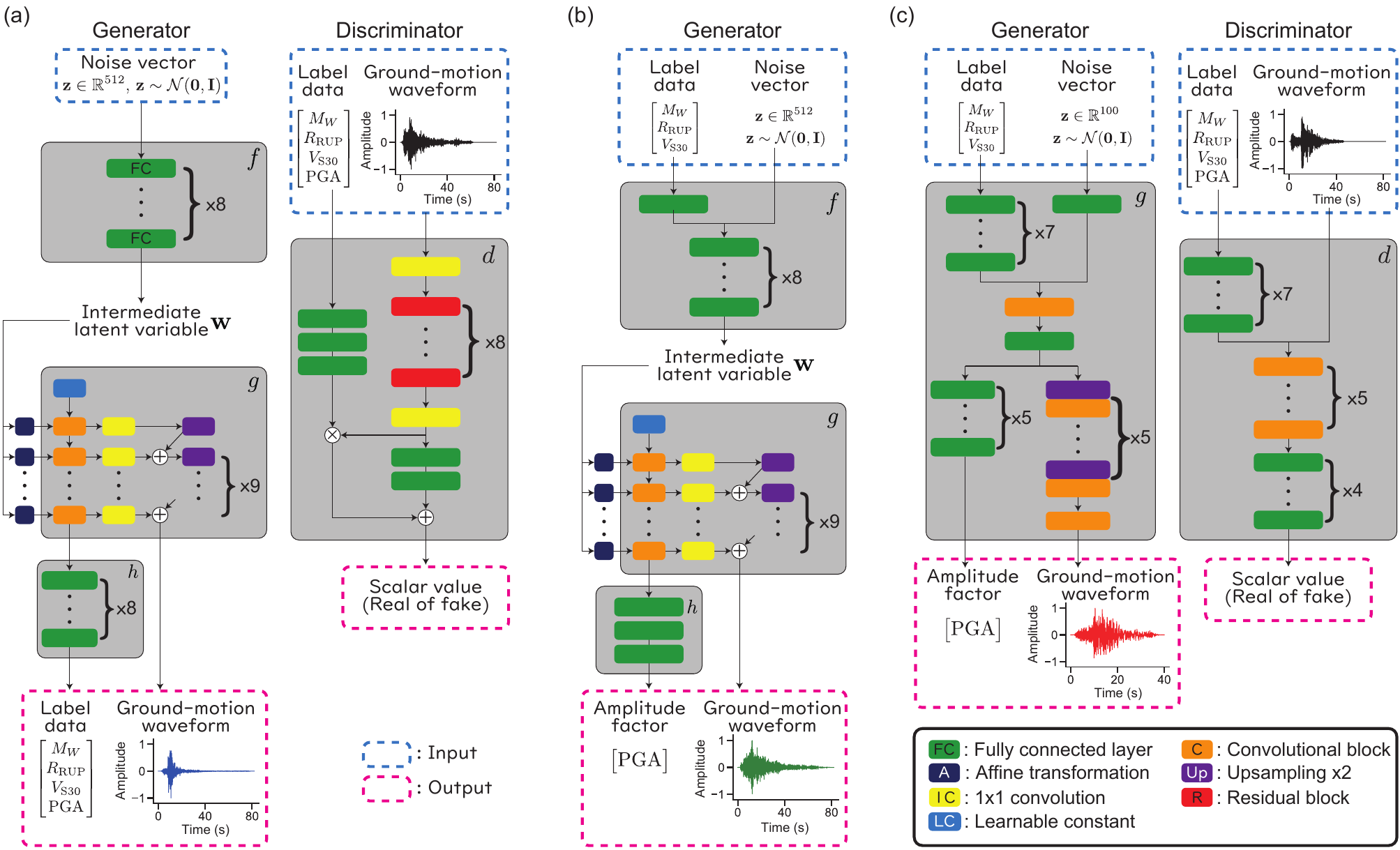}
  \caption{Diagrams of the DNN architectures of the GMGMs: (a) S-GMGM, (b) CS-GMGM, and (c) CW-GMGM.
  The discriminator of the CS-GMGM has the same architecture as that of the S-GMGM.}
  \label{fig:model_configurations}
\end{figure}

An overview of the DNN architectures of the S-GMGM, CW-GMGM, and CS-GMGM is shown in Figure \ref{fig:model_configurations}.
The DNN architecture of \cite{matsumoto2024ssgmgm} was employed for the S-GMGM, with only the dimensionality of the conditional label modified.
The generator of the S-GMGM takes a noise vector $\mathbf{z}\in\mathbb{R}^{512}$ as input and generates a ground-motion waveform $\mathbf{g}\in\mathbb{R}^{1\times 8192}$ and its conditional label $\mathbf{y}\in\mathbb{R}^4$.
The discriminator takes both $\mathbf{g}$ and $\mathbf{y}$ as input and outputs a scalar value.

The CW-GMGM adopts the same DNN architecture as \cite{Florez2022}.
We constructed the DNN using the program code provided in their GitHub repository.
The only modification was to the number of ground-motion components: while the original model handled three components,
the CW-GMGM was adjusted to handle a single component by modifying the corresponding DNN parameters.
The generator of the CW-GMGM takes a noise vector $\mathbf{z}\in\mathbb{R}^{100}$ and conditional label  $\mathbf{y}\in\mathbb{R}^4$ as inputs and generates a ground-motion waveform $\mathbf{g}\in\mathbb{R}^{1\times 4000}$.
The discriminator takes both the $\mathbf{g}$ and $\mathbf{y}$ as input and outputs a scalar value.
Although the discriminator in a WGAN is generally referred to as a critic, it is referred to as a discriminator in this study for consistency.

The CS-GMGM was constructed by modifying the generator of the S-GMGM into a cGAN framework.
Following \cite{Florez2022}, the conditional labels were embedded and concatenated with the noise vector before being input into the generator.
The discriminator architecture was identical to that of the S-GMGM.
The generator takes a noise vector $\mathbf{z}\in\mathbb{R}^{512}$ and conditional label $\mathbf{y}\in\mathbb{R}^4$ as inputs and generates a ground-motion waveform $\mathbf{g}\in\mathbb{R}^{1\times 8192}$,
while the discriminator takes $\mathbf{g}$ and $\mathbf{y}$ as inputs and outputs a scalar value.

The theoretical background of these GAN models can be found in \cite{Florez2022} and \cite{matsumoto2024ssgmgm},
and more detailed DNN configurations are available in our GitHub repository.

\subsection{Hyperparameters settings}

The hyperparameters for the S-GMGM and CS-GMGM were set to the same values as those used in \cite{matsumoto2024ssgmgm}.
The learning rate was 0.002, the batch size was 64, and the dimensions of both the noise vector and the intermediate latent variable were 512.
A standard normal distribution was used as the prior distribution of the noise vector.
The Adam optimizer \citep{kingma2017adam} was used for training.

For the CW-GMGM, several models were initially trained by modifying the hyperparameters proposed in \cite{Florez2022}, and their performance was compared.
As a result, only the batch size and the number of training epochs were changed, while the other hyperparameters were kept the same as in \cite{Florez2022}.
The learning rate was set to $1\times10^{-4}$, the batch size to 64, and the noise vector dimension to 100.
A standard normal distribution was used as the prior distribution of the noise vector.
The Adam optimizer \citep{kingma2017adam} was also used for training.

All deep learning processes and the construction of DNNs were implemented using the Python library PyTorch \citep{Paszke2019}.
Additional details on the hyperparameter settings are available in our GitHub repository.

\subsection{Optimal Epoch Selection Method}\label{subsec:optimal}

Since the loss function of GANs depends on the training state of the discriminator,
it is difficult to use the loss for relative comparison of the model performance across epochs.
When applying GMGMs to PSHA, it is therefore necessary to establish an objective method for determining the optimal number of epochs.
In this subsection, we extend the model evaluation method proposed by \cite{matsumoto2023fundamental} and propose a new approach for relative performance evaluation across different epochs.

The model performance is assessed based on how well the generator approximates the distribution of the training dataset.
Let the training dataset be $\mathcal{D}_o = \{\mathbf{g}_i^{(o)}\mid i= 1, \cdots, N_o\}$, and the set of ground-motion waveforms generated by the GMGM at epoch $l$ as $\mathcal{D}_{g, l} = \{\mathbf{g}_j^{(g, l)}\mid j=1, \cdots, N_g\}$.
Each ground-motion waveform is assumed to be an i.i.d. sample drawn from the probability distributions of the observed records $p_o$ and the generator $p_{g, l}$, respectively.
Using a feature vector $\boldsymbol{\xi}$ that characterizes the ground-motion waveform, the similarity between $p_o$ and 
$p_{g, l}$ is approximately evaluated by the Sinkhorn divergence (\cite{Cuturi2013}; \cite{Feydy2019}) as follows:
\begin{align}
\mathrm{S}_\epsilon(p_o, p_{g, l}) \simeq \mathrm{OT}_\epsilon(\hat{p}_{o}, \hat{p}_{g, l}) - \frac{1}{2}\mathrm{OT}_{\epsilon}(\hat{p}_o, \hat{p}_o)
- \frac{1}{2}\mathrm{OT}_\epsilon(\hat{p}_{g, l}, \hat{p}_{g, l})
\label{eq:sinkhorn}
\end{align}
in which $\hat{p}_o$ and $\hat{p}_{g, l}$ are empirical measures represented by the set of
feature vectors $\boldsymbol{\xi}$
calculated from $\mathcal{D}_o$ and $\mathcal{D}_{g, l}$, respectively.
The term $\mathrm{OT}_\epsilon$ is defined as:
\begin{equation}
    \mathrm{OT}_\epsilon(\alpha, \beta) := \min_{\pi_1=\alpha, \pi_2=\beta}
  \int_{\mathcal{X}^2}C\mathrm{d}\pi + \epsilon\int_{\mathcal{X}^2}\log\left(
    \frac{\mathrm{d}\pi}{\mathrm{d}\alpha\mathrm{d}\beta}
  \right)\mathrm{d}\pi
  \label{eq:ot_main}
\end{equation}
in which $\mathcal{X}$ is the data space, $\epsilon > 0$ is the temperature parameter for entropy regularization, and $C$ is the cost function. We set $C$ as follows:
\begin{equation}
  C(\mathbf{x}_1, \mathbf{x}_2) = \|\mathbf{x}_1 - \mathbf{x}_2\|_2
\end{equation}
The Sinkhorn divergence is an optimal transport-based metric that enables statistically stable and computationally efficient evaluation even for high-dimensional data.
The value $S_{\epsilon} \ge 0$ becomes zero when $p_o = p_{g, l}$, and smaller values indicate better approximation of the distribution.
Accordingly, by selecting the elements of $\boldsymbol{\xi}$ based on the purpose of the GMGM development and evaluating the approximation accuracy of their joint distribution,
the model performance at a given epoch can be quantitatively assessed.

\begin{table}[t]
  \centering
  \caption{List of the indices of ground-motion characteristics used in $\boldsymbol{\xi}$}
  \small
  \label{tab01}
  {\begin{tabular*}{\columnwidth}{@{\extracolsep{\fill}}lllll@{}} \toprule %
  \textbf{Parameter} & \textbf{Unit} & \textbf{Notation} & \textbf{Computational form} \\
  \midrule
  Peak ground acceleration (PGA) & cm/s${}^2$ & $\xi_1$ & $\log_{10}(\xi_1)$\\
  Peak ground velocity (PGV) & cm/s & $\xi_2$ & $\log_{10}(\xi_2)$ \\
  Ratio of PGV to PGA \citep{Malhotra1999} & s & $\xi_3$ & $\log_{10}(\xi_3)$ \\
    Instrumental seismic intensity \citep{jmaIntensity} & - & $\xi_4$ & $\xi_4$ \\
    Predominant frequency \citep{Rathje1998} & Hz & $\xi_5$ & $\xi_5$ \\
    Zero-level crossing rate \citep{Rezaeian2008} & /s & $\xi_6$ & $\xi_6$ \\
    Negative maxima and positive minima \citep{Rezaeian2010} & /s & $\xi_7$ & $\xi_7$ \\
    Significant duration, $D_{5-95}$ \citep{Trifunac1975duration} & s & $D_{5-95}$ & $D_{5-95}$ \\
    Significant duration, $D_{5-45}$ \citep{Rezaeian2010} & s & $D_{5-45}$ & $D_{5-45}$ \\
    Arias intensity \citep{arias1970measure} & cm/s & $I_A$ & $\log_{10}(I_A)$ \\
    Spectrum intensity \citep{Housner1961} & cm & $SI$ & $\log_{10}(SI)$ \\
    Mean period \citep{Rathje1998} & s & $T_m$ & $T_m$ \\
  \bottomrule
  \end{tabular*}}
\end{table}%


In this study, as an example, twelve indices that can be calculated from ground-motion waveforms were adopted as the elements of $\boldsymbol{\xi}$.
Table \ref{tab01} shows the list of indices.
These indices were selected to evaluate various characteristics of ground motions, including peak amplitude, power, duration, and frequency content.
The specific definitions and calculation methods for each index are provided in \nameref{app:char}.
The Sinkhorn divergence was computed using the Python library GeomLoss \citep{Feydy2019}, and details of the implementation are available in our GitHub repository.

We first evaluated the S-GMGM.
The maximum number of training iterations was set to 100,000, and the generator parameters were saved every 100 iterations (
$l = 100, 200, \cdots, 100000$), resulting in 1,000 models corresponding to different training states.
Here, an iteration refers to a single completion of the training loop.
Then, 100,000 random noise vectors $\mathbf{z}$ were sampled to generate the set $\mathcal{D}_{g, l}$
and the corresponding set of conditional labels $\mathcal{Y}_{g, l}$.
For both $\mathcal{D}_{o}$ and $\mathcal{D}_{g, l}$,
feature vectors $\{\boldsymbol{\xi}_i^{(o)}\}_{i=1}^{N_o}$ and $\{\boldsymbol{\xi}_{j}^{(g, l)}\}_{j=1}^{N_g}$ were computed. As shown in Table \ref{tab01},
several indices were transformed to the base-10 logarithmic scale, and each element was standardized using the mean and standard deviation of $\{\boldsymbol{\xi}_i^{(o)}\}$.
The similarity between distributions was then evaluated using equation (\ref{eq:sinkhorn}).
Among iterations with high similarity, we visually inspected the generated waveforms, spectra, and distributional characteristics following the procedure in \cite{matsumoto2024ssgmgm},
and finally selected the model at 35,300 iterations.

Next, the CW-GMGM and CS-GMGM were evaluated.
To ensure consistent analysis conditions, the conditional labels $\mathcal{Y}_{g, l}$ generated by the
optimal S-GMGM were used for data generation.
Thus, for both the CW-GMGM and CS-GMGM, $N_g = 100,000$.
For the CW-GMGM, the maximum number of epochs was set to 1,000, and the model was saved every 10 epochs.
For the CS-GMGM, the maximum number of iterations was set to 100,000, and the model was saved every 100 iterations.
The same evaluation procedures as those used for the S-GMGM were applied, and the final models were selected at 250 epochs for the CW-GMGM and 42,900 iterations for the CS-GMGM.

\begin{figure}[t]
  \centering
  \includegraphics[width=\columnwidth]{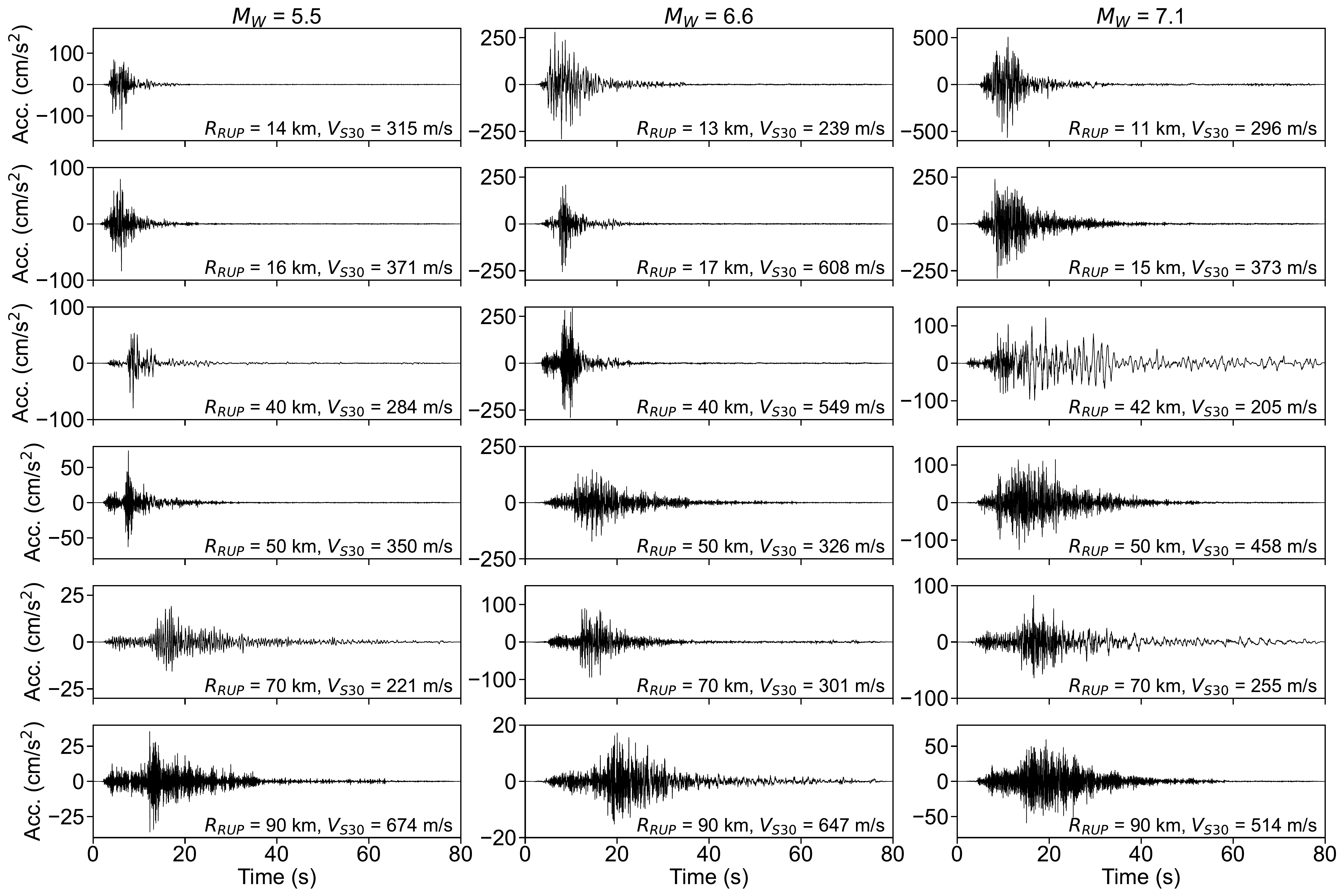}
  \caption{Examples of ground-motion acceleration waveforms generated by the S-GMGM.
    The waveform in each column correspond to the value of $M_W$ shown at the top.
    Each panel shows the associated $R_{\mathrm{RUP}}$ and $V_{\mathrm{S}30}$ values.}
  \label{fig:gen_wave_sgmgm_with_mw}
\end{figure}%

\begin{figure}[t]
  \centering
  \includegraphics[width=\columnwidth]{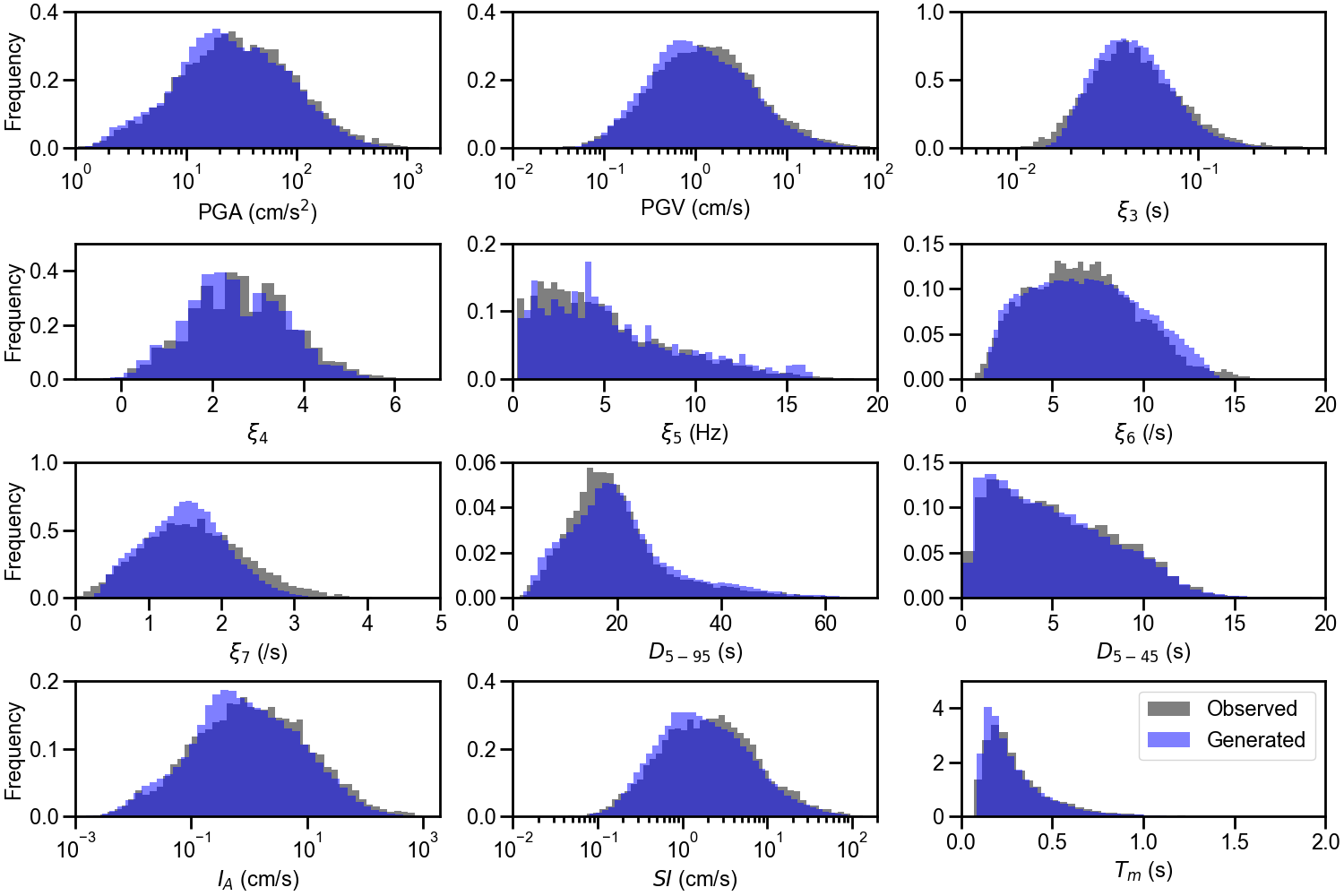}
  \caption{Comparison of the distributions of ground-motion characteristic indices in Table \ref{tab01} between the observed records and ground motions generated by the S-GMGM.}
  \label{fig:hist_feature_sgmgm}
\end{figure}%

\subsection{Training results}\label{subsec:training_r}

In this subsection,
we examine the generated results of the optimal GMGMs.
Although verifying the performance of the GMGMs is important for their application to PSHA, a detailed examination is not the main focus of this paper.
Therefore, only a portion of the generated results for the S-GMGM is presented in the main text, while additional results for the S-GMGM, as well as the results for the CW-GMGM and CS-GMGM, are provided in the supplemental material to this article.

For the S-GMGM, 100,000 samples of $\mathbf{z}$ were drawn from a standard normal distribution,
and 100,000 ground motions $\mathbf{g}$ and their corresponding conditional labels $\mathbf{y}$ were generated.
Several post-processing steps were applied to the ground motions.
After removing the offset so that the mean acceleration becomes zero,
a fourth-order Butterworth filter was applied with a 0.1 Hz low-frequency cutoff and a 20 Hz high-frequency cutoff.
A cosine taper was then applied to the first 100 steps and the last 200 steps of the waveform,
and the amplitude was restored by multiplying the PGA contained in the corresponding conditional label.
Examples of the acceleration waveforms with specific $M_W$, $R_{\mathrm{RUP}}$, and $V_{\mathrm{S}30}$ scenarios are shown in Figure \ref{fig:gen_wave_sgmgm_with_mw}.
Waveforms for different $M_W$, $R_{\mathrm{RUP}}$, and $V_{\mathrm{S}30}$ scenarios are shown in Figures S1 and S2. Waveforms generated by the CW-GMGM and CS-GMGM using the
same $M_W$, $R_{\mathrm{RUP}}$, and $V_{\mathrm{S}30}$ scenarios are also shown in Figures S3 to S5 and Figures S6 to S8, respectively.
The post-processing procedures for the waveforms are the same as those for the S-GMGM.
The generated waveforms capture characteristics such as the onset of P and S waves, envelope shapes, and the magnitude and distance scaling.
They also capture site effects, such as the appearance of long-period vibrations under relatively soft site conditions and short-period vibrations under relatively hard site conditions.
Statistical comparisons of the Fourier amplitude spectra between the generated waveforms and the observed records are shown in Figures S9 to S11.
Comparisons of the distributions of the ground-motion characteristic indices listed in Table \ref{tab01} are shown in Figure \ref{fig:hist_feature_sgmgm} and Figures S12 to S13.
The PGV values were computed by obtaining the velocity waveforms through numerical integration after removing the long-period components using a fourth-order Butterworth filter with a 0.2 Hz low-frequency cutoff.
The SI values were calculated by computing the 5\% damped velocity response spectra using the Nigam-Jennings method \citep{Nigam1969}.
As in the previous study \citep{matsumoto2024ssgmgm}, it can be seen that the distributions of the generated waveforms are consistent with those of the observed records in terms of both temporal and frequency characteristics.

\begin{figure}[t]
  \centering
  \includegraphics[width=\columnwidth]{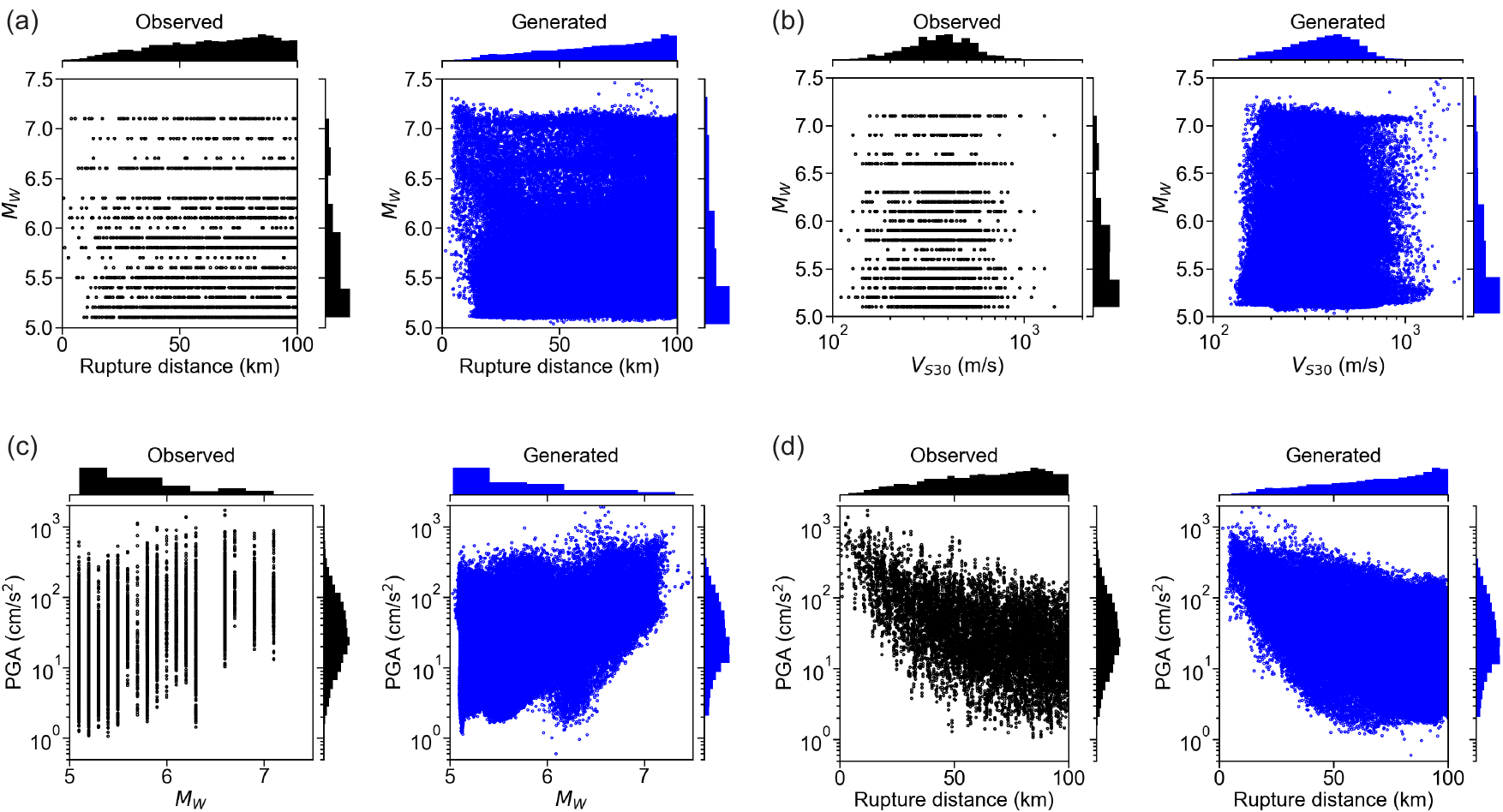}
  \caption{
    Comparison of the conditional label distributions between the observed records and those generated by the S-GMGM.
    (a) $R_{\mathrm{RUP}}$ and $M_W$, (b) $V_{\mathrm{S}30}$ and $M_W$, (c) $M_W$ and PGA, (d) $R_{\mathrm{RUP}}$ and PGA.
    In each case, the left panel shows the distribution of the observed records,
    and the right panel shows the distribution of the generated data.
  }
  \label{fig:scatter_label_dist2d}
\end{figure}%

\begin{figure}[t]
  \centering
  \includegraphics[width=\columnwidth]{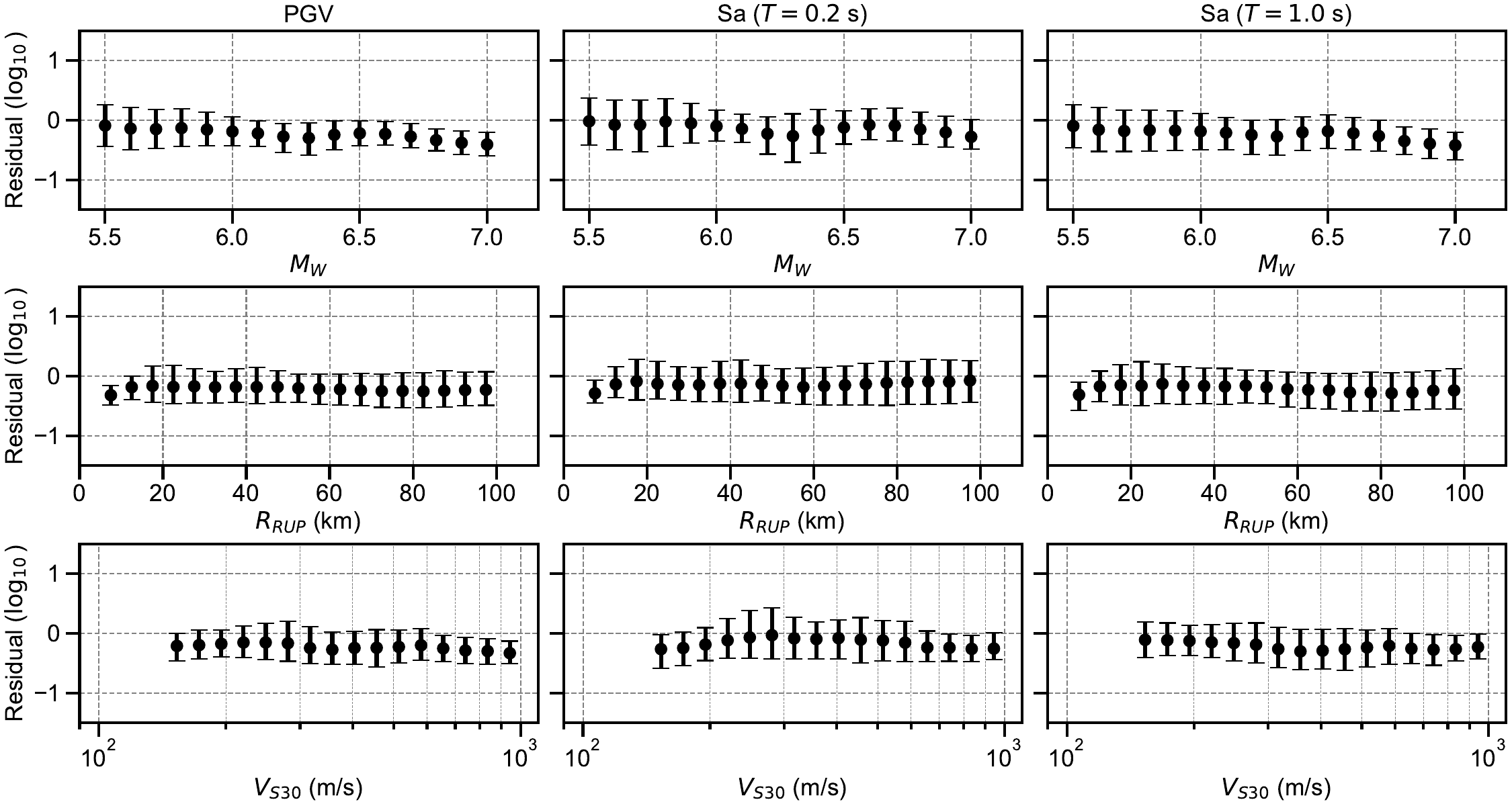}
  \caption{
    Residual plots comparing the S-GMGM with the MF13 GMM.
    The IM corresponding to each column is indicated at the top:
    PGV on the left, the 5\% damped spectral acceleration at a natural period of 0.2 s in the center,
    and the 5\% damped spectral acceleration at a natural period of 1.0 s on the right.
    Each bar shows the median and the 16th and 84th percentiles of the residuals.
  }
  \label{fig:residual_sgmgm_mf}
\end{figure}%

Next, for the 100,000 conditional labels generated by the S-GMGM, the relationships among the labels are compared
in Figure \ref{fig:scatter_label_dist2d}.
The distributions of the conditional labels are well matched with those of the observed records.
In addition, Figures \ref{fig:scatter_label_dist2d} (a) and (b) show the ranges of magnitude, distance, and $V_{\mathrm{S}30}$ that can be generated by the S-GMGM, and can be interpreted as figures indicating the applicability range for PSHA.
It is difficult to evaluate the applicability of the cGAN-based models, CW-GMGM and CS-GMGM,
to PSHA based on the range of their generated conditional labels,
because these models require the conditional labels to be specified during data generation and do not learn the distribution of the labels themselves.
One advantage of the S-GMGM is that its applicability to PSHA can be examined by checking the distribution of the generated data.

Finally, the residuals between the GMGMs and GMMs are compared for PGV and for the 5\% damped spectral acceleration at natural periods of $T = 0.2$ s and $T = 1.0$ s.
The residuals were defined as the base-10 logarithm of the GMGM divided by the GMM.
The \cite{Si1999} (hereafter, SM99) GMM and the \cite{Morikawa2013} (hereafter, MF13) GMM, which are used in the development of the national seismic hazard map for Japan \citep{Fujiwara2023}, were employed, and four NGA West2 GMMs (\cite{abrahamson2014summary}, ASK14; \cite{boore2014nga}, BSSA14; \cite{campbell2014nga}, CB14; \cite{chiou2014update}, CY14) were also used for comparison.
For each GMM, only the terms that can be computed using the $M_W$, $R_{\mathrm{RUP}}$, and $V_{\mathrm{S}30}$ were employed. The specific formulations are presented in \nameref{app:B}.
As the IM, the maximum amplitude of the two horizontal components was used in the SM99 GMM,
RotD100 \citep{boore2010} was used in the MF13 GMM, and RotD50 \citep{boore2010} was used in the NGA West2 GMMs.
Figure \ref{fig:residual_sgmgm_mf} shows the residual plots between the S-GMGM and the MF13 GMM.
The residual plots between the S-GMGM and the other GMMs are shown in Figures S14 to S18, those for the CW-GMGM in Figures S19 to S24, and those for the CS-GMGM in Figures S25 to S30.
Note that the GMGMs treat the two horizontal components independently, whereas the MF13 GMM uses RotD100,
which leads to an apparent underestimation.
The analysis results are therefore qualitatively consistent when this modeling difference is taken into account.
In addition, the median residual is approximately constant over a wide range of $M_W$, $R_{\mathrm{RUP}}$, and $V_{\mathrm{S}30}$, indicating that the magnitude and distance scaling, as well as the amplification effects of the surface soil, are appropriately represented.

\section{Numerical experiments}\label{sec:ne}

We examine the proposed waveform-based PSHA using two numerical experiments.
The first example considers a hypothetical single area source, whereas the second example focuses on an actual site and the surrounding faults in Japan.
For each case, we present the set of waveforms sampled by Algorithm 1 and compare the resulting hazard estimates with those obtained from a conventional PSHA using GMMs to examine the validity of the proposed approach.
The PSHA procedure based on GMMs follows the methodology described in \citet{Fujiwara2023}.
Following common settings in engineering applications \citep{Gerstenberger2020},
the seismic hazard is evaluated for a 50-year period in this study.

After presenting and examining the hazard analysis results,
nonlinear dynamic response analyses are conducted for a five-degree-of-freedom (5DOF) building model
using the ground-motion waveforms sampled in numerical example 2.
Based on the building response time histories, the exceedance probabilities of the EDPs are evaluated.
We also show the hazard deaggregation can be performed targeting EDPs with a straightforward manner.
All analyses were performed using Python 3.12 on a Ryzen 9 9950X 16-Core Processor and 64 GB RAM.
An NVIDIA RTX 4000 Ada GPU was used to generate the ground-motion waveforms with the GMGMs.

\begin{figure}[t]
  \centering
  \includegraphics[width=0.49\columnwidth]{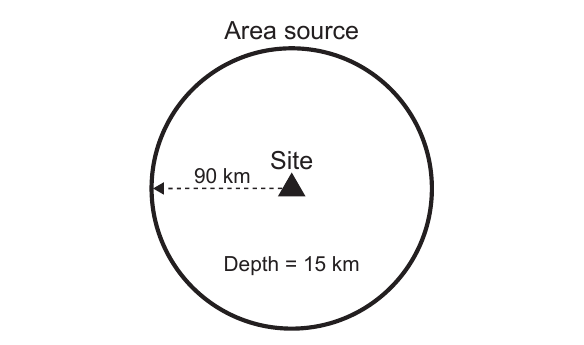}
  \caption{Seismic source geometry for the numerical example 1 with a single area source.}
  \label{fig:nu_example1}
\end{figure}%

\subsection{Example 1: single area source}

Figure \ref{fig:nu_example1} shows the setting of the target site and area source.
A circular source with a radius of 90 km is defined on a horizontal plane located 15 km beneath the site.
The value of $V_{\mathrm{S}30}$ at the site was set to 356 m/s.
The earthquake occurrence model is assumed to follow a stationary Poisson process,
with the minimum and maximum magnitudes set to $m_{\min} = 5.5$ and $m_{\max} = 6.8$, respectively.
The annual occurrence rate of earthquakes with magnitudes in the range $m_{\min} \le m \le m_{\max}$ is set to $\nu =0.5$.

Earthquakes are assumed to occur uniformly within the source.
Let $\hat{r}$ denote the horizontal distance from the center of the area source to the hypocenter.
The probability density function $p(\hat{r})$ is independent of magnitude and is given by:
\begin{equation}
  p(\hat{r}) = \frac{2\hat{r}}{R^2}, \quad 0 \le \hat{r} \le R \label{eq:ex1_dist_r}
\end{equation}
in which $R$ is the radius of the area source.
The hypocentral distance $r$ is computed as:
\begin{equation}
  r = \sqrt{\hat{r}^2 + 15^2}
\end{equation}
The probability density function $p(m)$ for magnitude $m$ is defined based on the Gutenberg-Richter (GR) law as a truncated exponential distribution with upper and lower bounds:
\begin{equation}
  p(m) = \frac{\beta}{1 - e^{-\beta(m_{\max} - m_{\min})}}e^{-\beta(m - m_{\min})}, \quad \beta = b\log 10
  \label{eq:ex1_dist_m}
\end{equation}
in which $b$ is the $b$-value of the GR law, taken as 0.9 following \cite{Fujiwara2005}.

Sampling from the distribution given in equations (\ref{eq:ex1_dist_r}) and (\ref{eq:ex1_dist_m}) was performed using inverse transform sampling \citep{Bishop2006}.
Let $u$ be a random variable uniformly distributed over the interval $[0, 1]$.
Then the samples of $\hat{r}$ and $m$ are obtained as:
\begin{align}
\hat{r} &= R\sqrt{u} \\
m &= m_{\min} - \frac{1}{\beta}\log\left[
    1 - u\left(1 - e^{-\beta(m_{\max} - m_{\min})}\right)
  \right]
\end{align}

\begin{figure}[t]
  \centering
  \includegraphics[width=\columnwidth]{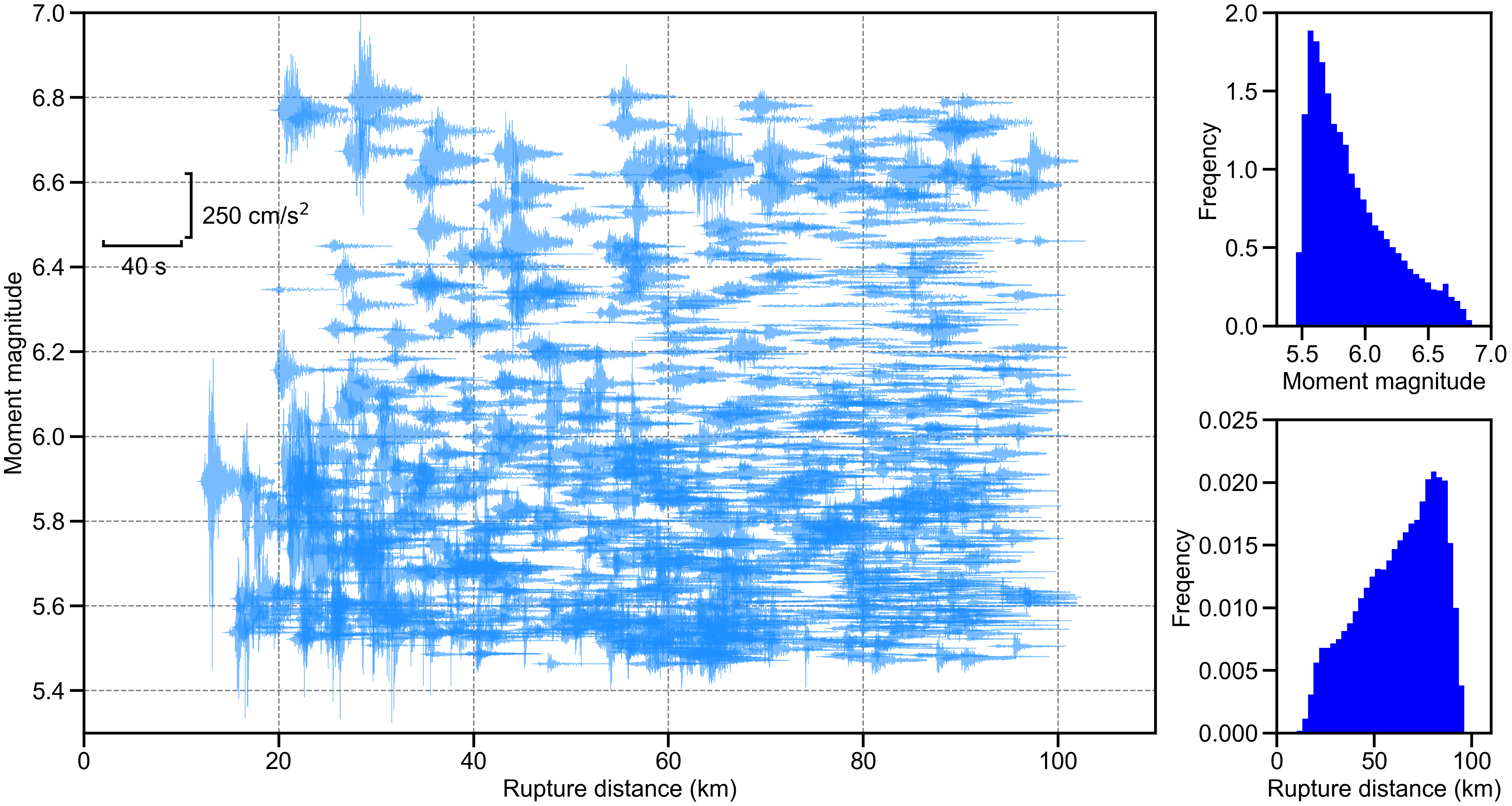}
  \caption{
    Hazard analysis results for numerical example 1 using the S-GMGM.
    The left panel shows 1,000 acceleration waveforms selected from the sampled ground motions.
    The beginning of each waveform corresponds to its associated $M_W$ and $R_{\mathrm{RUP}}$ values.
    For clarity, only the first 40 s of each waveform are shown.
    On the right hand side, the upper histogram presents the distribution of $M_W$ for all samples,
    and the lower histogram shows the distribution of $R_{\mathrm{RUP}}$.
  }
  \label{fig:gm_scat_loc_2_site}
\end{figure}%

Based on the above configuration, sampling of ground-motion waveforms was conducted using Algorithm 1.
We set the number of sampling iterations to $c_{sim} = 20,000$,
which resulted in 500,313 waveforms for the S-GMGM and 499,129 waveforms for the CW-GMGM and CS-GMGM.
For the S-GMGM, the tolerances were set to 0.05 for $M_W$, 5 km for $R_{\mathrm{RUP}}$, and 5 m/s for $V_{\mathrm{S}30}$.
The sampled ground motions were post-processed using the same procedure as described in \nameref{subsec:training_r} subsection.
The sampling required approximately 2,586 min for the S-GMGM, about 2 min for the CW-GMGM, and about 10 min for the CS-GMGM.
Because the computational cost of generating ground-motion waveforms is very low,
the sampling for both the CW-GMGM and CS-GMGM took approximately 82 min and 20 min, respectively, when executed using only a CPU.

\begin{figure}[t]
  \centering
  \includegraphics[width=\columnwidth]{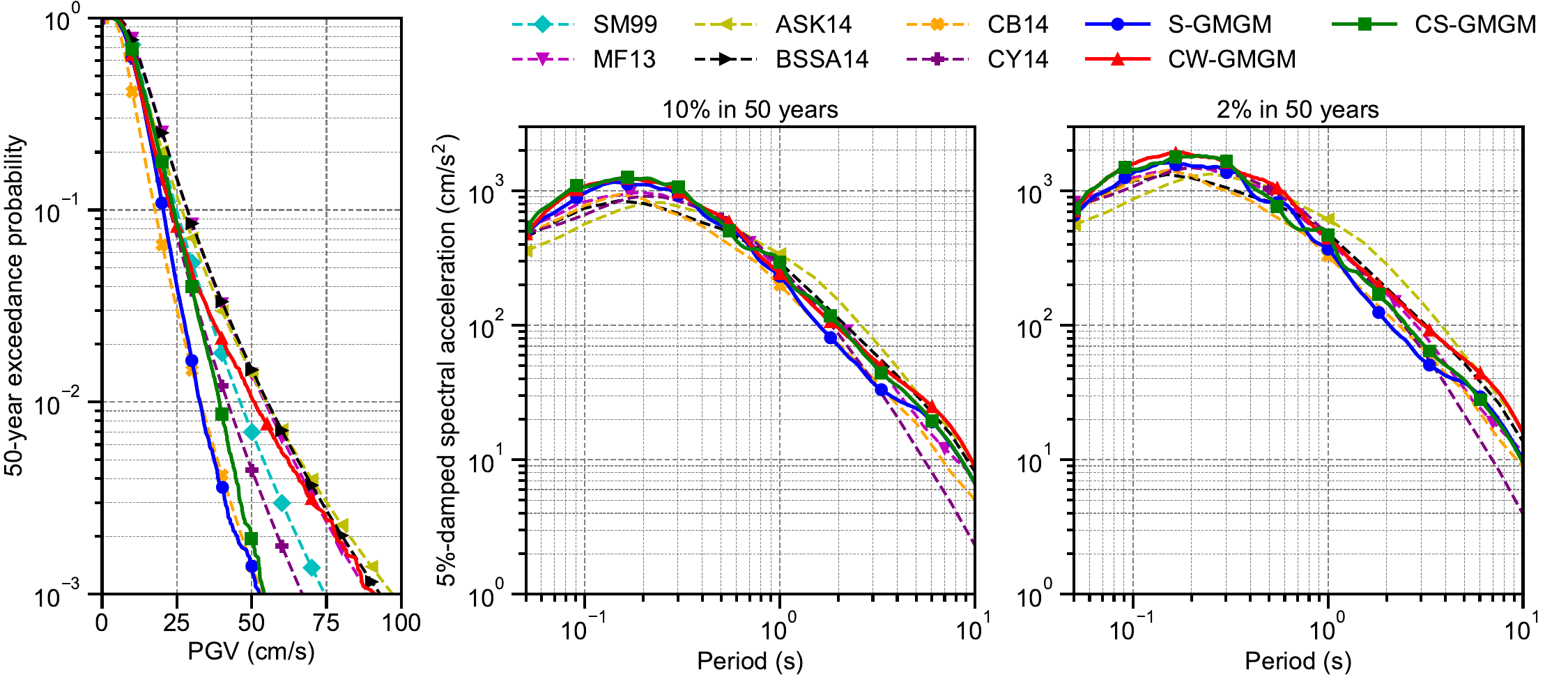}
  \caption{
    Comparison of the hazard analysis results obtained from the GMGMs and GMMs for numerical example 1.
    The left panel shows the 50-year exceedance probability of PGV.
    The middle panel presents the 5\%-damped uniform hazard spectra for a 10\% exceedance probability in 50 years,
    and the right panel shows the uniform hazard spectra for a 2\% exceedance probability in 50 years.
    The three solid lines correspond to the GMGMs, and the six dashed lines correspond to the GMMs.
    The circle, triangle, and square markers represent S-GMGM, CW-GMGM, and CS-GMGM, respectively.
  }
  \label{fig:hazard_analysis_results_poisson}
\end{figure}%

Figure \ref{fig:gm_scat_loc_2_site} shows the hazard analysis results obtained using the S-GMGM.
The results obtained using the CW-GMGM and CS-GMGM are presented in Figures S31 and S32, respectively.
The sampled waveforms follow the assumed distributions $p(\hat{r})$ and $p(m)$,
and the waveforms exhibit the expected trend of larger amplitudes at shorter distances and for larger magnitudes.
These observations confirm that the seismic hazard is appropriately represented through the collection of sampled waveforms.

To verify the validity of the distribution of the sampled waveforms,
we computed the PGV hazard curve and the 5\%-damped uniform hazard spectra corresponding to 10\% and 2\% exceedance probabilities in 50 years, following the procedure described in the \nameref{subsec:IM_exceed} subsection.
These results were compared with those obtained from the conventional PSHA using GMMs,
as shown in Figure \ref{fig:hazard_analysis_results_poisson}.
The analysis results are generally consistent,
and the hazard analysis results obtained using the GMGMs fall within the inter-model variability of the GMM-based evaluations.
Thus, the distribution of waveforms obtained using the proposed method can be considered appropriate.

\begin{figure}[t]
  \centering
  \includegraphics[width=0.49\columnwidth]{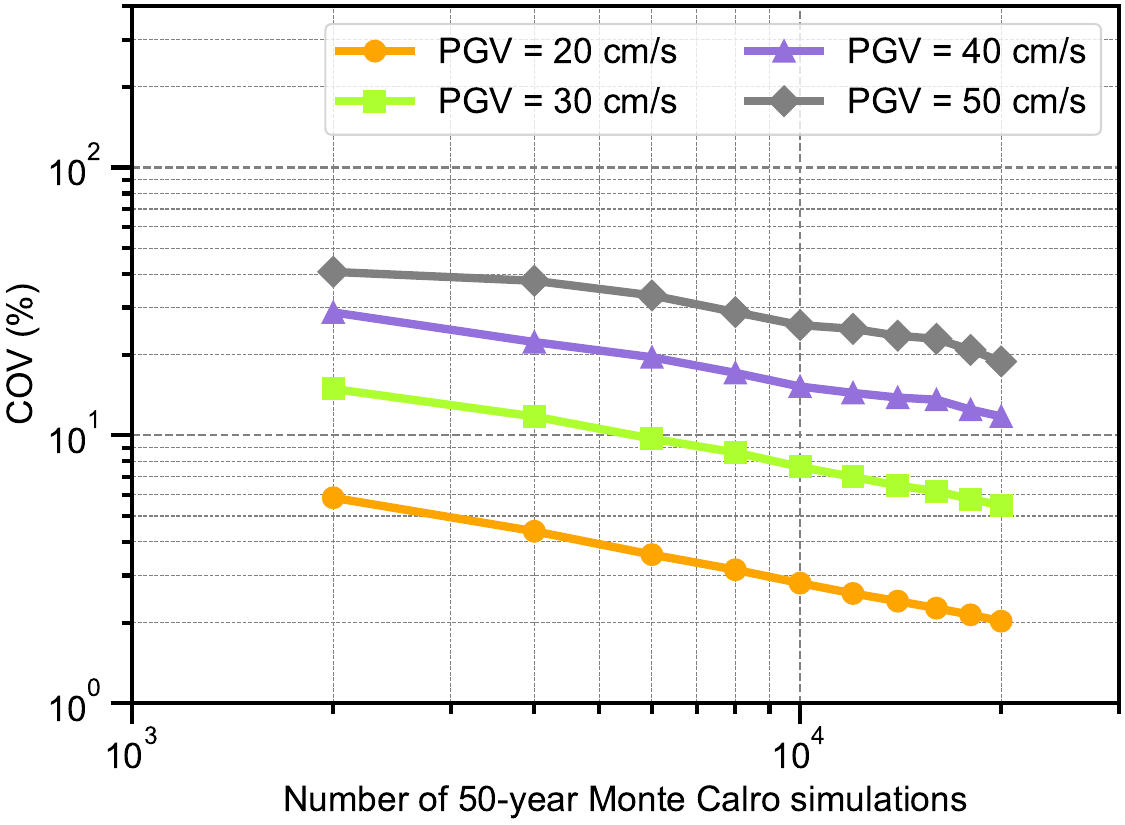}
  \caption{
    COV of the evaluated 50-year exceedance probability for different PGV levels as a function of the number of
    Monte Carlo simulations in numerical example 1 for the S-GMGM.
    The horizontal axis represents the number of Monte Carlo simulations, $c_{sim}$, in Algorithm 1.
    The circle, square, triangle, and diamond markers denote PGV levels of 20, 30, 40, and 50 cm/s,
    respectively, which correspond to 50-year exceedance probabilities of approximately
    0.11, 0.016, 0.0036, and 0.0014.
  }
  \label{fig:cov_pgv50_poisson_sgmgm}
\end{figure}%

Next, we examine whether the number of samples is sufficient to evaluate the seismic hazard in a Monte Carlo sense.
For the hazard analysis results of the S-GMGM, the evolution of the COV for each PGV level is
shown in Figure \ref{fig:cov_pgv50_poisson_sgmgm}.
With $c_{sim} = 20,000$,
the COV at the PGV level of 40 cm/s decreases to approximately 10\%.
A PGV of 40 cm/s corresponds to a 50-year exceedance probability of 0.0036 (i.e., a return period of approximately 13,900 years),
indicating that the seismic hazard can be evaluated with reasonable accuracy up to a level sufficient for engineering applications.
On the other hand, when hazard levels at lower exceedance probabilities are to be evaluated, or when a target COV of approximately 1\% is required for the MCS,
the current sample size of about 500,000 waveforms is insufficient.
In particular, high-accuracy estimation in the low-exceedance probability range requires a substantially larger-scale analysis, and improving computational efficiency remains an important issue for future work.
This aspect is discussed in more detail in \nameref{sec:conc} section.

The CW-GMGM and CS-GMGM can directly generate waveforms by specifying the required conditional labels, whereas the S-GMGM must repeatedly generate samples until waveforms corresponding to the required conditional labels are obtained.
Since the distributions of conditional labels generated by the S-GMGM generally match those of the observed records,
the occurrence frequency of waveforms with large magnitudes and short distances becomes extremely low.
Consequently, a large number of generations is required before obtaining waveforms that satisfy the target conditions, 
resulting in a significant increase in computational cost.
This advantage in computational efficiency is one of the key characteristics of the cGAN-based GMGMs.

\begin{figure}[th]
  \centering
  \includegraphics[width=0.49\columnwidth]{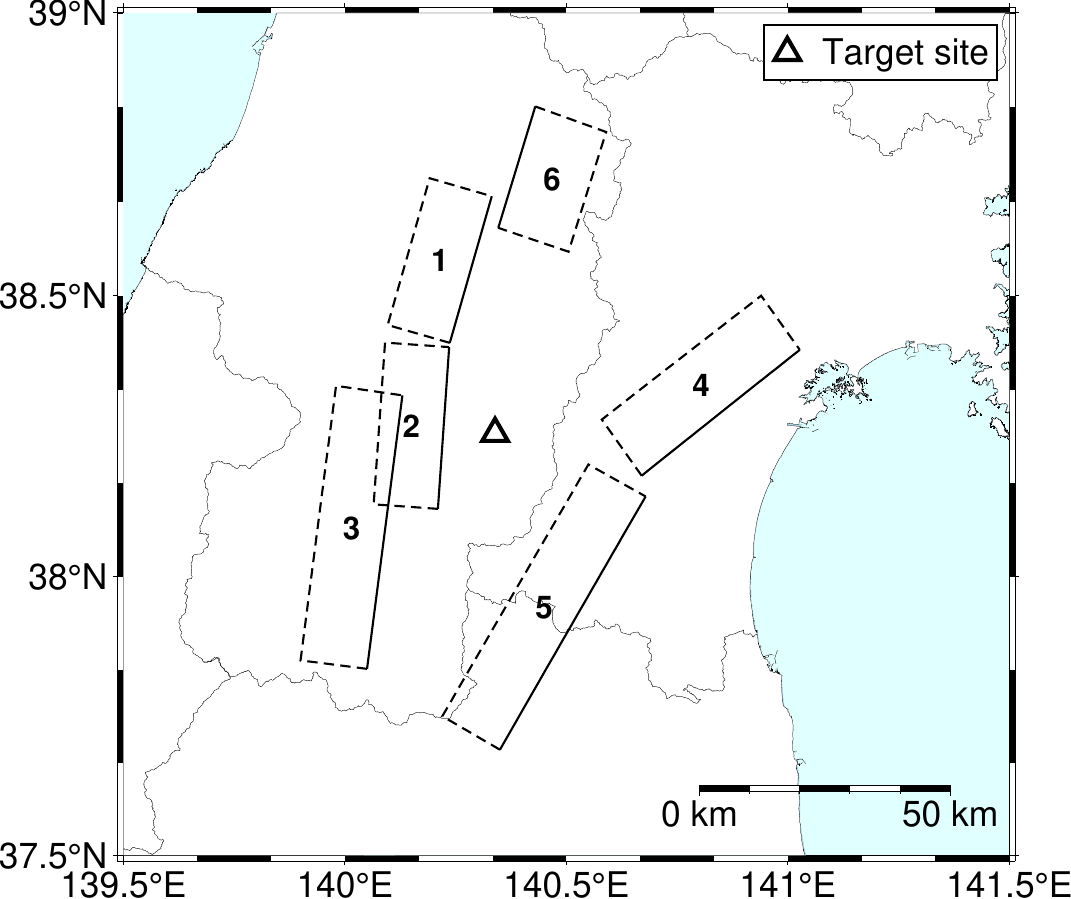}
  \caption{Location of the target site and source faults.
  The triangle represents the site, and the six rectangles represent the source fault planes.
  The solid line indicates the upper edge of each fault.
  The number shown at the center of each rectangle corresponds to the fault number in Table \ref{tab02}.
}
  \label{fig:map_ymg_fault}
\end{figure}%

\begin{table}[ht]
  \centering
  \caption{List of earthquake occurrence models for the source faults shown in Figure \ref{fig:map_ymg_fault}.
    $\mu$ and $\alpha$ are the parameters related to the mean and variability of the BPT distribution, respectively;
    $\nu$ is the annual earthquake occurrence rate for the model following a Poisson process;
    and $T_{cur}$ is the number of years that have elapsed since the most recent earthquake on the source fault.}
  \label{tab02}
  \small
  {\begin{tabular*}{\columnwidth}{@{\extracolsep{\fill}}lllllllll@{}} \toprule %
      No. & Name & $M_W$ & $R_{\mathrm{RUP}}$ (km)
      & Model & $\mu$ or $1 / \nu$ (yr) & $T_{cur}$ (yr) & $\alpha$ \\
  \midrule
    1 & Yamagata-bonchi fault zone (Northern segment) & 6.8 & 20.1 & BPT & 3250 & 2755 & 0.24 \\
    2 & Yamagata-bonchi fault zone (Southern segment) & 6.8 & 10.9 & Poisson & 2500 & - & -  \\
    3 & Nagai-bonchi-seien fault zone & 7.1 & 19.2 & BPT & 5650 & 1200 & 0.24 \\
    4 & Nagamachi-Rifu-sen fault zone & 6.9 & 23.0 & Poisson & 5000 & - & -  \\
    5 & Fukushima-bonchi-seien fault zone & 7.1 & 14.0 & BPT & 8000 & 1955 & 0.24  \\
    6 & Shinjo-bonchi fault zone (Eastern part) & 6.6 & 40.7 & BPT & 4000 & 3100 & 0.24 \\
  \bottomrule
  \end{tabular*}}
\end{table}%


\subsection{Example 2: actual site and source faults}\label{subsec:ex2}

Figure \ref{fig:map_ymg_fault} shows the spatial distribution of the target site and the source faults considered in this numerical example.
The target site is located in the Tohoku region of Japan.
The $V_{\mathrm{S}30}$ value at the site is 356 m/s and was obtained
from the Japan Seismic Hazard Information Station (J-SHIS) database.
Six source faults located near the site were selected from the active fault evaluations by the Headquarters for Earthquake Research Promotion (HERP) \citep{Headquarterschouki}.
The earthquake occurrence models of the selected faults are shown in Table \ref{tab02}.
When information on past earthquake occurrences is available for a given source fault,
the BPT distribution is used as the occurrence model; otherwise, a stationary Poisson process is used.
In this study, the earthquake occurrence models, their parameters, and the values of $M_W$ were determined based on the evaluations published by HERP.
Source faults 1, 3, 5, and 6 follow the BPT distribution, while the others follow the stationary Poisson process.
The values of $R_{\mathrm{RUP}}$ were computed based on the locations of the specified source faults.

In the earthquake occurrence model based on the BPT distribution,
the probability $P(E; T_{cur}, \Delta t)$ that an earthquake $E$ will occur within the next $\Delta t$ years is computed as:
\begin{align}
P(E; T_{cur}, \Delta t) &= \frac{\phi(T_{cur} + \Delta t) - \phi(T_{cur})}{1 - \phi(T_{cur})} \label{eq:bpt_dist_prob_chap4} \\
\phi(t) &= \Phi\left[
    \frac{1}{\alpha}\left\{\sqrt{\frac{t}{\mu}} - \sqrt{\frac{\mu}{t}}\right\}
  \right] + e^{2/\alpha^2}\Phi\left[
    -\frac{1}{\alpha}\left\{\sqrt{\frac{t}{\mu}} + \sqrt{\frac{\mu}{t}}\right\}
  \right] \label{eq:bpt_dist_prob_2_chap4}
\end{align}
in which $T_{cur}$ is the number of years elapsed since the most recent earthquake,
and $\mu$ and $\alpha$ are the parameters of the BPT distribution.

\begin{figure}[t]
  \centering
  \includegraphics[width=\columnwidth]{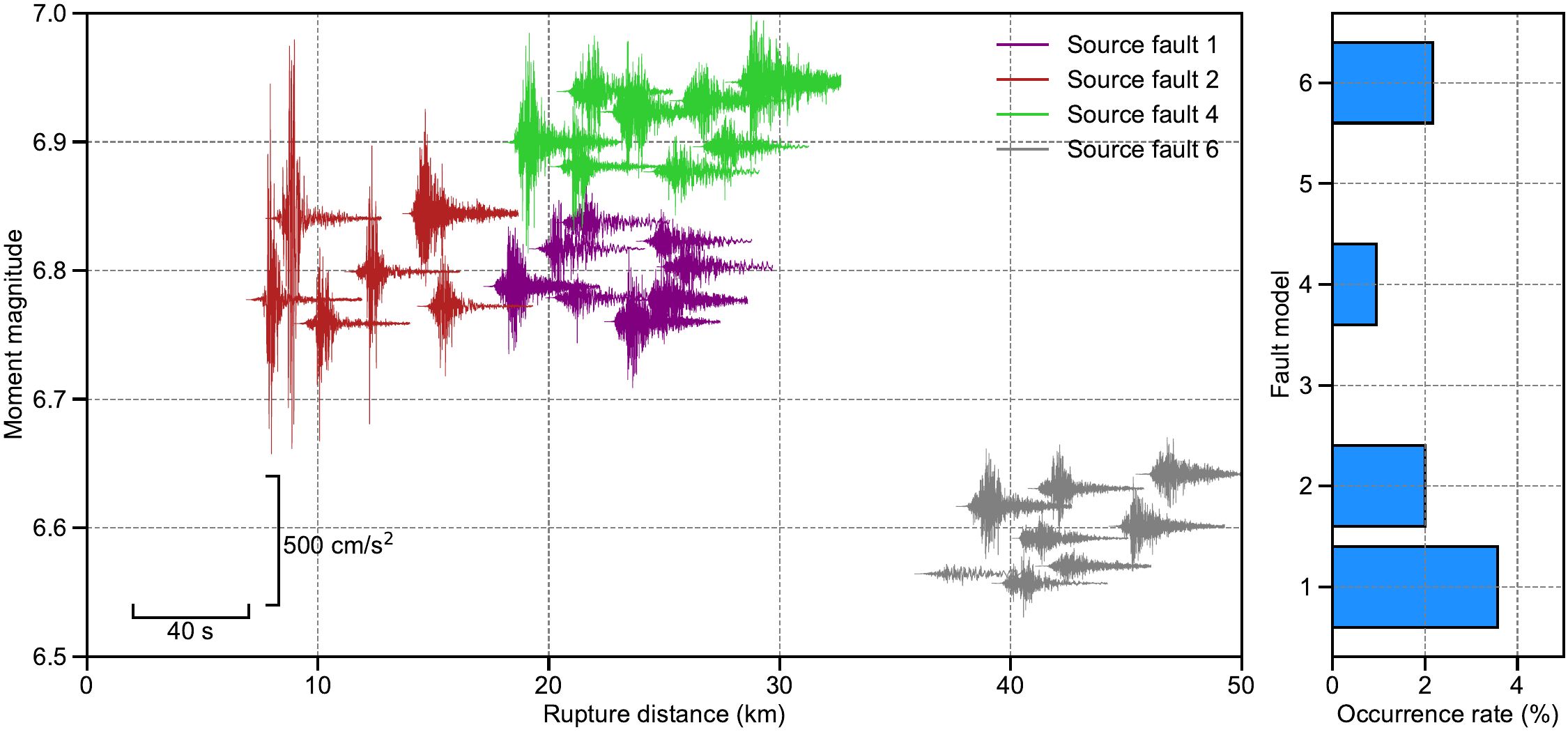}
  \caption{
    Hazard analysis results for numerical example 2 using the S-GMGM.
    The left panel shows examples of ground-motion waveforms obtained through MCS,
    with colors indicating the corresponding source faults.
    The beginning of each waveform corresponds to its associated $M_W$ and $R_{\mathrm{RUP}}$ values.
    The right panel shows the 50-year earthquake occurrence probabilities for each source fault, computed from the MCS results.
  }
  \label{fig:gm_scat_loc_bpt_site}
\end{figure}%

We assumed that the probability of multiple earthquake occurrences within the 50 years is negligible when the
BPT distribution is used,
and earthquake occurrence simulations were conducted by setting $\Delta t = 50$.
No restriction is imposed on the number of earthquakes that may occur within 50-year period in the occurrence model based on the stationary Poisson process.
Earthquake occurrences were assumed to be independent across the source faults.
We set delta functions as the probability distributions of $r$ and $m$,
and the values of $M_W$ and $R_{\mathrm{RUP}}$ listed in Table \ref{tab02} were used as fixed parameters.
Based on these settings, ground-motion waveforms were sampled using Algorithm 1 with a target number of samples
$N_w = 20,000$.
As a result, 20,000 waveforms were obtained for the S-GMGM, CW-GMGM, and CS-GMGM, respectively.
The corresponding values of $c_{sim}$ were 230,449, 226,172, and 228,059, respectively.
For the S-GMGM, the same tolerances as those used in numerical example 1 were applied.
The sampled ground motions were post-processed following the same procedure described in the \nameref{subsec:training_r} subsection.
The computational environment was identical to that used in numerical example 1.

\begin{figure}[th]
  \centering
  \includegraphics[width=\columnwidth]{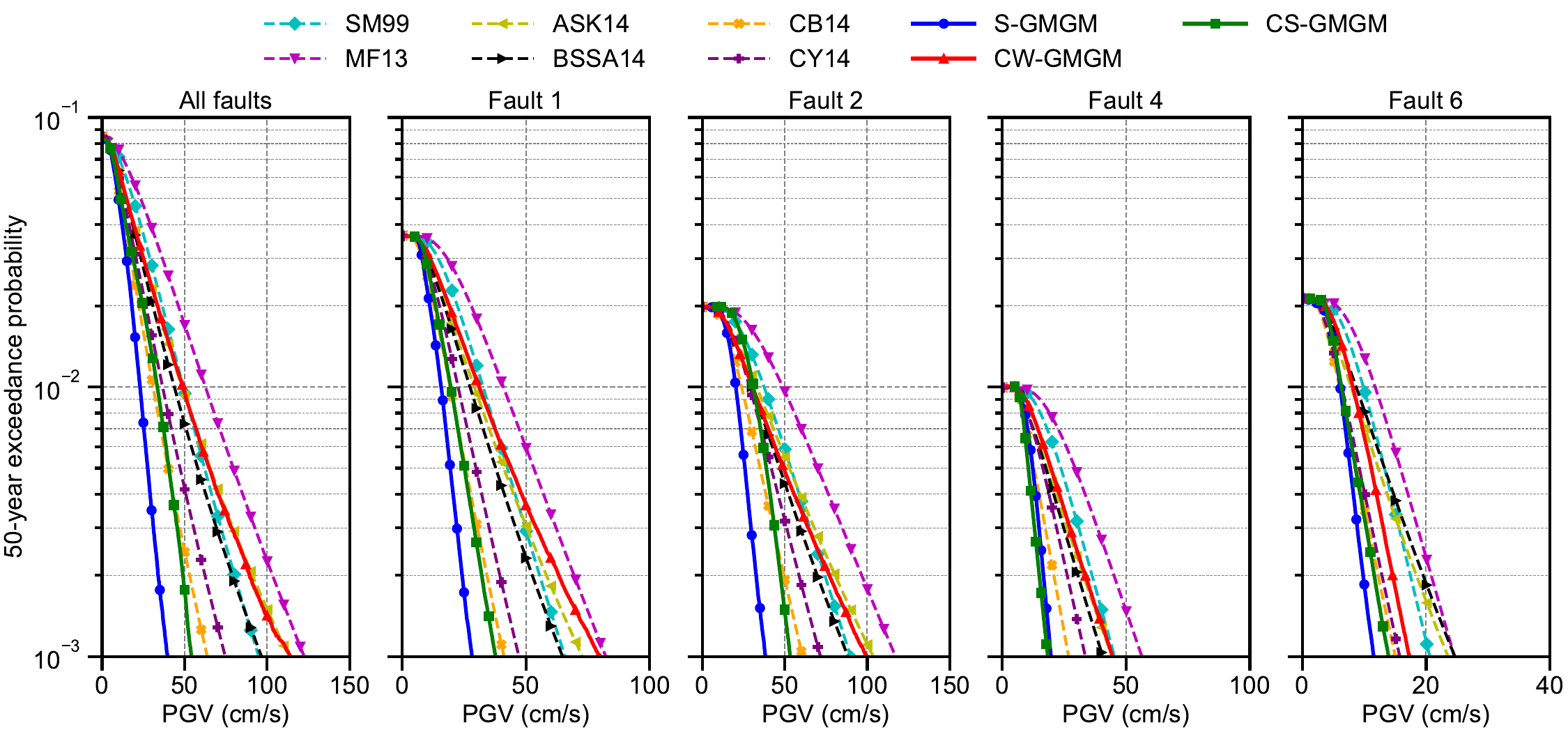}
  \caption{Comparison of the hazard analysis results obtained from the GMGMs and GMMs for numerical example 2.
    Each panel shows the 50-year exceedance probability of PGV, with the leftmost panel presenting the results
    considering all target source faults, and the remaining four panels showing the results for each source fault individually.
    The three solid lines correspond to the GMGMs, and the six dashed lines correspond to the GMMs.
    The circle, triangle, and square markers represent S-GMGM, CW-GMGM, and CS-GMGM, respectively.
  }
  \label{fig:hazard_analysis_results_bpt_pgv}
\end{figure}%

Figure \ref{fig:gm_scat_loc_bpt_site} shows the hazard analysis results obtained using the S-GMGM.
The results obtained with the CW-GMGM and CS-GMGM are shown in Figures S33 and S34, respectively.
The earthquake occurrence probabilities in the MCS for each source fault are generally consistent with the probabilities computed from the parameters listed in Table \ref{tab02}.
Note that source faults 3 and 5 did not generate any events throughout the MCS because their occurrence probabilities are extremely low.
The sampled waveforms show relatively large amplitudes in the near-fault setting and for larger $M_W$,
indicating that reasonable waveforms were sampled for each source fault.

To examine the validity of the distribution of the sampled waveforms,
we computed PGV values from the sampled ground motions and evaluated the 50-year exceedance probability of PGV
using the method described in \nameref{subsec:IM_exceed} subsection.
In addition, to examine the influence of each source fault,
we sampled 20,000 waveforms for each source fault using Algorithm 1 and computed PGV hazard curves in the same manner.
These results were compared with those obtained using the GMMs,
as shown in Figure \ref{fig:hazard_analysis_results_bpt_pgv}.
The hazard curves derived from the GMGMs and GMMs exhibit generally good agreement,
indicating that the distribution of the sampled waveforms is also reasonable.

On the other hand, when examining the GMGM results individually,
it can be seen that the S-GMGM and CS-GMGM tend to exhibit lower PGV evaluations than the GMMs,
whereas the CW-GMGM tends to produce relatively larger PGV values.
To consider the cause of this discrepancy, we first compare the results of the GMGMs with the SM99 and MF13 GMMs, both developed for Japan.
Because the GMGMs treat the two horizontal components independently while the GMMs use the maximum amplitude of the two horizontal components or RotD100 as the IM, the PGV values from the GMGMs should, qualitatively, be smaller.
Consistent with this expectation, the S-GMGM and CS-GMGM always produce lower estimates than the SM99 and MF13 GMMs.
However, the CW-GMGM yields PGV values larger than those of the SM99 GMM in some cases.
For the NGA-West2 GMMs, because RotD50 is used for the IMs, the PGV level should qualitatively be comparable to that of the GMGMs.
However, the S-GMGM and CS-GMGM still yield lower estimates.

These differences may arise from the distinct DNN architectures used in the GMGMs and from differences in how variability is modeled between the GMGMs and the GMMs.
Faults 1, 2, and 4 correspond to conditions involving $M_W$ 6.8 or 6.9 and relatively short distances,
where the number of observed records is limited and model training becomes challenging.
Although the S-GMGM and CS-GMGM share relatively similar DNN architectures, they differ from that of CW-GMGM;
thus, the learning behavior in such data-sparse regions may vary depending on the model architecture.
Regarding the variability in these data-scarce regions, GMMs model variability using a lognormal distribution,
allowing the variability to be specified based on seismological and earthquake engineering knowledge even when data are limited.
In contrast, GMGMs learn to approximate the empirical distribution of the training dataset itself.
Therefore, in magnitude-distance bins with very few observed records, the variability of the ground-motion waveforms may be underestimated due to the scarcity of data.

Based on the above considerations, hazard analysis results obtained using GMGMs may exhibit variability
arising from factors different from those in GMM-based evaluations.
Therefore, in hazard analysis using GMGMs, it is desirable to develop multiple GMGMs using different approaches or by different researchers, and to integrate the results using a logic-tree framework similar to that employed in conventional PSHA.

\begin{figure}[t]
    \centering
    \includegraphics[width=0.45\columnwidth]{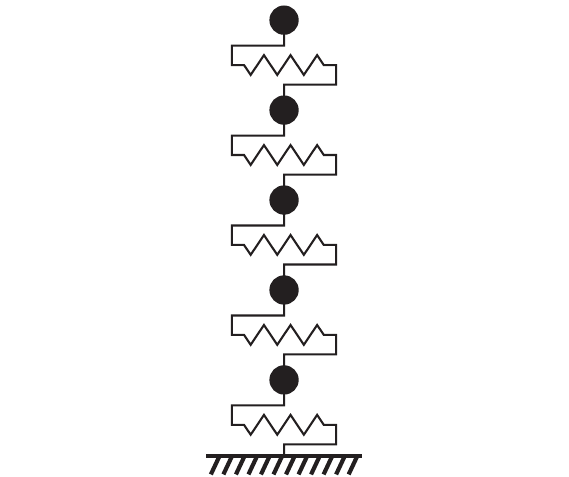}
    \caption{Diagram of the building model.}
    \label{fig:building_model_mdof_main}
\end{figure}%

\begin{table}[ht]
  \centering
  \caption{Parameters of the building model.}
  \label{tab03}
  {\begin{tabular*}{\columnwidth}{@{\extracolsep{\fill}}lllll@{}} \toprule %
  Story & Height (m) & Mass ($\times 10^6$ kg) & Initial stiffness (kN/m) \\
  \midrule
    5 & 3.95 & 673 & 32000 \\
    4 & 3.95 & 564 & 34500 \\
    3 & 3.95 & 564 & 35000 \\
    2 & 3.95 & 565 & 36000 \\
    1 & 4.60 & 568 & 37000 \\
  \bottomrule
  \end{tabular*}}
\end{table}%

\begin{figure}[th]
  \centering
  \includegraphics[width=0.49\columnwidth]{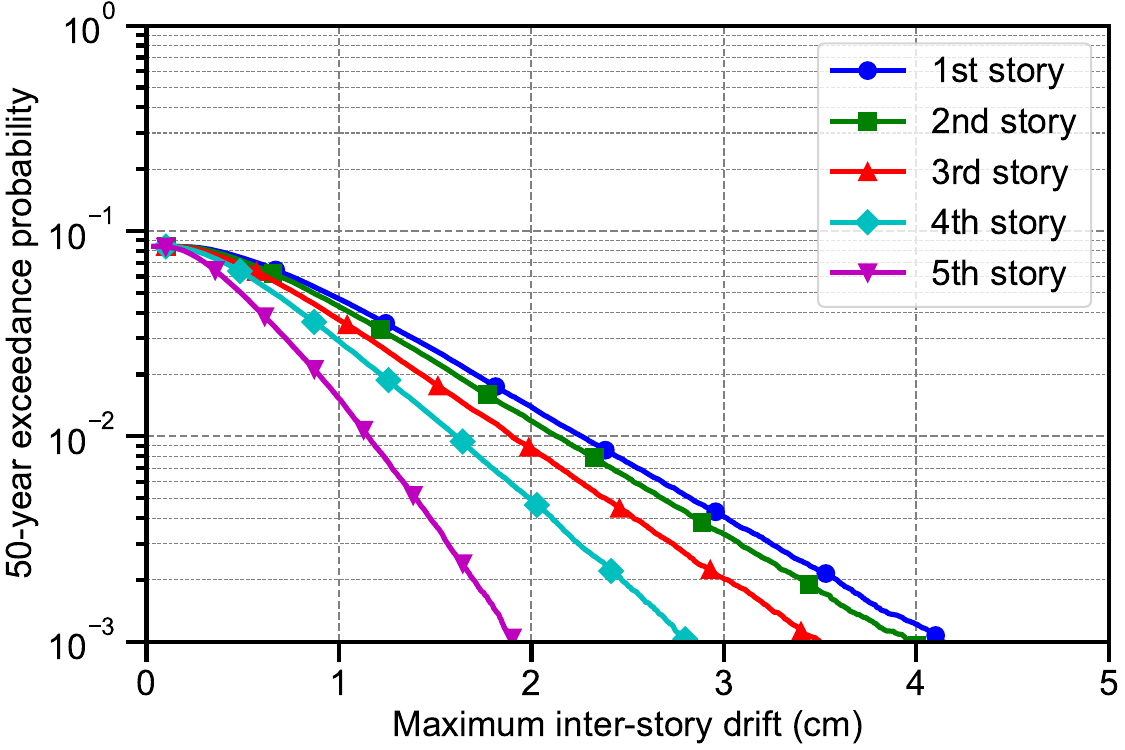}
  \caption{50-year exceedance probability of the maximum interstory drift for each story,
  based on the ground motions evaluated as the seismic hazard for numerical example 2 using the S-GMGM.}
  \label{fig:edp_50y_poe_bpt_sgmgm}
\end{figure}%

\subsection{Evaluations targeting building response}

Assuming that a building exists at the target site considered in numerical example 2,
nonlinear dynamic response analyses were conducted using the sampled ground motions as input.
The building model is shown in Figure \ref{fig:building_model_mdof_main}, and the model parameters are summarized in Table \ref{tab03}.
A 5DOF system was considered, and a bilinear hysteresis model was adopted.
The yield displacement was defined as the displacement corresponding to an inter-story drift angle of 1/100,
and the stiffness degradation ratio was set to 0.05.
The damping ratio was set to $\zeta = 2\%$, and initial stiffness proportional damping was used.
The first and second natural periods of the building model are approximately 0.91 s and 0.31 s, respectively.
As the EDP, the maximum inter-story drift of each story is used.

\begin{figure}[th]
  \centering
  \includegraphics[width=\columnwidth]{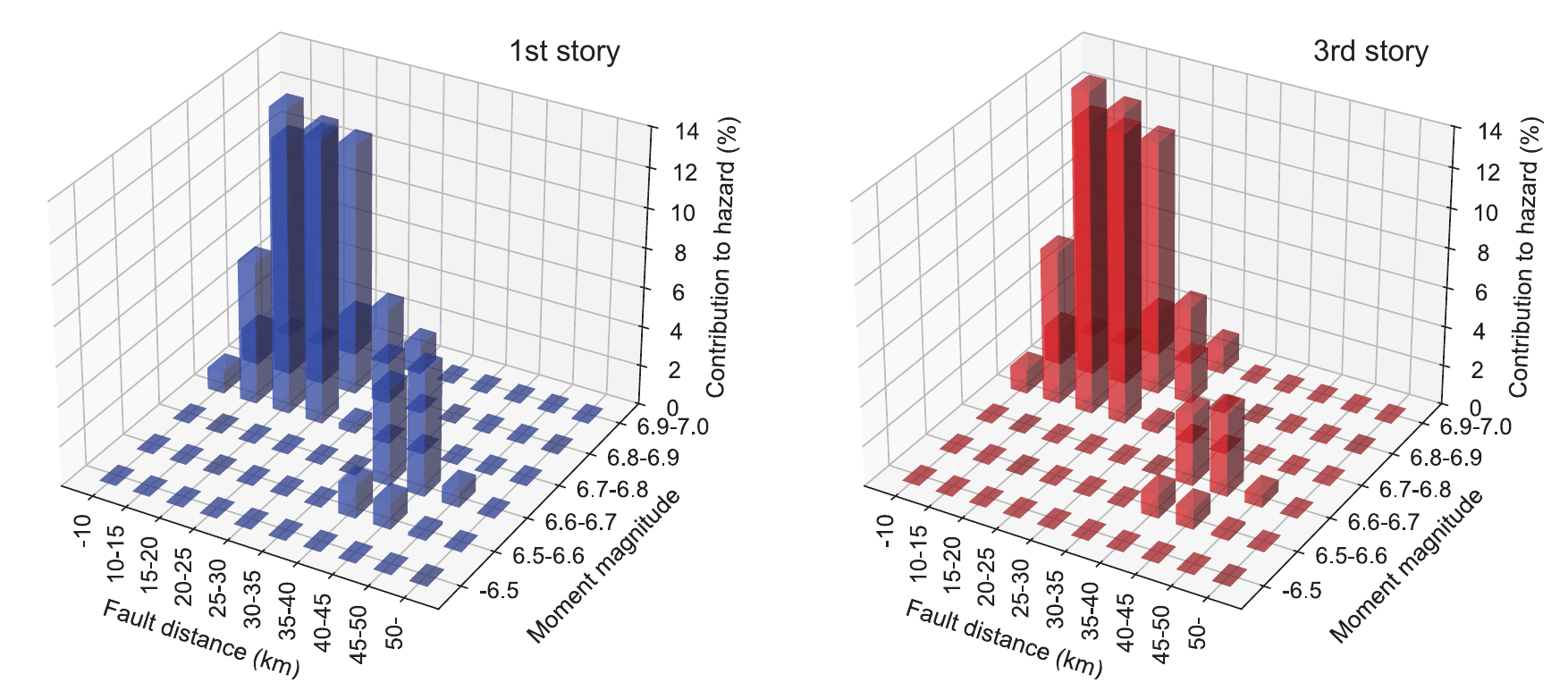}
  \caption{
    Hazard disaggregation results corresponding to Figure \ref{fig:edp_50y_poe_bpt_sgmgm} for the case where the maximum inter-story drift reaches 0.5 cm.
The left panel shows the results for the first story, and the right panel shows the results for the third story.
}
  \label{fig:hazard_disag_sgmgm}
\end{figure}%

Using the 20,000 ground motions sampled by the S-GMGM, $\mathcal{G}_{wav} = \{\mathbf{g}_{i, j, k}\}$,
the relative displacement response waveforms for each story, $\mathbf{d}$, were computed to obtain the set $\mathcal{R} = \{\mathbf{d}_{i, j, k, l}\mid l = 1, \cdots, 5\}$.
Here, the subscript $l$ denotes the story index.
Since the sampled ground-motion waveforms follow the seismic hazard,
the set $\mathcal{R}$ also represents the probability distribution of building responses that reflect the seismic hazard.
Therefore, the exceedance probability of the maximum inter-story drift, $EDP_{i, j, k, l}$,
computed from each waveform in $\mathcal{R}$,
can be evaluated using the same procedure described in \nameref{subsec:IM_exceed} subsection as:
\begin{equation}
  P_l(EDP > a; t) \simeq 1 - \prod_{i}^{N_s}\left\{
    1 - \frac{1}{c_{sim}}\sum_{k=1}^{c_{sim}}\left(
      1 - \prod_{j\in J_i(k)}\mathbbm{1}[EDP_{i, j, k, l} \le a]
    \right)
  \right\}
\end{equation}
in which $P_l(EDP > a; t)$ is the $t$-year exceedance probability of the EDP for story $l$.
Figure \ref{fig:edp_50y_poe_bpt_sgmgm} presents the 50-year exceedance probabilities of
the maximum inter-story drift for each story.
These results demonstrate that the evaluation of EDPs directly from the seismic hazard
is straightforward using the proposed method.

In the proposed waveform-based PSHA, each waveform in $\mathcal{G}_{wav}$ retains the corresponding information from the earthquake catalog of MCS.
Therefore, each building response waveform $\mathbf{d}_{i, j, k, l}$ and each $EDP_{i, j, k, l}$ is associated one-to-one with magnitude and distance information.
This enables hazard disaggregation targeting EDPs.
For a case in which an EDP exceeds a specific level $a$ in $t$-year time period,
the disaggregation of PSHA can be formulated as:
\begin{align}
P_l(M, R\mid EDP > a; t) \simeq \frac{
  \displaystyle\sum_{i\in B_{M, R}}\sum_{k=1}^{c_{sim}}\sum_{j\in J_i(k)}\mathbbm{1}\left[EDP_{i, j, k, l} > a\right]
}{
  \displaystyle\sum_{M}\sum_{R}\sum_{k=1}^{c_{sim}}\sum_{j\in J_i(k)}\mathbbm{1}\left[EDP_{i, j, k, l} > a\right]
}
\end{align}
in which $B_{M, R}$ is the set of source indices $i$ whose magnitude $M$ and distance $R$ values fall within the bin corresponding to $(M, R)$.
Figure \ref{fig:hazard_disag_sgmgm} shows the disaggregation results for cases in which the maximum inter-story drift exceeds 0.5 cm in 50 years.
It can be seen that the hazard contributions are different between the first and third story.
For example, the histogram bins near $M_W = 6.6$ in Figure \ref{fig:hazard_disag_sgmgm} corresponds to the
contribution from the response analyses excited by ground motions associated with Fault 6.
Note that the variation in the magnitude-distance values arises from the acceptance tolerance in the sampling
procedure of the S-GMGM.
It can be seen that Fault 6 may cause larger damage on the first story compared with the third story.
Thus, the proposed method, which conducts PSHA based on ground-motion waveforms, allows the seismic hazard results to be analyzed not only in terms of IMs but also in terms of building response.
It should be noted, however, that the GMGM used in this study considers only $M_W$ as the source characteristic.
Therefore, the results presented in Figure \ref{fig:hazard_disag_sgmgm} do not reflect
fault-specific characteristics beyond magnitude.
Future development of GMGMs incorporating more detailed source characteristics is expected to enable
seismic design and risk assessment that more closely integrate SSC considerations with building seismic response.

\section{Conclusion and discussion}\label{sec:conc}

We proposed a novel framework for PSHA, named waveform-based PSHA, that can directly evaluate
the distribution of ground-motion waveforms.
We presented the mathematical formulation of the waveform-based PSHA as well as an MCS algorithm for its implementation.
As a method for modeling the probability distribution of ground-motion waveforms,
we adopted GAN-based deep generative models and developed three GMGMs.
We proposed a method for quantitatively evaluating the performance of GMGMs
and confirmed that the trained GMGMs are capable of generating high-quality ground-motion waveforms
and accurately capturing their probability distributions.
Subsequently, hazard analyses using the proposed method were performed for two numerical examples: one involving a hypothetical area source and another involving an actual site and source faults in Japan.
We demonstrated that seismic hazard can be represented as a set of ground-motion waveforms
and validated the analysis results by comparing the IM-based hazard derived from these waveforms with the results of conventional PSHA using GMMs.
Finally, nonlinear dynamic response analyses of a building model were conducted using the ground motions from the evaluated seismic hazard,
and it was demonstrated that the exceedance probabilities of EDPs, as well as hazard disaggregation with respect to EDPs,
can be analyzed in a straightforward manner.

Although the proposed hazard evaluation requires MCS using deep learning models,
the GAN-based generative models employed in this study have low computational cost for data generation,
and the MCS computation is relatively inexpensive for the cGAN-based models.
On the other hand, for the GMGM that model joint distribution of ground-motion waveform and its conditional label,
analyses become inefficient in regions where observed records are sparse.
In this study, MCS was performed so that the COV was approximately 10\%,
but achieving a COV on the order of 1\% would require substantially more iterations of MCS.
Improving the computational efficiency of the proposed method by using importance sampling (e.g., \cite{Houng2025})
is an important direction for future work.

The three GMGMs used in this study were trained based on the same dataset,
however, relatively large differences were observed in the resulting hazard curves.
As a possible factor of such differences,
we identified in \nameref{subsec:ex2} subsection the influence of variations in the DNN architectures used in the GMGMs.
Another potential factor is the effect of the training state of the GMGMs.
When hazard analyses were conducted using GMGMs at different training epochs,
the resulting hazard curves differed significantly in some cases.
Since the optimization algorithms used in deep learning do not guarantee convergence to a global optimum,
a model trained using the same dataset and DNN architecture does not necessarily produce identical hazard estimates.
In other words, while extending GMMs to deep learning-based models has enabled the evaluation of ground-motion distributions at the waveform level rather than at the IM level,
it has also increased the sources of epistemic uncertainty.
Accordingly, further advancement of quantitative performance evaluation methods for GMGMs,
such as the approach proposed in \nameref{subsec:optimal} subsection, is particularly important.

In this study, only three explanatory variables, $M_W$, $R_{\mathrm{RUP}}$, and $V_{\mathrm{S}30}$,
were used to characterize ground-motion waveforms.
While incorporating additional explanatory variables is an important direction for future development,
such modeling requires a larger number of observed records for training.
Thus, the compilation of a large-scale waveform database is another essential task for future research.
Moreover, because this paper focuses solely on ground-motion modeling,
further advancement of SSC and GMC within the proposed framework also constitutes an important topic for future investigation.

\section*{Data and Resources}
The strong-motion observed records and the shear-wave velocity values of the surface soil used in this study can be downloaded through the website of the
National Research Institute for Earth Science and Disaster Resilience (NIED; \url{https://www.kyoshin.bosai.go.jp/kyoshin/}, last accessed June 2025).
The moment magnitude values were obtained from the NIED F-net database (\url{https://www.fnet.bosai.go.jp/}, last accessed June 2025).
The source-fault information used in the numerical experiments of PSHA can be obtained from the website of the 
Headquarters for Earthquake Research Promotion (\url{https://www.jishin.go.jp/evaluation/}, last accessed June 2025).
The program code used in deep learning is available in the GitHub repository (\url{https://github.com/Mat-main-00/psha-gmgm.git}, last accessed November 2025).
The list of earthquakes and observation stations used for training is also available on this GitHub repository.

\section*{Declaration of Competing Interests}

The authors acknowledge that there are no conflicts of interest recorded.

\section*{Acknowledgments}
This study was supported by Grants-in-Aid for Scientific Research from the Japan Society for the Promotion of Science (JSPS KAKENHI) Grant Number JP22J23006 and JP25K23485.

\appendix

\section{Appendix A}\label{app:char}

\subsection{Definition of the indices of ground-motion characteristics used in this study}

We used the JMA instrumental seismic intensity proposed by the Japan Meteorological Agency (JMA) as a measure of seismic intensity.
While the original JMA instrumental seismic intensity is defined using all three components of ground motion,
it was calculated using only a single horizontal component in this study.
First, a bandpass filter $F(f)$ defined by the following equations is applied to the acceleration time-history data:
\begin{align}
  F(f) &= F_l(F) \times F_h(f) \times F_t(f) \\
  F_l(f) &= \left\{1 - \exp\left(-\left(\frac{f}{f_0}\right)^3\right)\right\}^{0.5} \\
  F_h(f) &= \left[1 + 0.694\left(\frac{f}{f_c}\right)^2 + 0.241\left(\frac{f}{f_c}\right)^4 + 0.0557\left(\frac{f}{f_c}\right)^6 + 0.009664\left(\frac{f}{f_c}\right)^8 \right. \notag \\
  &\quad \left. + 0.00134\left(\frac{f}{f_c}\right)^{10} + 0.000155\left(\frac{f}{f_c}\right)^{12}\right]^{-0.5} \\
  F_t(f) &= \left(\frac{1}{f}\right)^{0.5}
\end{align}
where $F_l$, $F_h$, and $F_t$ represent the low-cut filter, high-cut filter, and the filter for the effect of period, respectively.
$f_0$ and $f_c$ are constants and we set to 0.5 and 10, respectively.
$\xi_4$ is then calculated as:
\begin{equation}
  \xi_4 = 2\log_{10}a + 0.94
\end{equation}
in which $a$ is the threshold level for which the total duration of ground-motion data exceeding $|a|$ is 0.3 s.
Following the JMA definition, the computed value of $\xi_4$ was rounded to the third decimal place and then truncated at the second decimal place.

The predominant frequency $\xi_5$ was calculated from the Fourier amplitude spectrum smoothed using a Parzen window with a bandwidth of 0.5 Hz.

The zero-level crossing rate, $\xi_6$,
was computed as the average of the zero-level up-crossing and down-crossing rates.
Since the ground-motion waveforms used in this study were clipped to fixed time windows (81.92 s or 40 s) to serve as inputs for the deep-learning models, the calculation of $\xi_6$ was limited to the portion corresponding to the time interval of $D_{5-95}$ in order to reduce the effects of waveform truncation.

The negative maxima and positive minima, originally introduced by \cite{Rezaeian2010} as indices corresponding to the bandwidth of ground motions, indicates that ground motion with wider bandwidth tends to have a larger number of negative maxima and positive minima.
In this study, the average number of negative maxima and positive minima per unit time was used as $\xi_7$.
Similar to the calculation of $\xi_6$,
the computation was limited to the vibrations within the time interval corresponding to $D_{5-95}$.

Significant duration, $D_{5-95}$, is known as a measure of the duration of ground motion and is calculated as follows:
\begin{align}
  \frac{\displaystyle\int_0^{t_5}a^2(t)\mathrm{d}t}{\displaystyle\int_0^{t_r}a^2(t)\mathrm{d}t} &= \frac{5}{100}, &
  \frac{\displaystyle\int_0^{t_{95}}a^2(t)\mathrm{d}t}{\displaystyle\int_0^{t_r}a^2(t)\mathrm{d}t} &= \frac{95}{100}, &
  D_{5-95} &= t_{95} - t_{5}
\end{align}
in which $a(t)$ is the acceleration at time $t$, $t_r$ is the total duration of the records, and $t_5$ and $t_{95}$ is the time that the cumulative power of the ground motion to reach 5\% and 95\% of the total cumulative power, respectively.
The value of $D_{5-45}$ is known to correspond to the time at the middle of the strong-shaking phase \citep{Rezaeian2010},
and is calculated as follows:
\begin{align}
  \frac{\displaystyle\int_0^{t_{45}}a^2(t)\mathrm{d}t}{\displaystyle\int_0^{t_r}a^2(t)\mathrm{d}t} &= \frac{45}{100}, &
  D_{5-45} &= t_{45} - t_{5}
\end{align}
where $t_{45}$ is the time that the cumulative power of the ground motion to reach the 45\% of the total cumulative power.

Arias intensity is a measure of total energy contained in the ground motion.
Although it is generally defined over the entire duration of the waveform,
it was calculated only for the portion corresponding to $D_{5-95}$ as follows for the same reason described in the calculation of $\xi_6$:
\begin{equation}
  I_A = \frac{\pi}{2g}\int_{t_5}^{t_{95}}a^2(t)\mathrm{d}t
\end{equation}
in which $g$ is the gravitational acceleration and we set $g = 980.665$ cm/s$^{\text{2}}$.

There are several definitions of spectrum intensity.
In this study, we adopted that proposed by \cite{Housner1961}:
\begin{equation}
  SI = \frac{1}{2.4}\int_{0.1}^{2.5}S_V(T, \zeta)\mathrm{d}T \label{eq:si_mod}
\end{equation}
in which $S_V(T, \zeta)$ is the spectral velocity, $T$ is the natural period, and $\zeta$ is the damping ration, which was set to 5\%.

The mean period, representing the overall frequency characteristics of the waveform, was computed following the definition of \cite{Rathje1998}:
\begin{equation}
  T_m = \frac{\sum_i C_i^2\times\left(\frac{1}{f_i}\right)}{\sum_{i}C_i^2}\quad \text{for}\quad 0.25\:\text{Hz}\:\le f_i\le\:20\:\text{Hz} \label{eq:mean_period}
\end{equation}
in which $f_i$ is the discretized frequency and $C_i$ is the Fourier amplitude at $f_i$.

\section{Appendix B}\label{app:B}
\subsection{GMM formulations used in this paper}
\subsubsection{SM99}

The SM99 GMM targets only PGA and PGV, and for crustal earthquake, PGV is evaluated by the following equation:
\begin{equation}
  \log_{10}\mathrm{PGV} = 0.58M_W + 0.0038 D - \log_{10}(R_{\mathrm{RUP}} + 0.0028\times 10^{0.50M_W}) - 0.002R_{\mathrm{RUP}} - 1.29
\end{equation}
in which $D$ is the hypocenter depth (unit: km).
In the \nameref{subsec:training_r} subsection,
considering that the GMGMs are developed with the observed records of crustal earthquakes, we set $D = 10$ km.
The SM99 GMM evaluates the IMs on the hard rock site condition with an average shear-wave velocity of 600 km/s.
To account for the amplification of the surface soil, the PGV values were corrected using the empirical relationship of  \cite{Fujimoto2003}:
\begin{equation}
  \log_{10}\widehat{\mathrm{PGV}} = \log_{10}\mathrm{PGV} + 0.66\log_{10}\frac{600}{V_{\mathrm{S}30}}
\end{equation}
For the variability, the following equation, which was used in \cite{Fujiwara2023}, was employed:
\begin{equation}
  \sigma = \begin{cases}
    0.23 & R_{\mathrm{RUP}} \le 20 \text{km} \\
    0.23 - 0.03\times\frac{\log_{10}(R_{\mathrm{RUP}} / 20)}{\log_{10}(30 / 20)} & 20\text{km} < R_{\mathrm{RUP}} \le 30 \text{km} \\
    0.20 & 30\text{km} < R_{\mathrm{RUP}}
  \end{cases}\label{eq:var_gmm_jp}
\end{equation}
in which $\sigma$ is the base-10 logarithmic standard deviation of PGV.

\subsubsection{MF13}

Among the model proposed in \cite{Morikawa2013},
a model with a quadratic magnitude term was adopted.
The median value of IMs was evaluated using the following equations:
\begin{align}
  \log_{10}IM &= a\left(M_{W^\prime} - M_{W_1}\right)^2 + bR_{\mathrm{RUP}} + c - \log_{10}\left(R_{\mathrm{RUP}} + d\cdot 10^{eM_{W^\prime}}\right) + G_s \\
  M_{W^\prime} &= \min(M_{W}, M_{W_0}) \label{eq:mf13_main_egmgm} \\
  G_s &= p_s\log_{10}\left(\frac{\min(V_{\mathrm{S}\max}, V_{\mathrm{S}30})}{V_0}\right)
\end{align}
in which $a$, $b$, $c$, $d$, $e$, $M_{W_0}$, $M_{W_1}$, $p_s$, $V_0$, and $V_{\mathrm{S}\max}$ are coefficients.
$G_s$ is the correction term for amplification by shallow soft soils.
The variability of PGV and Sa was evaluated in equation (\ref{eq:var_gmm_jp}) as in the practice of \cite{Fujiwara2023}.

\subsubsection{ASK14}

The median value of IMs was evaluated using the following equation:
\begin{align}
  \log IM &= f_1(M_W, R_{\mathrm{RUP}}) + f_5(\widehat{Sa}_{1180}, V_{\mathrm{S}30}) + \mathrm{Regional}(V_{\mathrm{S}30}, R_{\mathrm{RUP}})
\end{align}
in which $\widehat{Sa}_{1180}$ is the median spectral acceleration on hard rock.
The variability of the IMs was evaluated as:
\begin{equation}
  \sigma = \sqrt{\phi^2(T, M_W, \widehat{Sa}_{1180}, V_{\mathrm{S}30}) + \tau^2(T, M_W, \widehat{Sa}_{1180}, V_{\mathrm{S}30})
}
\end{equation}
in which $\phi$ and $\tau$ are the intra-event and inter-event standard deviations, respectively.
The formulation, coefficients, and symbol definitions of $f_1$, $f_5$, $\mathrm{Regional}(\cdot)$, $\phi$, and $\tau$ are the same as those in \cite{abrahamson2014summary}.

\subsubsection{BSSA14}

The median value of IMs was evaluated using the following equation:
\begin{equation}
  \log IM = F_E(M_W, mech) + F_P(R_{JB}, M_W, region) + F_S(V_{\mathrm{S}30}, R_{JB}, M_W, region, z_1)
\end{equation}
in which $mech$ is the parameter
specifying the focal mechanism, $R_{JB}$ is the Joyner-Boore distance, $region$ is the parameter specifying the region, and $z_1$ is the depth to the layer where the shear-wave velocity is 1.0 km/s.
In the calculations, $mech$ was set to ``unspecified'', and the value corresponding to Japan was used for $region$.
The value of $R_{\mathrm{RUP}}$ was directly used as $R_{JB}$.
In the site-effect term $F_S$, the amplification term associated with $z_1$ was removed in the calculations.

The variability was evaluated using the following equation:
\begin{equation}
  \sigma = \sqrt{\phi^2(M_W, R_{JB}, V_{\mathrm{S}30}) + \tau^2(M_W)}
\end{equation}
The formulation, coefficients, and symbol definitions of $F_E$, $F_P$, $R_S$, $\phi$, and $\tau$ are the same as those in \cite{boore2014nga}.

\subsubsection{CB14}

The median value of IMs was evaluated using the following equation:
\begin{equation}
  \log Y = \begin{cases}
    \log \mathrm{PGA} & \mathrm{PSA} < \mathrm{PGA} \;\text{and}\; T < 0.25\;\text{s} \\
    f_{mag} + f_{dis} + f_{site} + f_{atn} & \text{otherwise}
  \end{cases}
  \label{eq:cb14_main}
\end{equation}
in which PSA is the pseudo spectral acceleration.
$f_{mag}$, $f_{dis}$, $f_{site}$, and $f_{atn}$ are the terms for magnitude, geometric attenuation, shallow site response, and anelastic attenuation, respectively.
The formulation, coefficients, and symbol definitions are the same as those in \cite{campbell2014nga}.
The variability was also evaluated using the same equation as in \cite{campbell2014nga}, calculated as:
\begin{equation}
  \sigma = \sqrt{\tau^2 + \phi^2}
\end{equation}

\subsubsection{CY14}

The median value of IMs was evaluated using the following equation:
\begin{align}
  \begin{split}
    \log IM_{ref} &= c_2(M_W - 6) + \frac{c_2 - c_3}{c_n}\log\left(1 + e^{c_n(c_M - M_W)}\right) \\
  &{} \quad + c_4\log\left(
    R_{\mathrm{RUP}} + c_5\cosh\left(
      c_6\cdot\max(M_W - c_{HM}, 0)
    \right)
  \right) \\
  &{} \quad + (c_{4a} - c_{4})\log\left(
    \sqrt{R_{\mathrm{RUP}}^2 + c_{RB}^2}
  \right) \\
  &{} \quad + \left\{
    c_{\gamma1} + \frac{c_{\gamma2}}{\cosh\left(\max(M_W - c_{\gamma3}, 0)\right)}
  \right\}R_{\mathrm{RUP}} 
  \end{split}
  \label{eq:cy14_base_model} \\
  \begin{split}
    \log IM &= \log IM_{ref} \\
      &{} \quad + \phi_1\cdot\min\left(\log\left(\frac{V_{\mathrm{S}30}}{1130}\right), 0\right) \\
    &{} \quad + \phi_2\left(
      e^{\phi_3\left(
        \min(V_{\mathrm{S}30}, 1130) - 360
      \right)} - e^{\phi_3(1130 - 360)}
    \right)\log\left(
      \frac{IM_{ref} + \phi_4}{\phi_4}
    \right)
  \end{split}
\end{align}
in which $IM_{ref}$ is the median IM under the reference condition,
and $c_2$, $c_3$, $c_4$, $c_5$, $c_6$, $c_n$, $c_M$, $c_{HM}$, $c_{4a}$, $c_{RB}$, $c_{\gamma 1}$, $c_{\gamma 2}$, $c_{\gamma 3}$, $\phi_1$, $\phi_2$, $\phi_3$, and $\phi_4$ are coefficients.
The notation for the coefficients follows \cite{chiou2014update}.
The variability was evaluated using the following equation:
\begin{equation}
  \sigma = \sqrt{
    (1 + NL_0)^2\tau^2 + \sigma_{NL_0}^2
  }
\end{equation}
Here, the formulations, coefficients, and symbol definitions for $\tau$, $NL_0$, and $\sigma_{NL_0}$ are the same as those in \cite{chiou2014update}.

\section{Supplementary Material}
\setcounter{figure}{0}
\renewcommand{\thefigure}{S\arabic{figure}.}

\begin{figure}[h]
  \centering
  \includegraphics[width=\columnwidth]{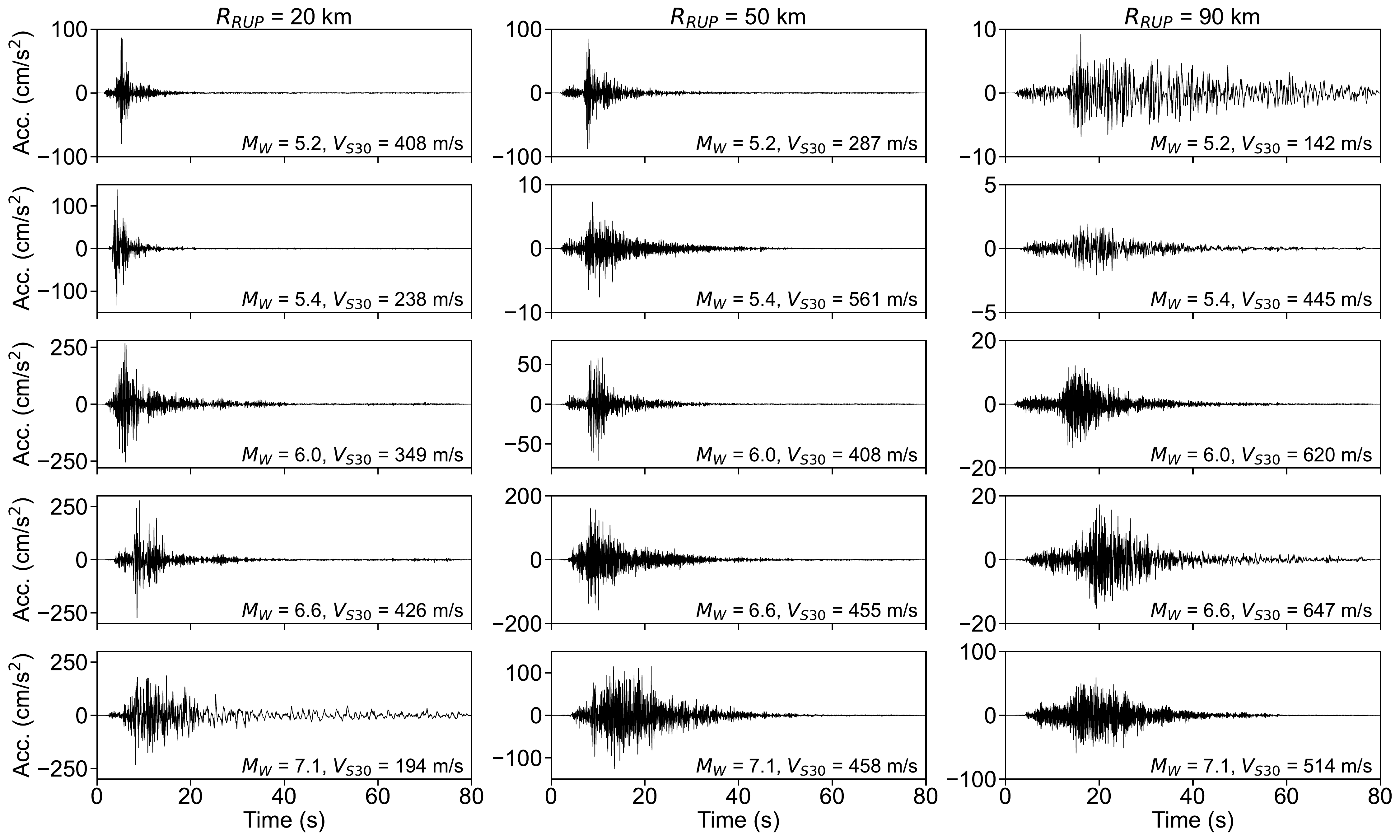}
  \caption{Examples of ground-motion waveforms generated by the S-GMGM.
  The waveforms in each column correspond to the value of $R_{\mathrm{RUP}}$ shown at the top.
Each panel shows the associated $M_W$ and $V_{\mathrm{S}30}$ values.}
\end{figure}%
\clearpage

\begin{figure}[h]
  \centering
  \includegraphics[width=\columnwidth]{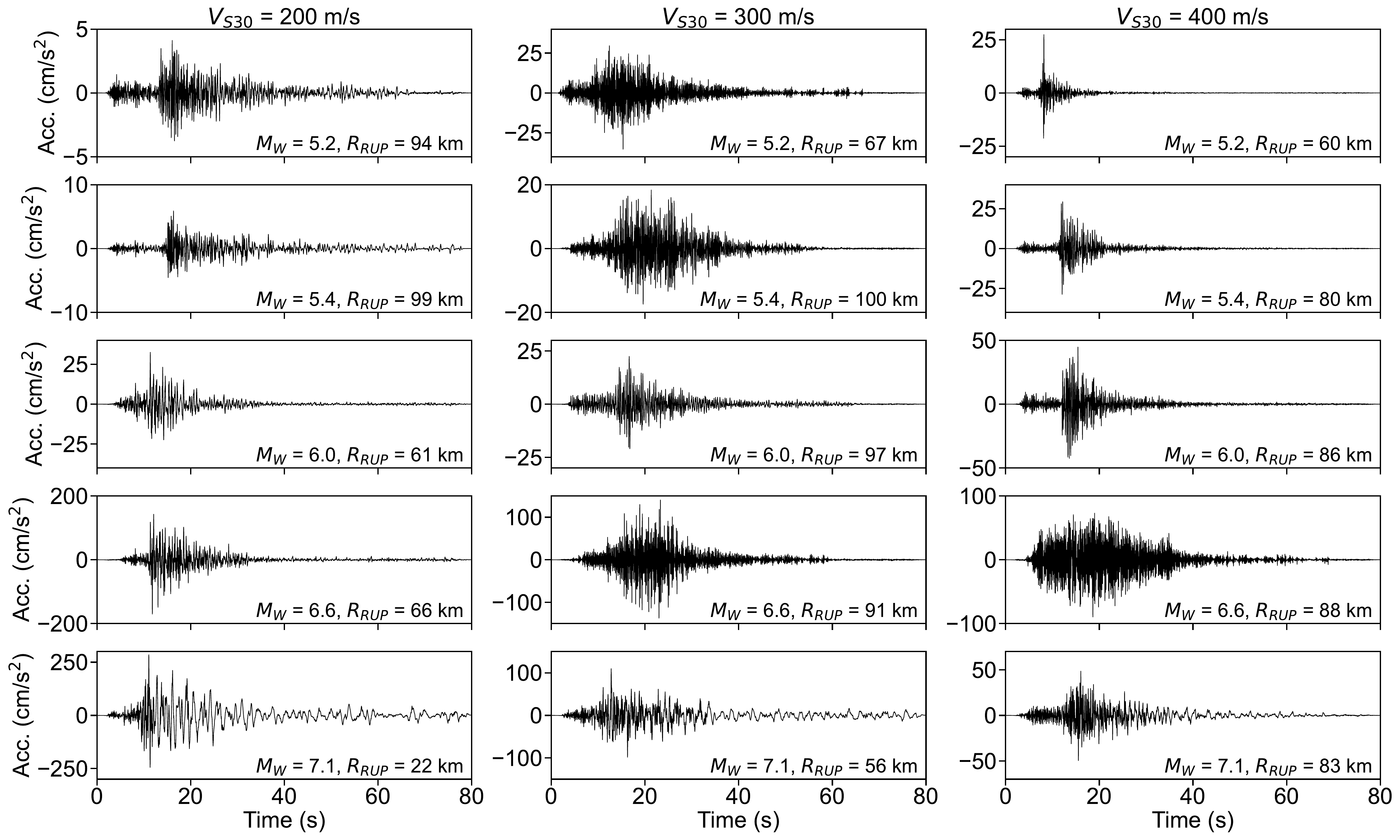}
  \caption{Examples of ground-motion waveforms generated by the S-GMGM.
    The waveforms in each column correspond to the value of $V_{\mathrm{S}30}$ shown at the top.
  Each panel shows the associated $M_W$ and $R_{\mathrm{RUP}}$ values.}
\end{figure}%

\clearpage

\begin{figure}[h]
  \centering
  \includegraphics[width=\columnwidth]{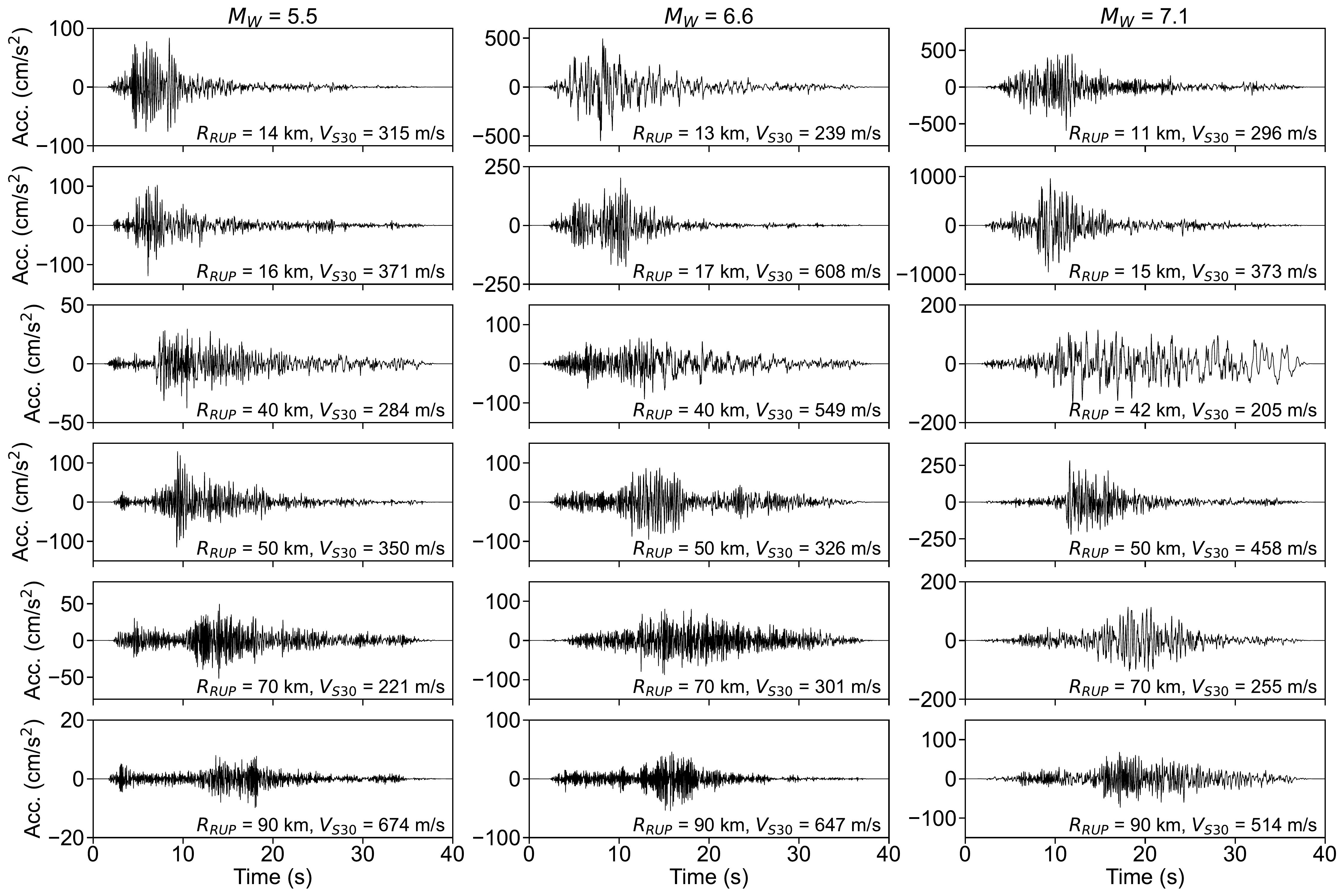}
  \caption{Examples of ground-motion waveforms generated by the CW-GMGM.
    The waveforms in each column correspond to the value of $M_W$ shown at the top.
  Each panel shows the associated $R_{\mathrm{RUP}}$ and $V_{\mathrm{S}30}$ values.}
\end{figure}%

\clearpage

\begin{figure}[h]
  \centering
  \includegraphics[width=\columnwidth]{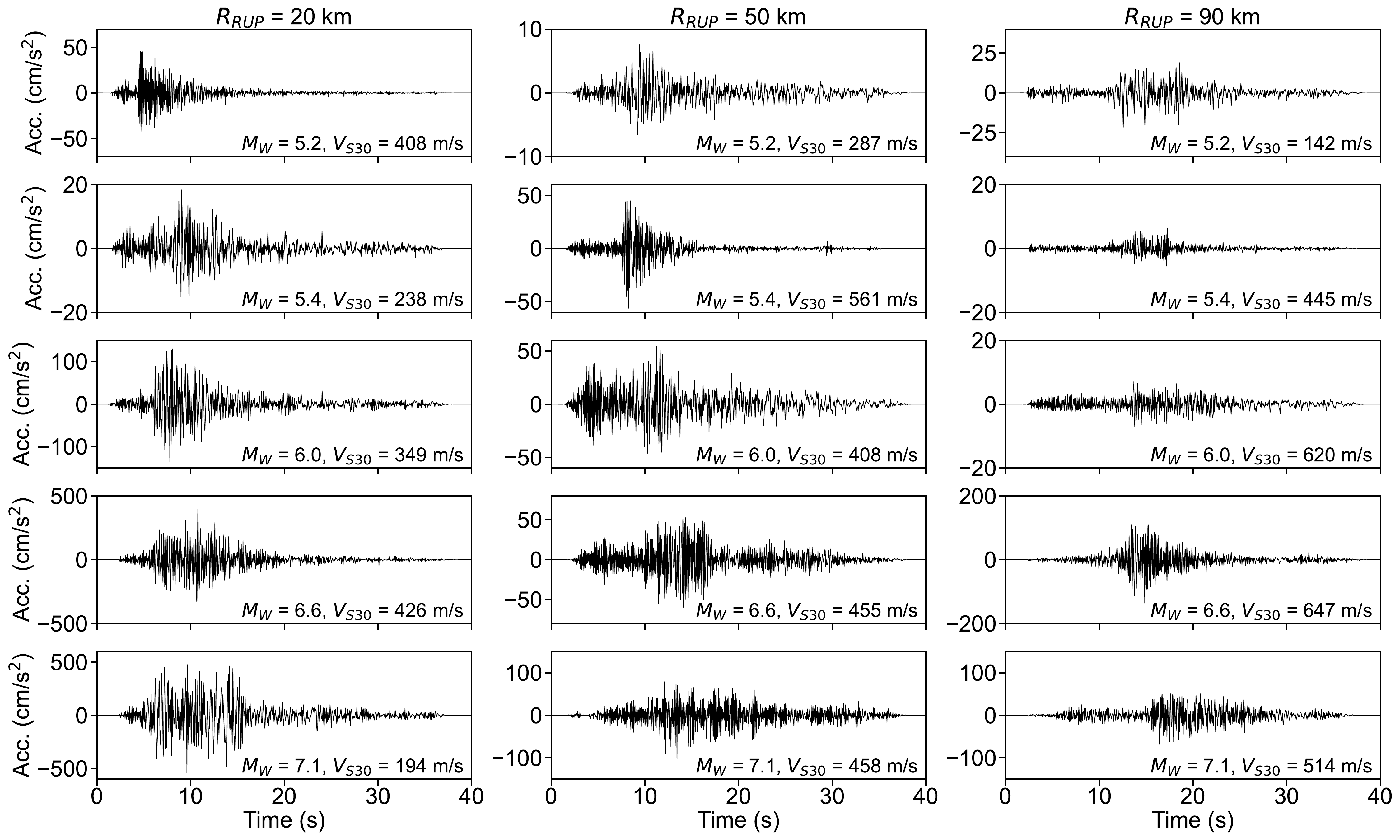}
  \caption{Examples of ground-motion waveforms generated by the CW-GMGM.
    The waveforms in each column correspond to the value of $R_{\mathrm{RUP}}$ shown at the top.
  Each panel shows the associated $M_W$ and $V_{\mathrm{S}30}$ values.}
\end{figure}%

\clearpage

\begin{figure}[h]
  \centering
  \includegraphics[width=\columnwidth]{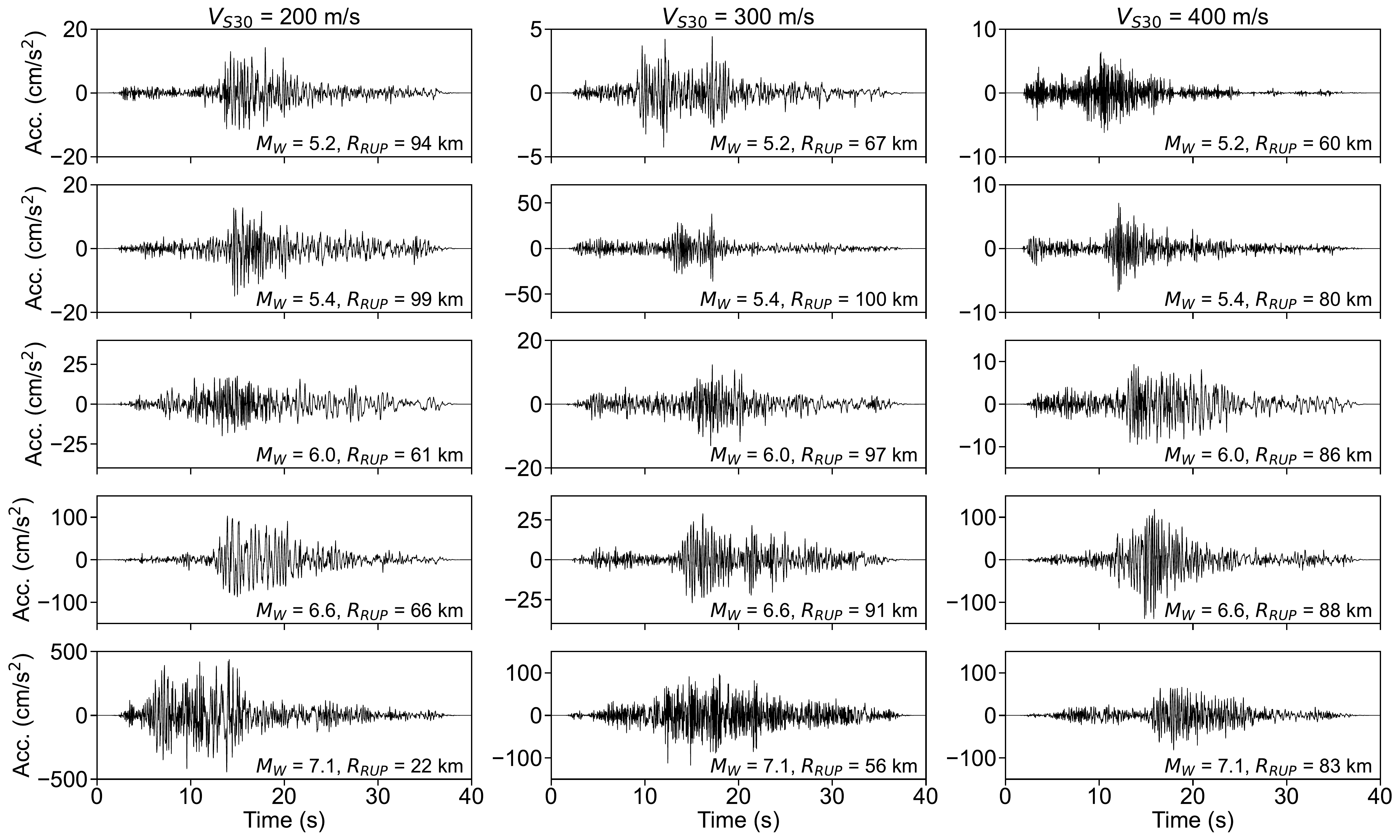}
  \caption{Examples of ground-motion waveforms generated by the CW-GMGM.
    The waveforms in each column correspond to the value of $V_{\mathrm{S}30}$ shown at the top.
  Each panel shows the associated $M_W$ and $R_{\mathrm{RUP}}$ values.}
\end{figure}%

\clearpage

\begin{figure}[h]
  \centering
  \includegraphics[width=\columnwidth]{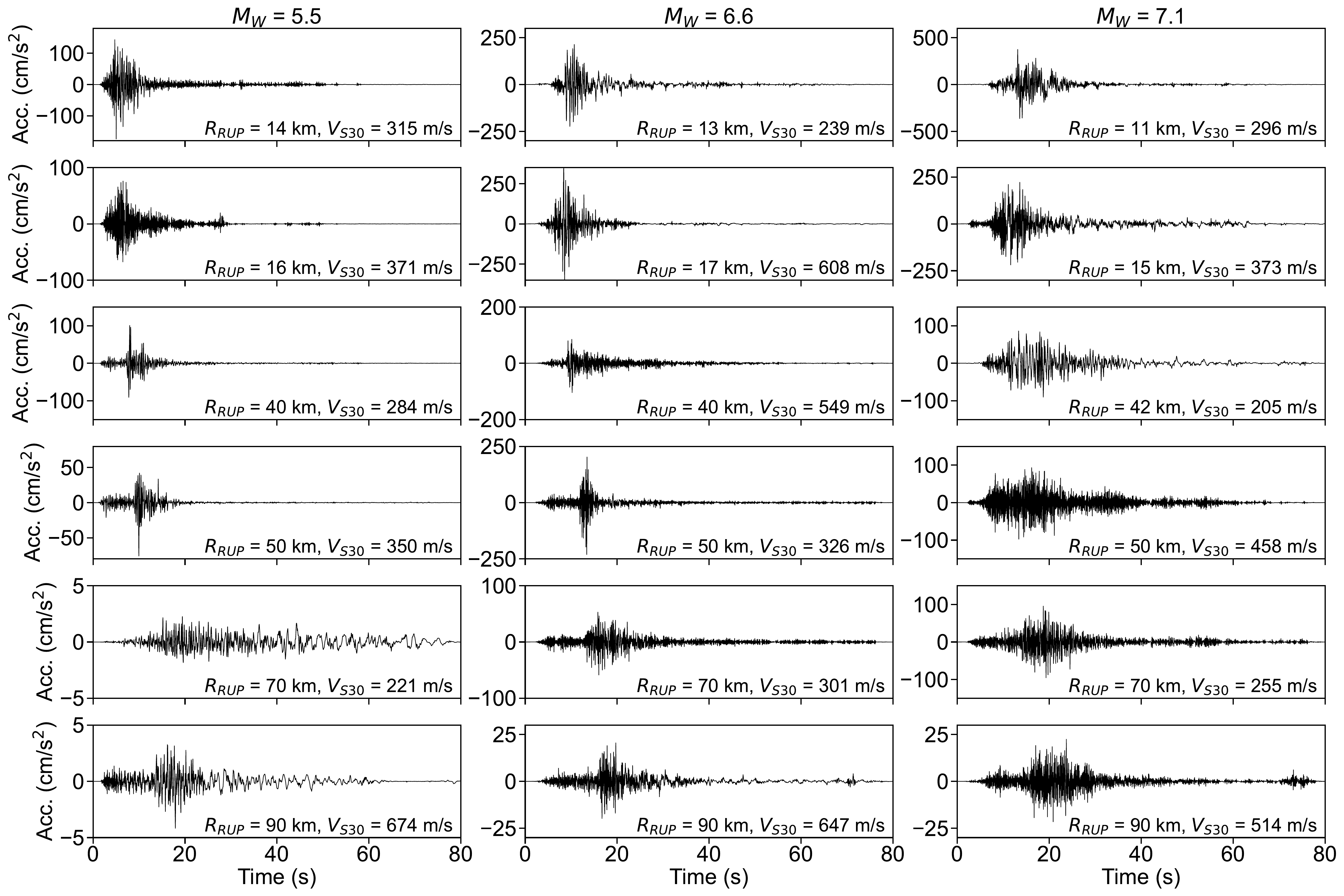}
  \caption{Examples of ground-motion waveforms generated by the CS-GMGM.
    The waveforms in each column correspond to the value of $M_W$ shown at the top.
  Each panel shows the associated $R_{\mathrm{RUP}}$ and $V_{\mathrm{S}30}$ values.}
\end{figure}%

\clearpage

\begin{figure}[h]
  \centering
  \includegraphics[width=\columnwidth]{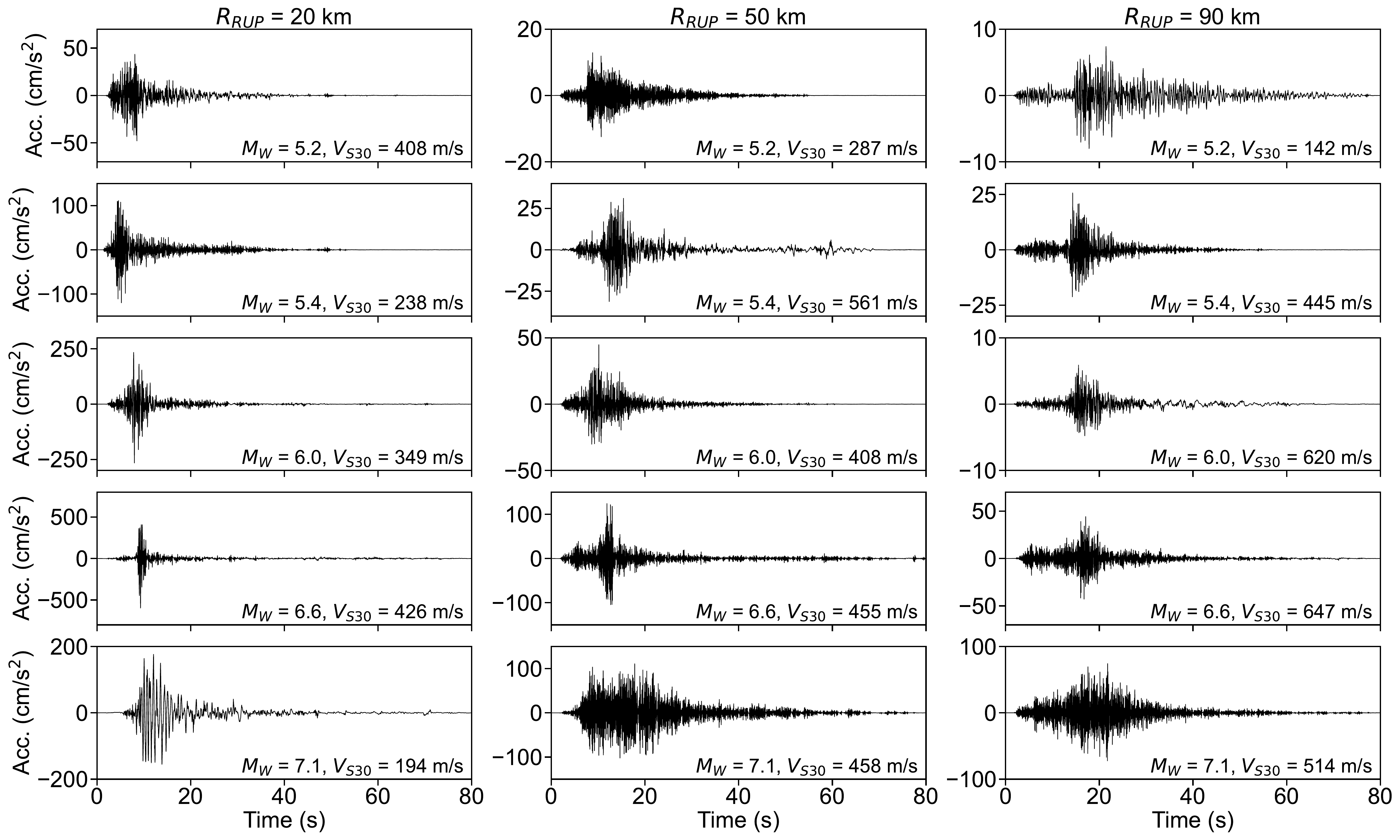}
  \caption{Examples of ground-motion waveforms generated by the CS-GMGM.
    The waveforms in each column correspond to the value of $R_{\mathrm{RUP}}$ shown at the top.
  Each panel shows the associated $M_W$ and $V_{\mathrm{S}30}$ values.}
\end{figure}%

\clearpage

\begin{figure}[h]
  \centering
  \includegraphics[width=\columnwidth]{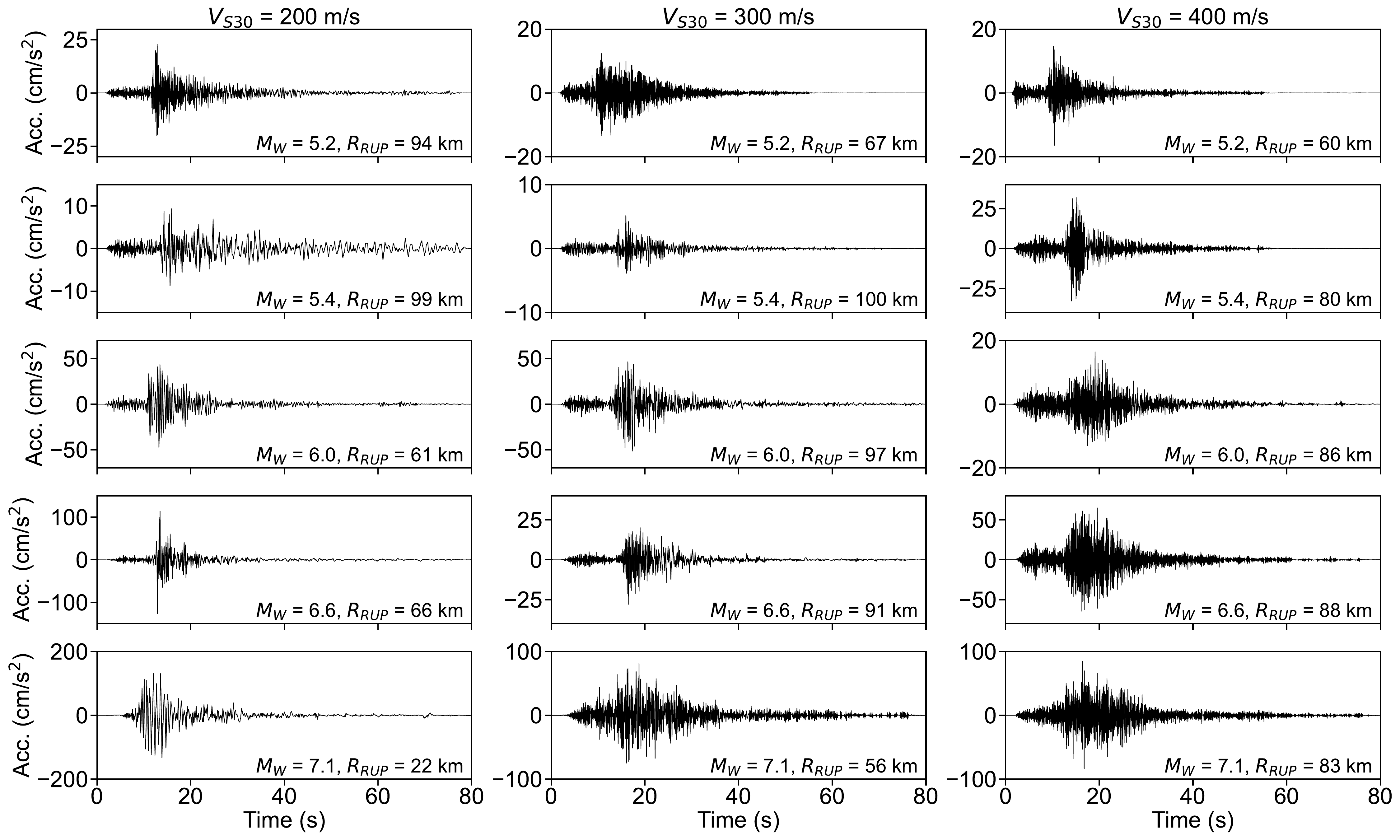}
  \caption{Examples of ground-motion waveforms generated by the CS-GMGM.
    The waveforms in each column correspond to the value of $V_{\mathrm{S}30}$ shown at the top.
  Each panel shows the associated $M_W$ and $R_{\mathrm{RUP}}$ values.}
\end{figure}%

\clearpage

\begin{figure}[ht]
  \centering
  \begin{subfigure}{\columnwidth}
    \centering
    \includegraphics[width=\columnwidth]{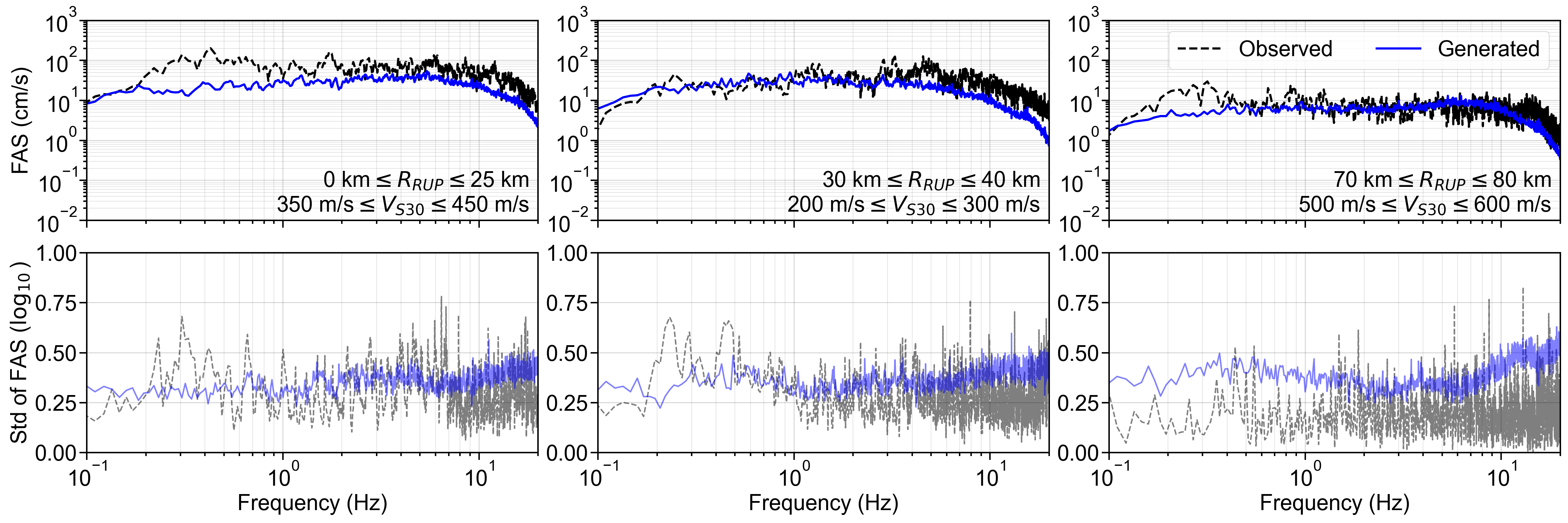}
    \caption{$6.7 \le M_W \le 6.9$}
  \end{subfigure} \\%
  \begin{subfigure}{\columnwidth}
    \centering
    \includegraphics[width=\columnwidth]{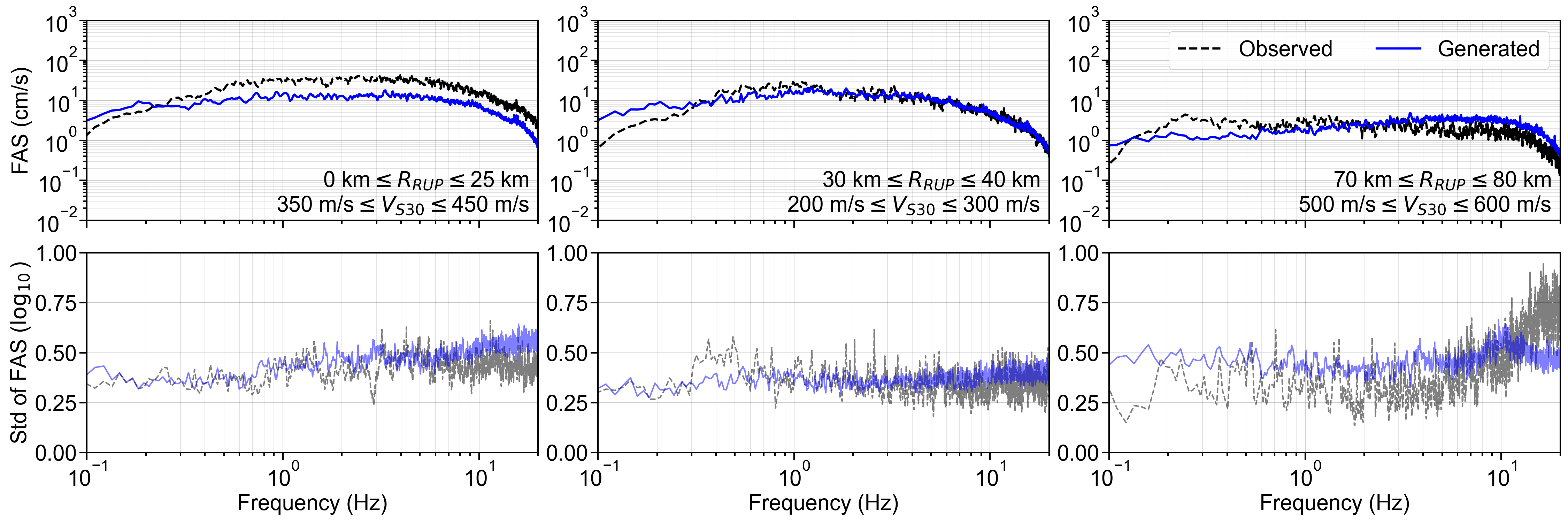}
    \caption{$6.1 \le M_W \le 6.3$}
  \end{subfigure} \\ %
  \begin{subfigure}{\columnwidth}
    \centering
    \includegraphics[width=\columnwidth]{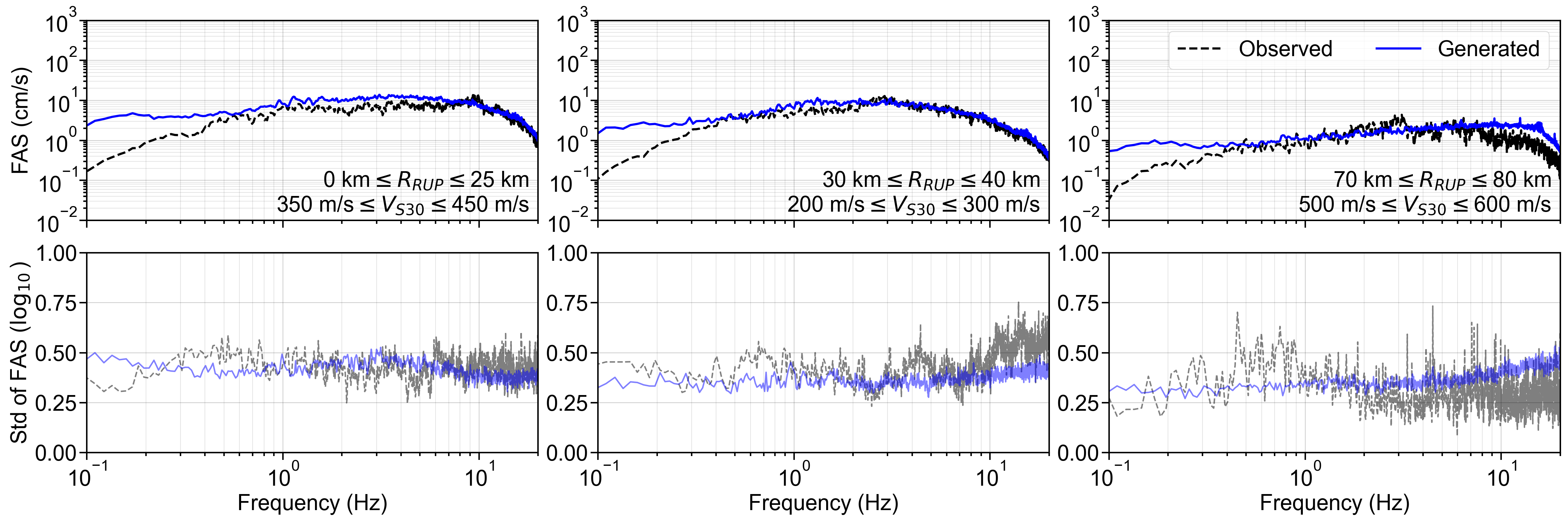}
    \caption{$5.5 \le M_W \le 5.7$}
  \end{subfigure} \\ %
  \caption{
    Comparison of the logarithmic means and logarithmic standard deviations of the Fourier amplitude spectra (FAS)
    generated by the S-GMGM across three different $M_W$ ranges.
    For each panel, the dashed black line represents the observed records,
    and the solid blue line represents the generated ground motions.
  }
  \label{fig:f_amp_comp_sgmgm_case_0}
\end{figure}%
\clearpage

\begin{figure}[ht]
  \centering
  \begin{subfigure}{\columnwidth}
    \centering
    \includegraphics[width=\columnwidth]{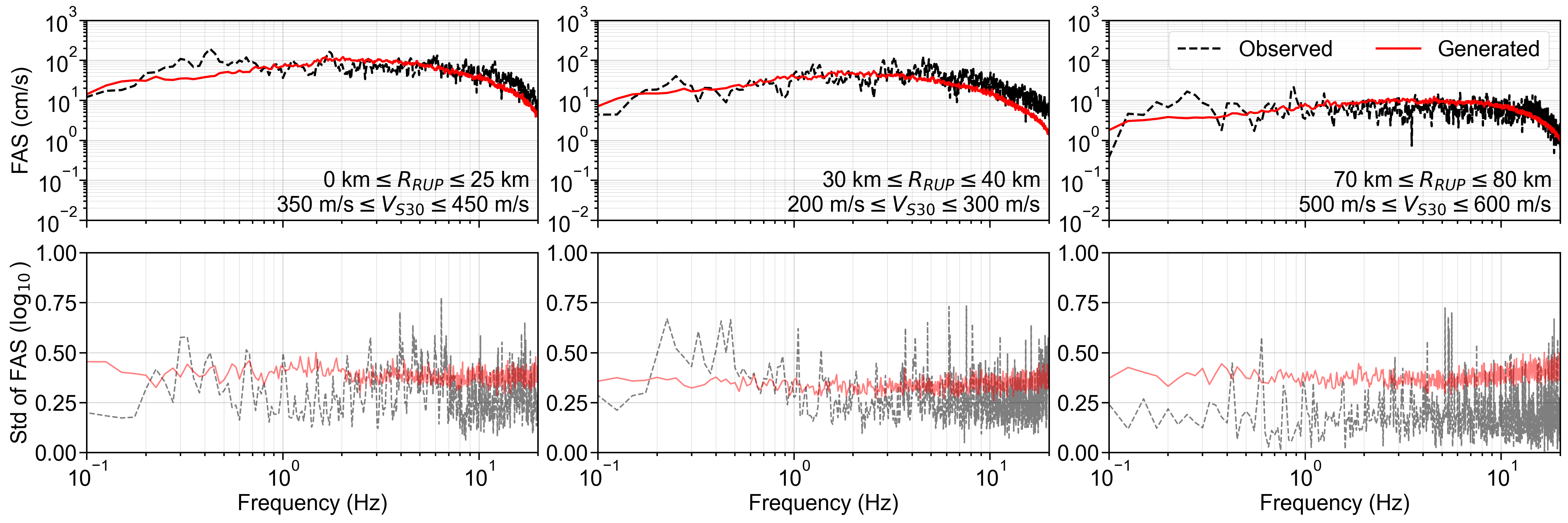}
    \caption{$6.7 \le M_W \le 6.9$}
  \end{subfigure} \\%
  \begin{subfigure}{\columnwidth}
    \centering
    \includegraphics[width=\columnwidth]{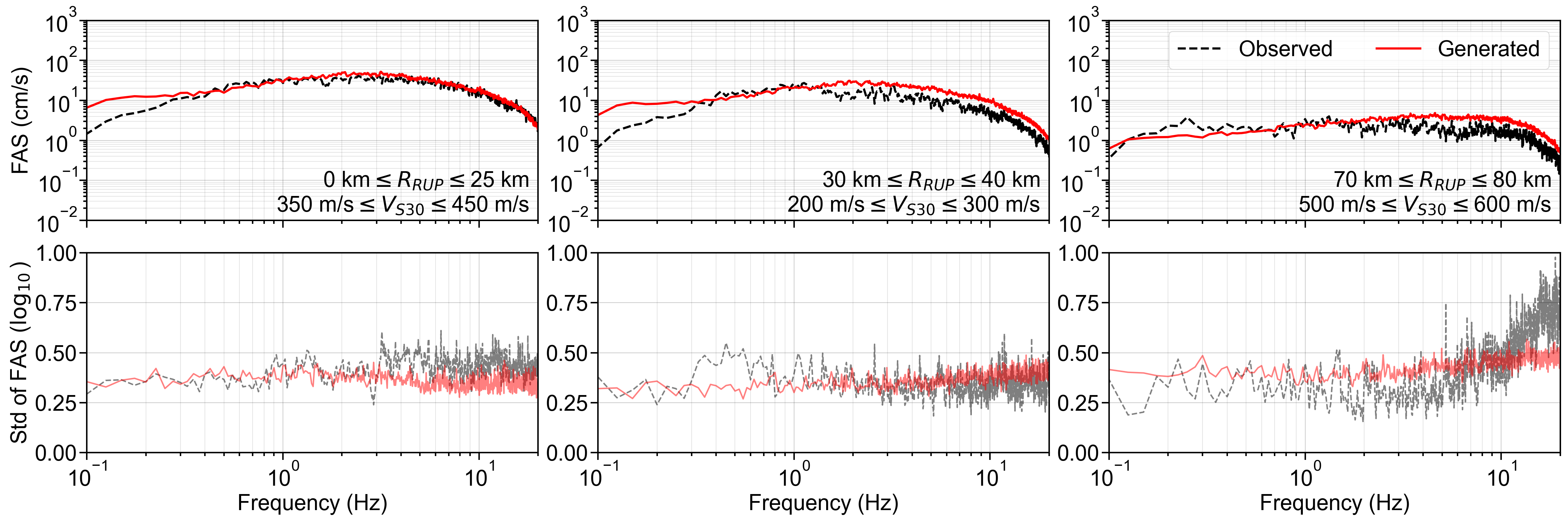}
    \caption{$6.1 \le M_W \le 6.3$}
  \end{subfigure} \\ %
  \begin{subfigure}{\columnwidth}
    \centering
    \includegraphics[width=\columnwidth]{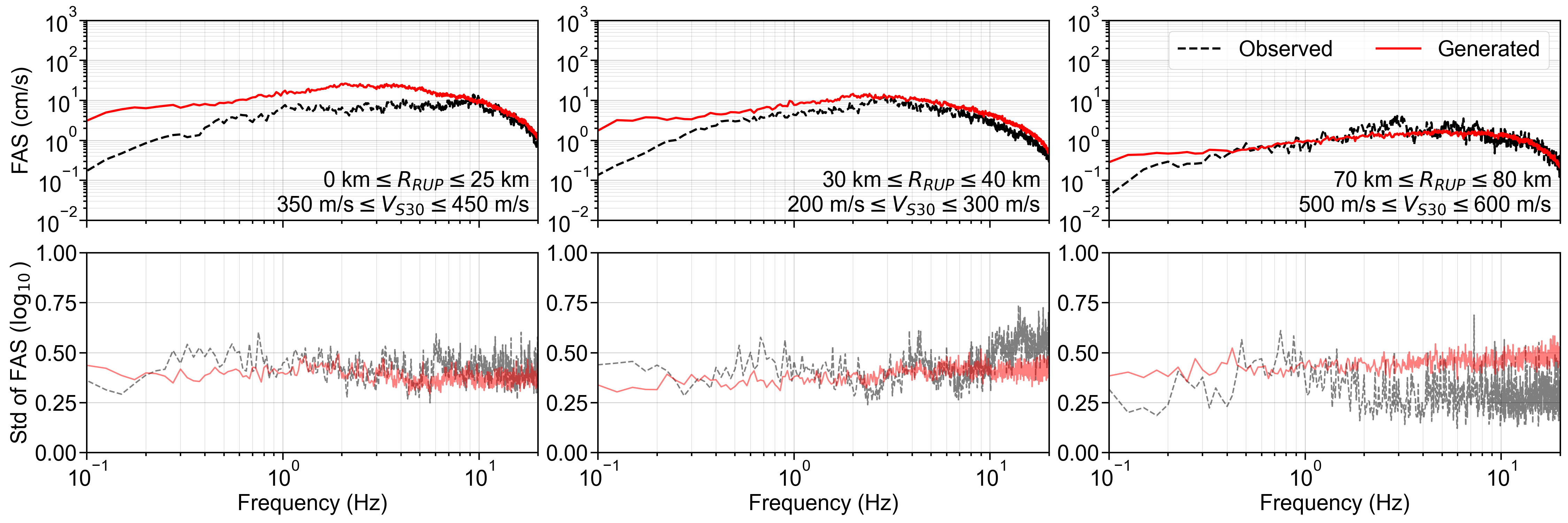}
    \caption{$5.5 \le M_W \le 5.7$}
  \end{subfigure} \\ %
  \caption{
    Comparison of the logarithmic means and logarithmic standard deviations of the Fourier amplitude spectra (FAS)
    generated by the CW-GMGM across three different $M_W$ ranges.
    For each panel, the dashed black line represents the observed records,
    and the solid red line represents the generated ground motions.
  }
  \label{fig:f_amp_comp_cwgmgm_case_0}
\end{figure}%
\clearpage

\begin{figure}[ht]
  \centering
  \begin{subfigure}{\columnwidth}
    \centering
    \includegraphics[width=\columnwidth]{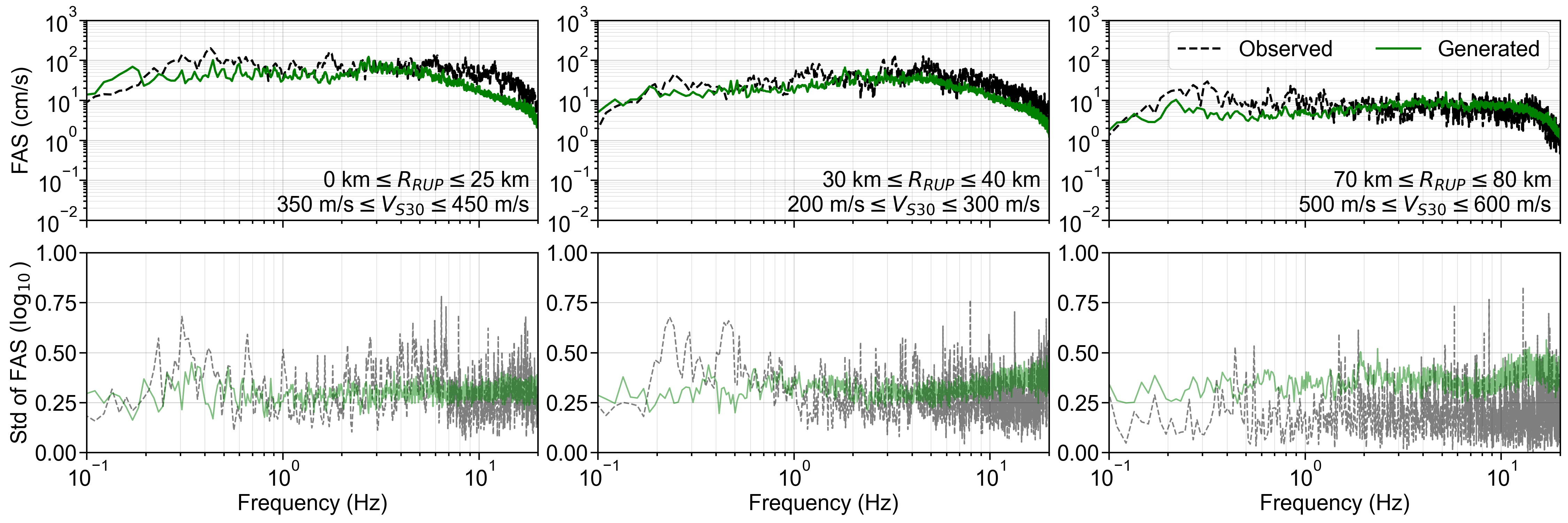}
    \caption{$6.7 \le M_W \le 6.9$}
  \end{subfigure} \\%
  \begin{subfigure}{\columnwidth}
    \centering
    \includegraphics[width=\columnwidth]{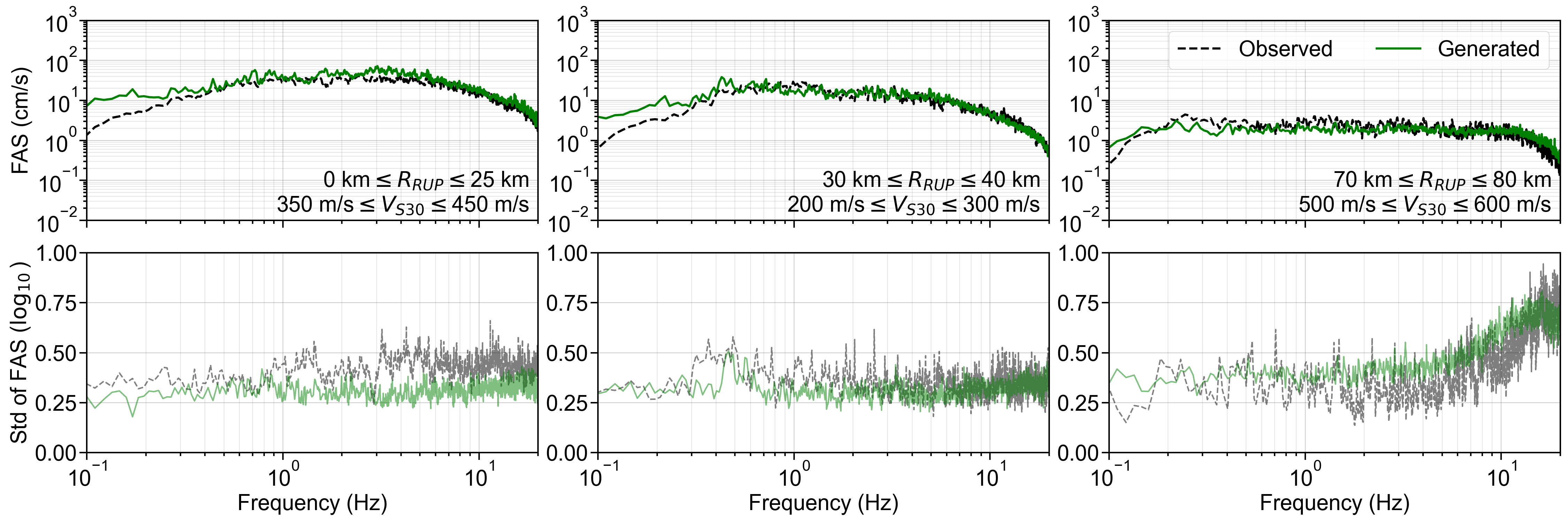}
    \caption{$6.1 \le M_W \le 6.3$}
  \end{subfigure} \\ %
  \begin{subfigure}{\columnwidth}
    \centering
    \includegraphics[width=\columnwidth]{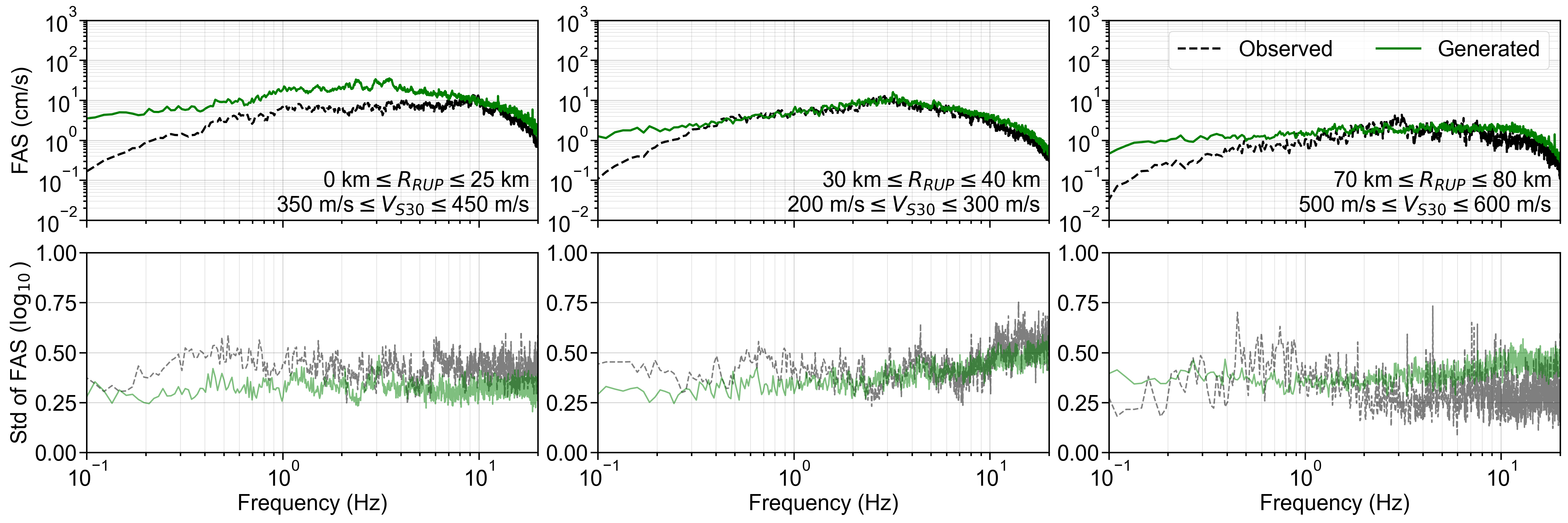}
    \caption{$5.5 \le M_W \le 5.7$}
  \end{subfigure} \\ %
  \caption{
    Comparison of the logarithmic means and logarithmic standard deviations of the Fourier amplitude spectra (FAS)
    generated by the CS-GMGM across three different $M_W$ ranges.
    For each panel, the dashed black line represents the observed records,
    and the solid green line represents the generated ground motions.}
  \label{fig:f_amp_comp_csgmgm_case_0}
\end{figure}%
\clearpage

\begin{figure}[ht]
  \centering
  \includegraphics[width=\columnwidth]{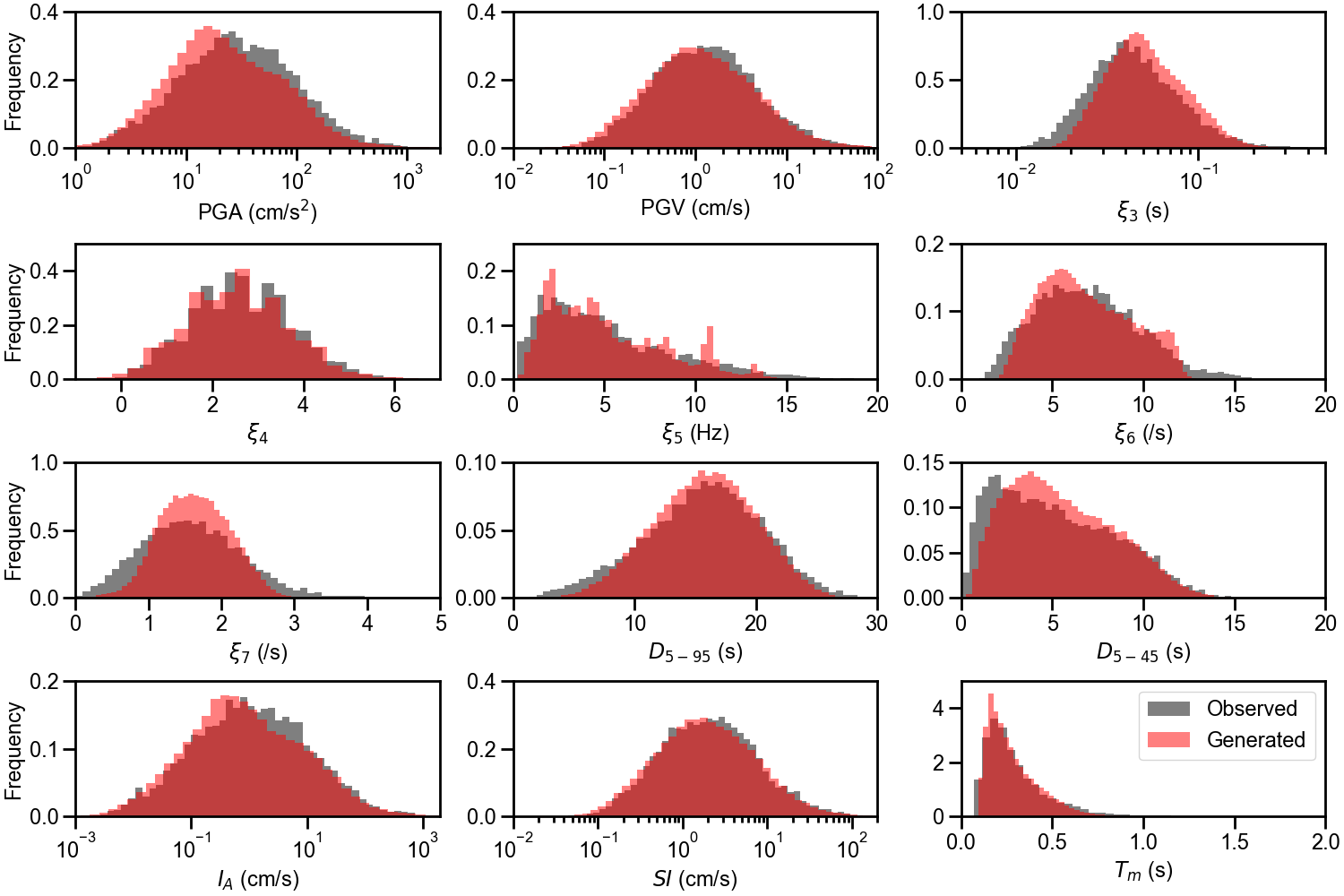}
  \caption{Comparison of the distributions of ground-motion characteristic indices between the observed records and
  the ground motions generated by the CW-GMGM.}
  \label{fig:hist_feature_cwgmgm}
\end{figure}%
\clearpage

\begin{figure}[ht]
  \centering
  \includegraphics[width=\columnwidth]{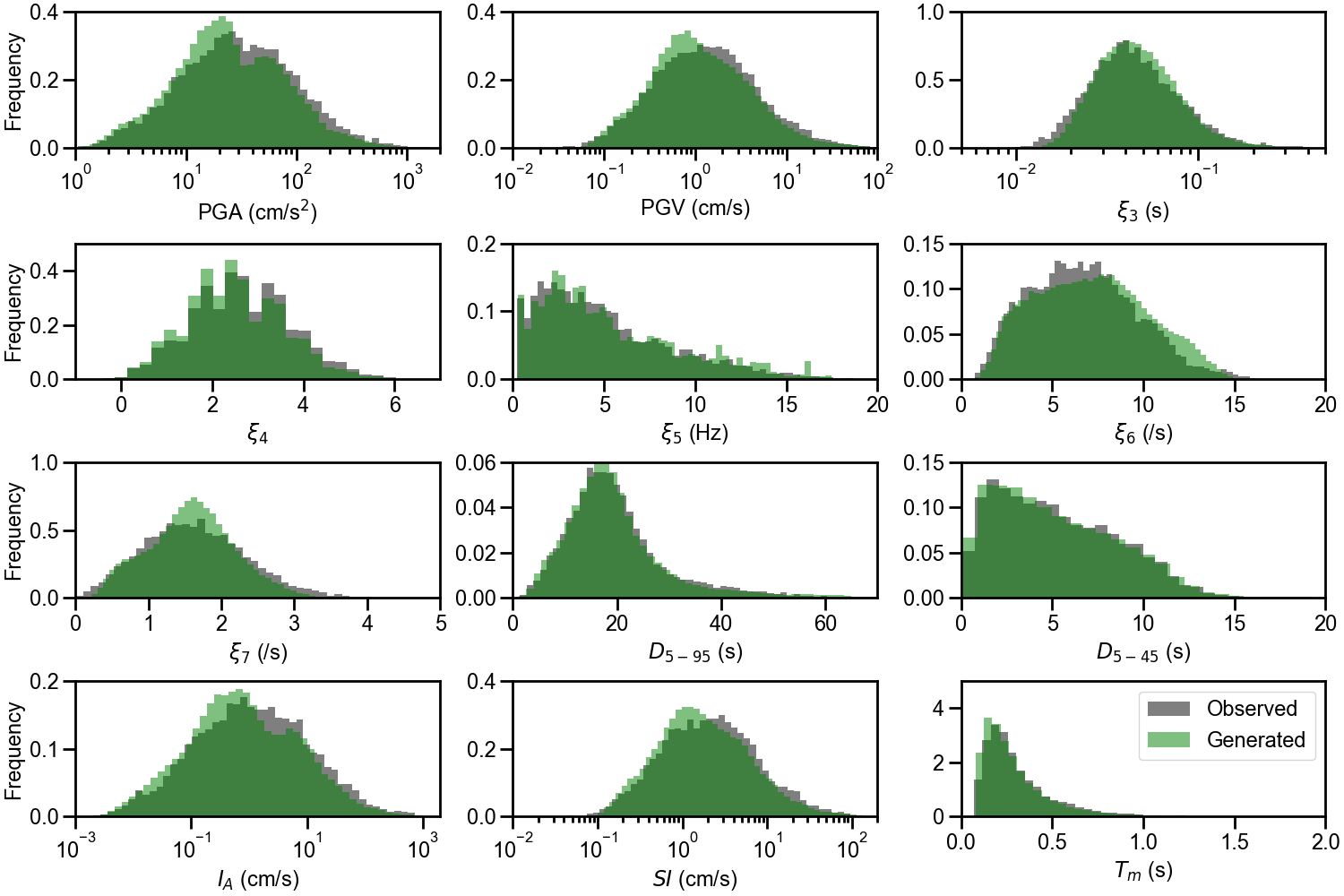}
  \caption{Comparison of the distributions of ground-motion characteristic indices between the observed records and
  the ground motions generated by the CS-GMGM.}
  \label{fig:hist_feature_csgmgm}
\end{figure}%
\clearpage

\begin{figure}[th]
  \centering
  \begin{minipage}{\columnwidth}
    \centering
    \includegraphics[width=0.4\columnwidth]{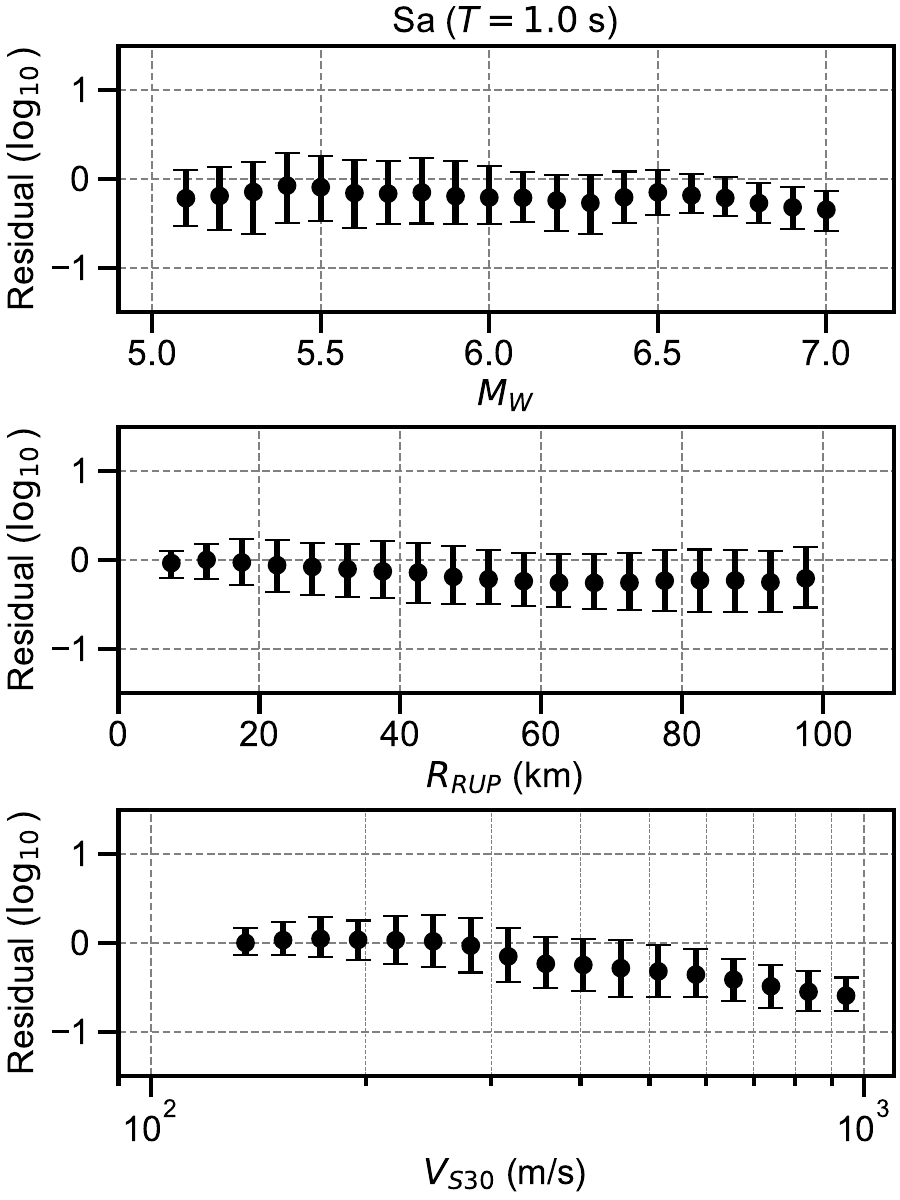}
    \caption{
      Residual plots for PGV comparing the S-GMGM with the SM99 GMM.
      Each bar shows the median and the 16th and 84th percentiles of the residuals.
    }
    \label{fig:residual_sgmgm_sm}
  \end{minipage}%

  \vspace{1.0cm}

  \begin{minipage}{\columnwidth}
    \centering
    \includegraphics[width=\columnwidth]{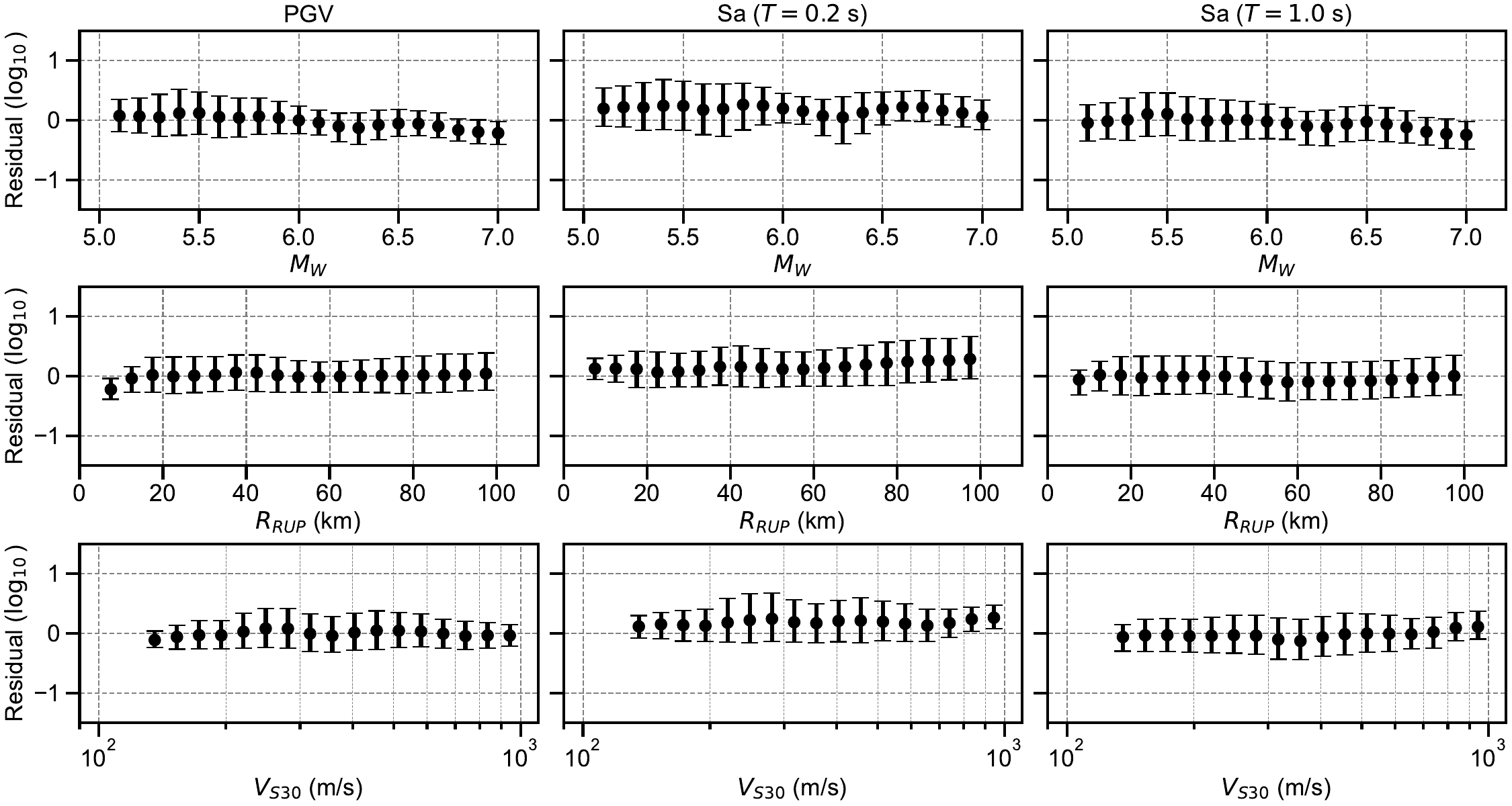}
    \caption{
      Residual plots comparing the S-GMGM with the ASK14 GMM.
      The IM corresponding to each column is indicated at the top:
      PGV on the left, the 5\% damped spectral acceleration at a natural period of 0.2 s in the center,
      and the 5\% damped spectral acceleration at a natural period of 1.0 s on the right.
      Each bar shows the median and the 16th and 84th percentiles of the residuals.
    }
    \label{fig:residual_sgmgm_ask}
  \end{minipage}%
\end{figure}%
\clearpage

\begin{figure}[th]
  \centering
  \begin{minipage}{\columnwidth}
    \centering
    \includegraphics[width=\columnwidth]{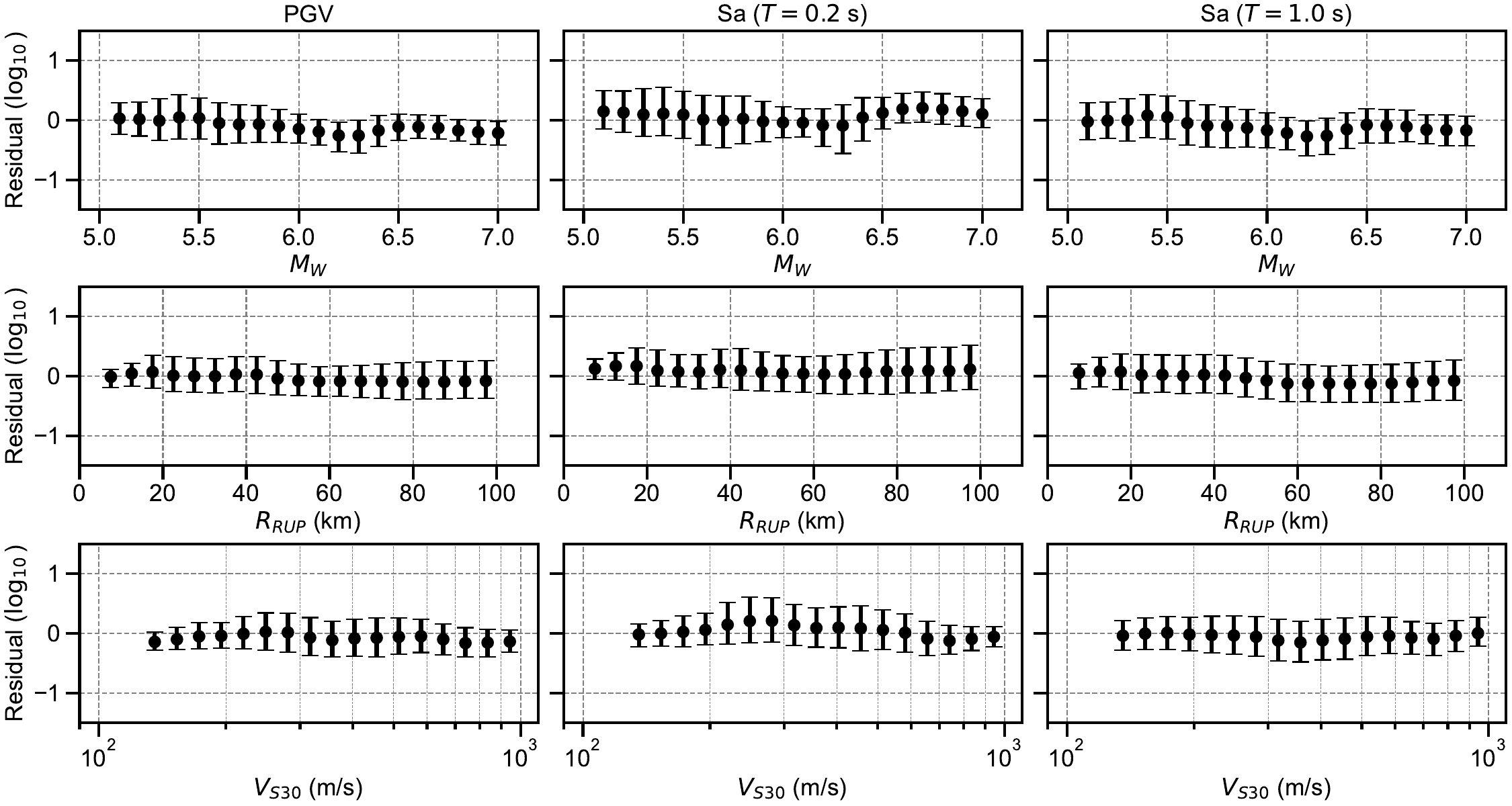}
    \caption{
      Residual plots comparing the S-GMGM with the BSSA14 GMM.
      Each bar shows the median and the 16th and 84th percentiles of the residuals.
    }
    \label{fig:residual_sgmgm_bssa}
  \end{minipage}%

  \vspace{2.0cm}

  \begin{minipage}{\columnwidth}
    \centering
    \includegraphics[width=\columnwidth]{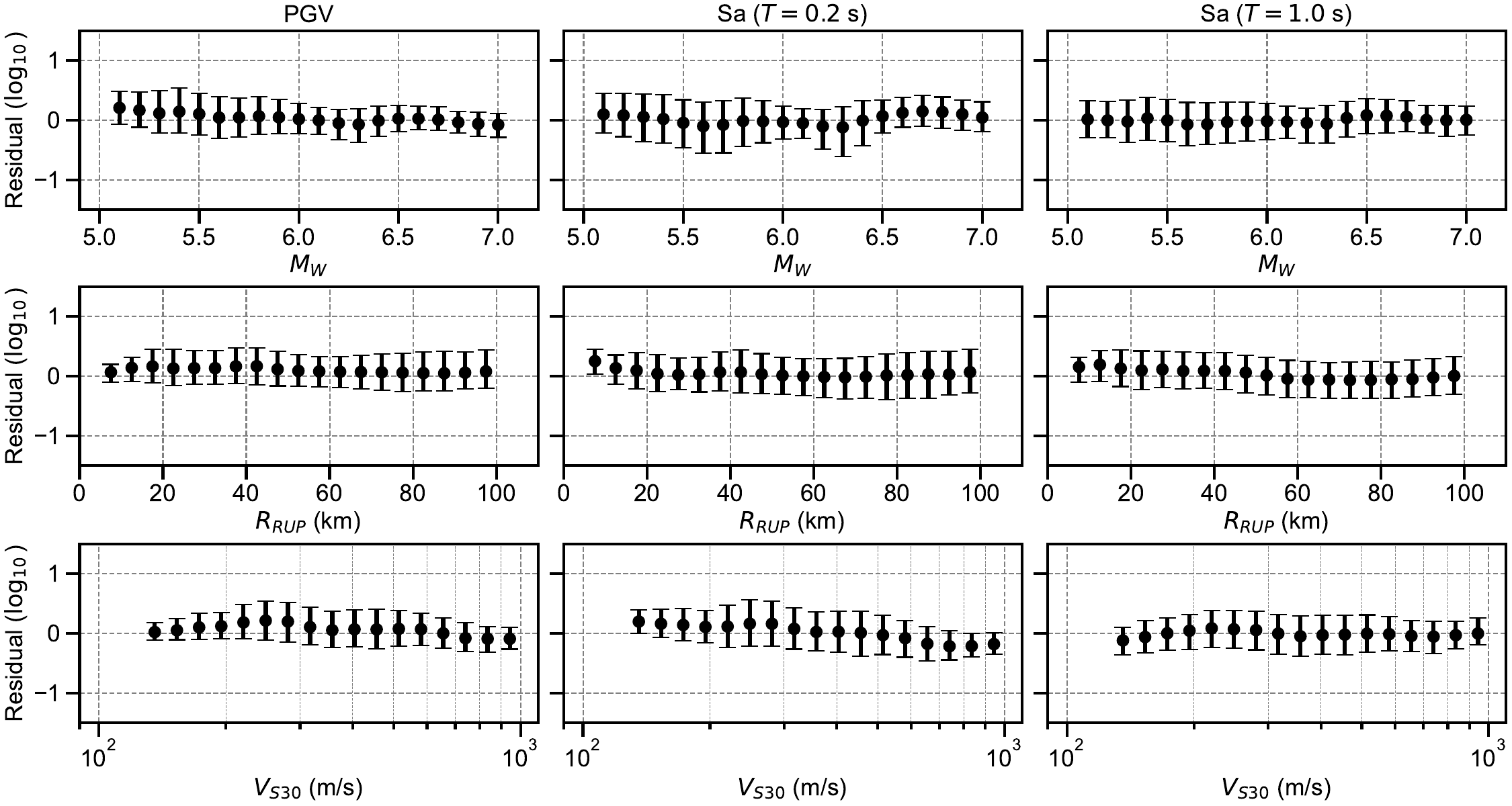}
    \caption{
      Residual plots comparing the S-GMGM with the CB14 GMM.
      Each bar shows the median and the 16th and 84th percentiles of the residuals.
    }
    \label{fig:residual_sgmgm_cb}
  \end{minipage}%
\end{figure}%
\clearpage

\begin{figure}[th]
  \centering
  \begin{minipage}{\columnwidth}
    \centering
    \includegraphics[width=\columnwidth]{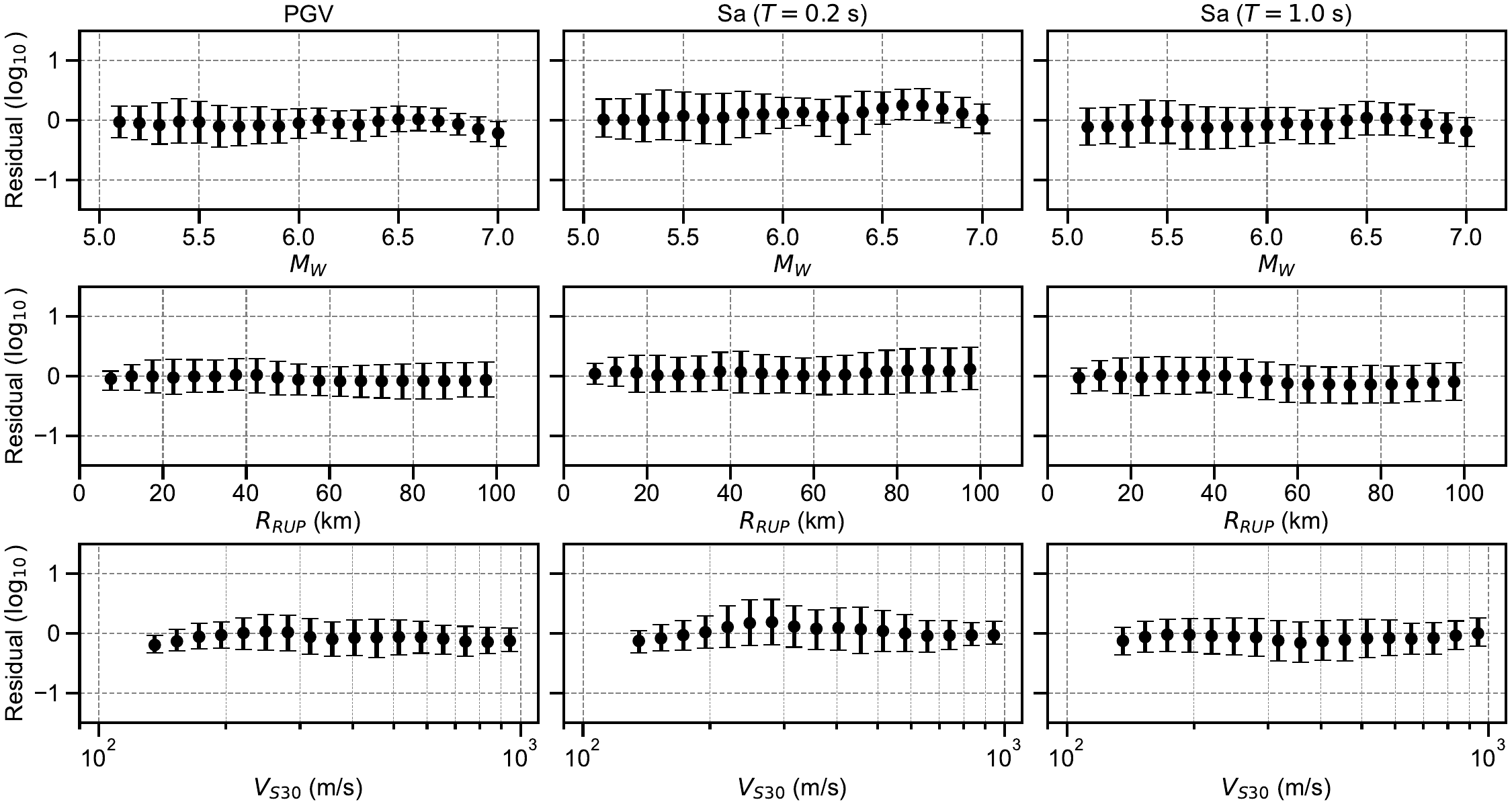}
    \caption{
      Residual plots comparing the S-GMGM with the CY14 GMM.
      Each bar shows the median and the 16th and 84th percentiles of the residuals.
    }
    \label{fig:residual_sgmgm_cy}
  \end{minipage}%
\end{figure}%
\clearpage

\begin{figure}[th]
  \centering
  \begin{minipage}{\columnwidth}
    \centering
    \includegraphics[width=0.4\columnwidth]{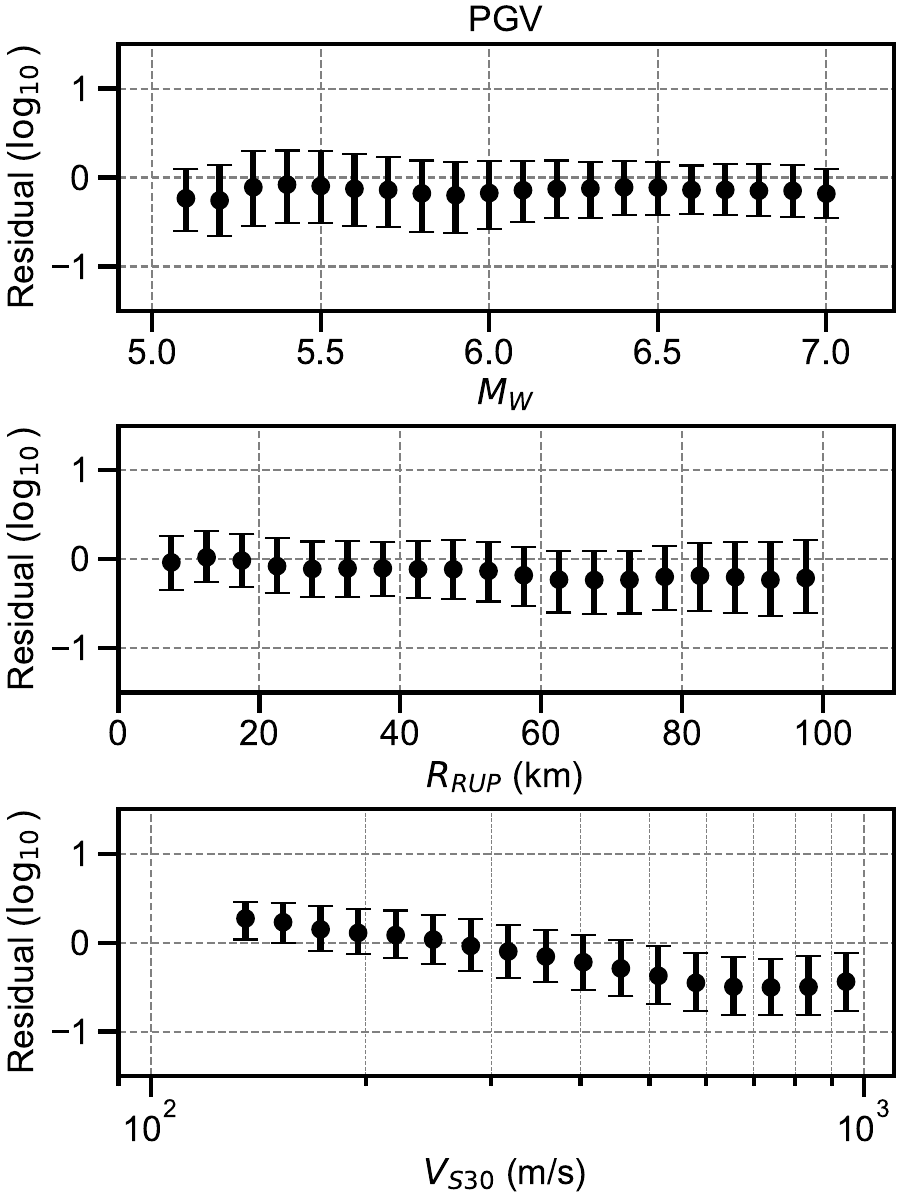}
    \caption{
      Residual plots comparing the CW-GMGM with the SM99 GMM.
      Each bar shows the median and the 16th and 84th percentiles of the residuals.
    }
    \label{fig:residual_cwgmgm_sm}
  \end{minipage}%

  \vspace{2.0cm}

  \begin{minipage}{\columnwidth}
    \centering
    \includegraphics[width=\columnwidth]{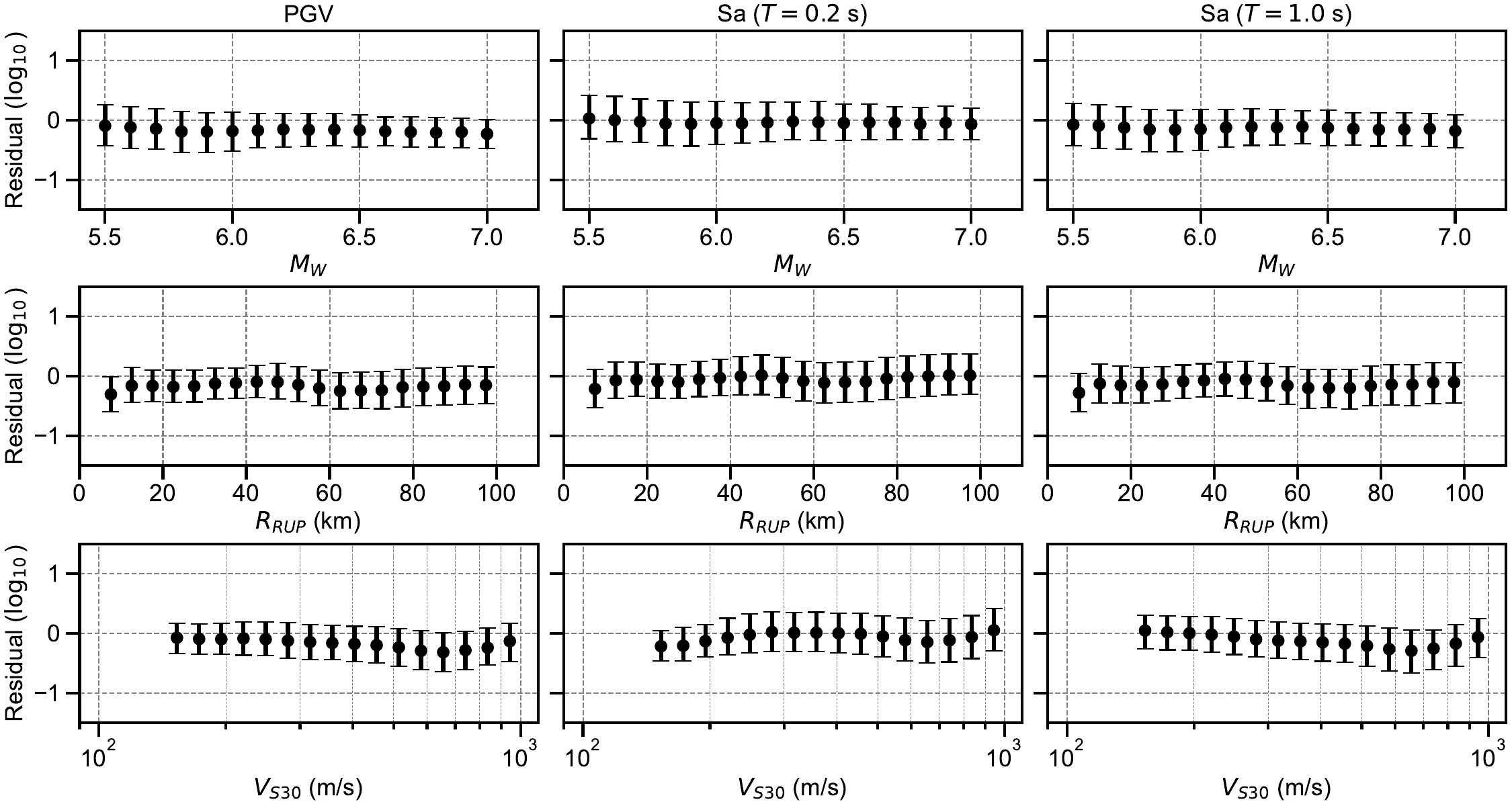}
    \caption{
      Residual plots comparing the CW-GMGM with the MF13 GMM.
      Each bar shows the median and the 16th and 84th percentiles of the residuals.
    }
    \label{fig:residual_cwgmgm_mf}
  \end{minipage}%
\end{figure}%
\clearpage

\begin{figure}[th]
  \centering
  \begin{minipage}{\columnwidth}
    \centering
    \includegraphics[width=\columnwidth]{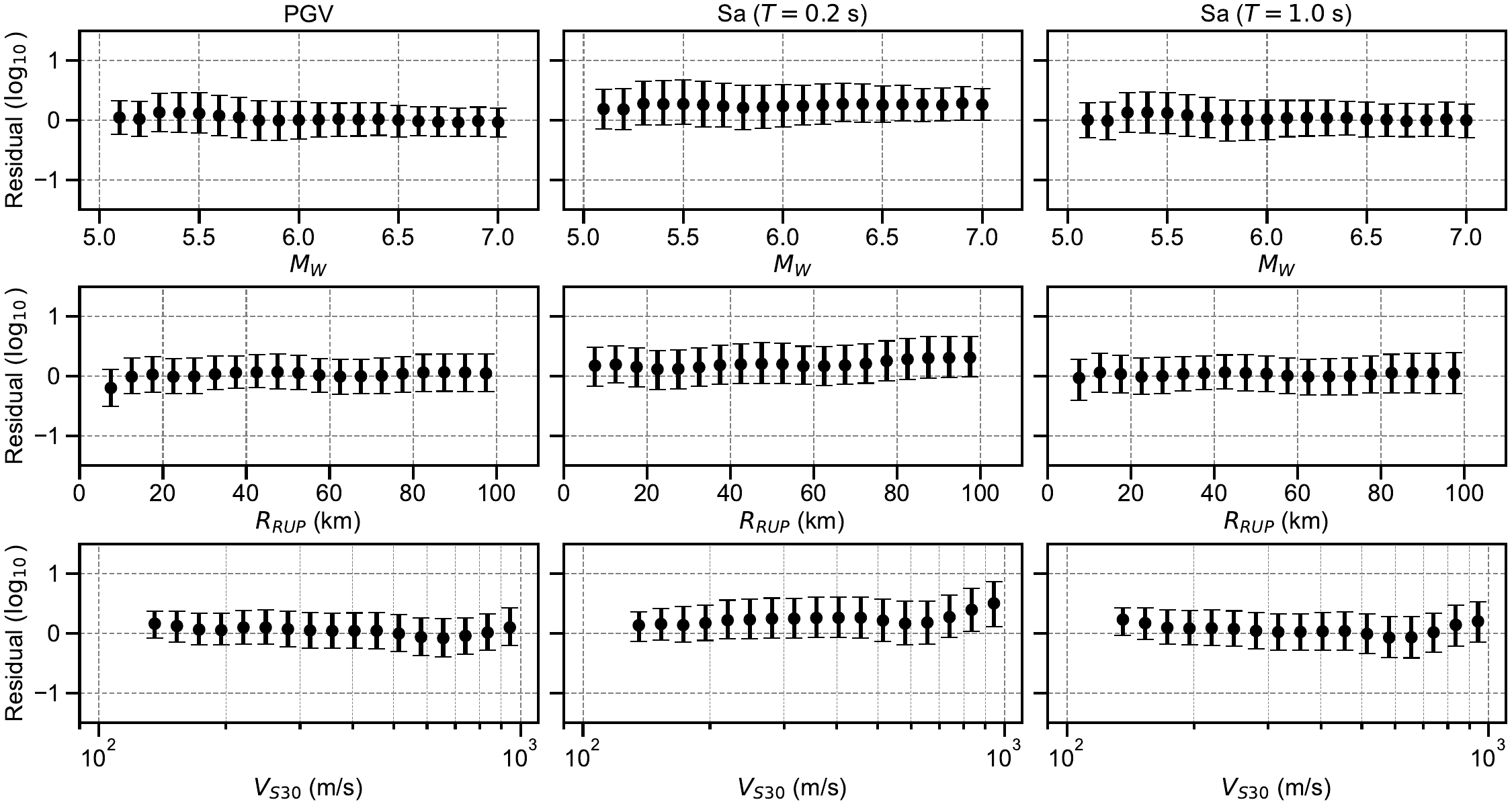}
    \caption{
      Residual plots comparing the CW-GMGM with the ASK14 GMM.
      Each bar shows the median and the 16th and 84th percentiles of the residuals.
    }
    \label{fig:residual_cwgmgm_ask}
  \end{minipage}%

  \vspace{2.0cm}

  \begin{minipage}{\columnwidth}
    \centering
    \includegraphics[width=\columnwidth]{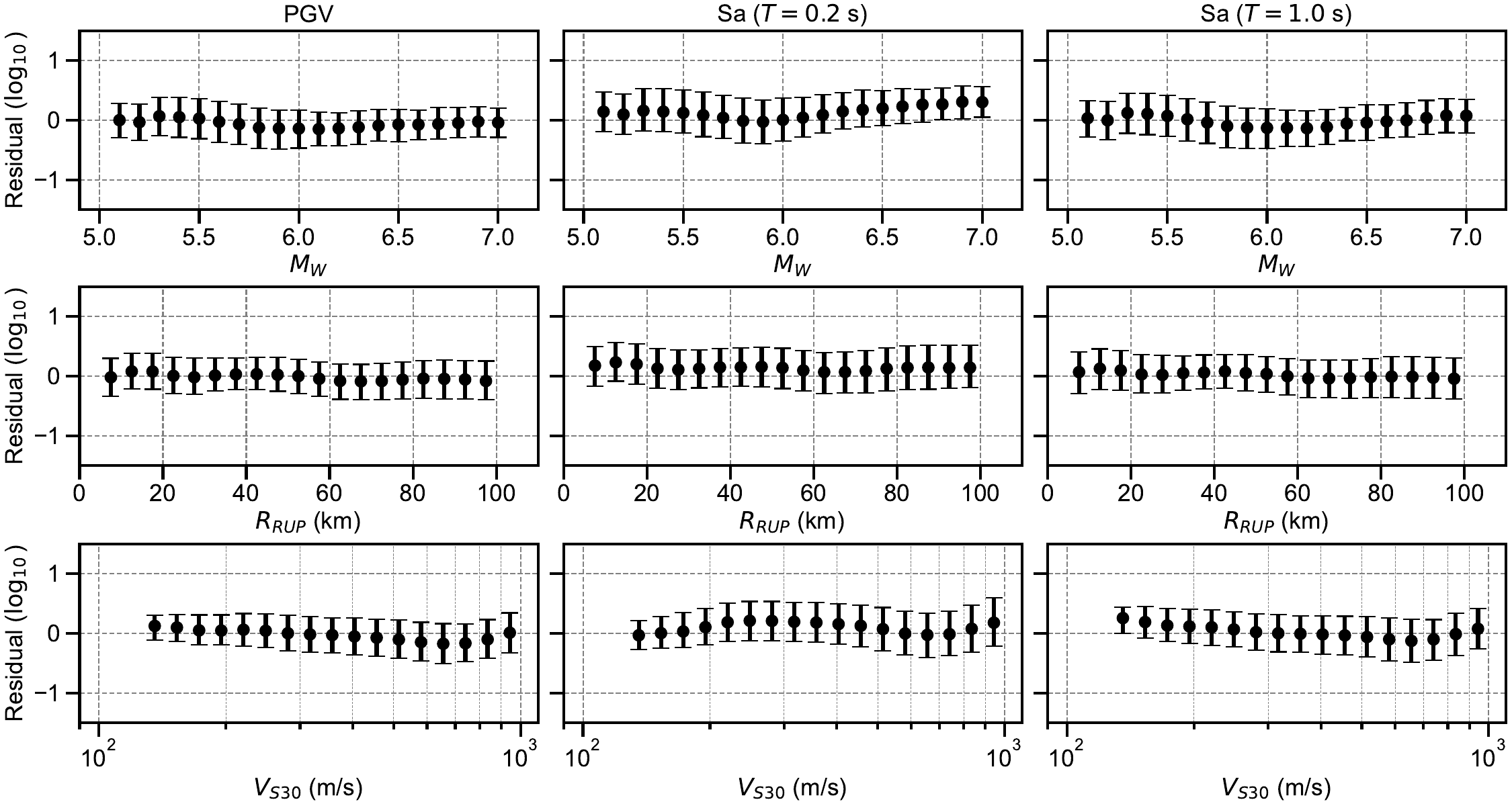}
    \caption{
      Residual plots comparing the CW-GMGM with the BSSA14 GMM.
      Each bar shows the median and the 16th and 84th percentiles of the residuals.
    }
    \label{fig:residual_cwgmgm_bssa}
  \end{minipage}%
\end{figure}%
\clearpage

\begin{figure}[th]
  \centering
  \begin{minipage}{\columnwidth}
    \centering
    \includegraphics[width=\columnwidth]{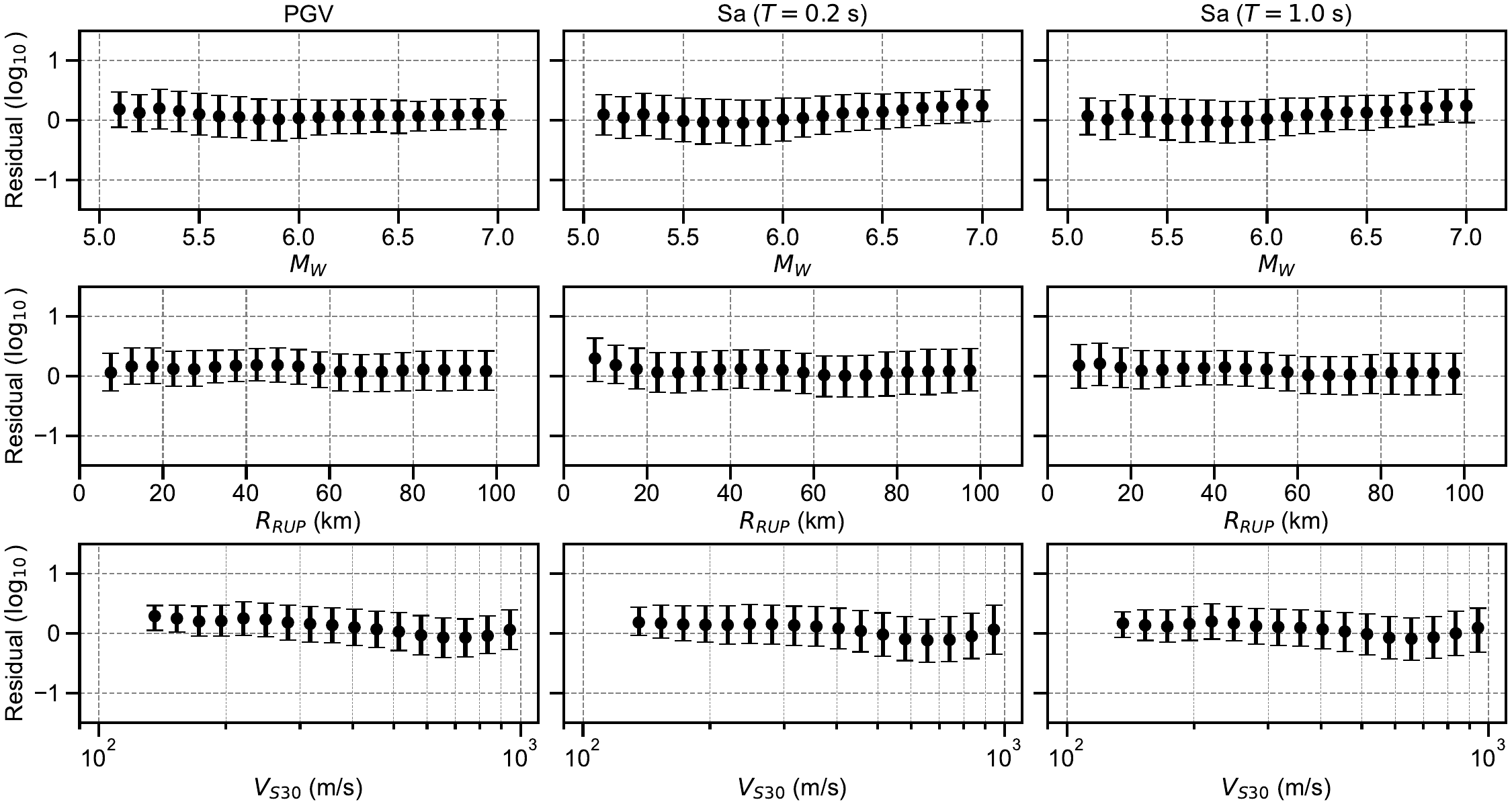}
    \caption{
      Residual plots comparing the CW-GMGM with the CB14 GMM.
      Each bar shows the median and the 16th and 84th percentiles of the residuals.
    }
    \label{fig:residual_cwgmgm_cb}
  \end{minipage}%

  \vspace{2.0cm}

  \begin{minipage}{\columnwidth}
    \centering
    \includegraphics[width=\columnwidth]{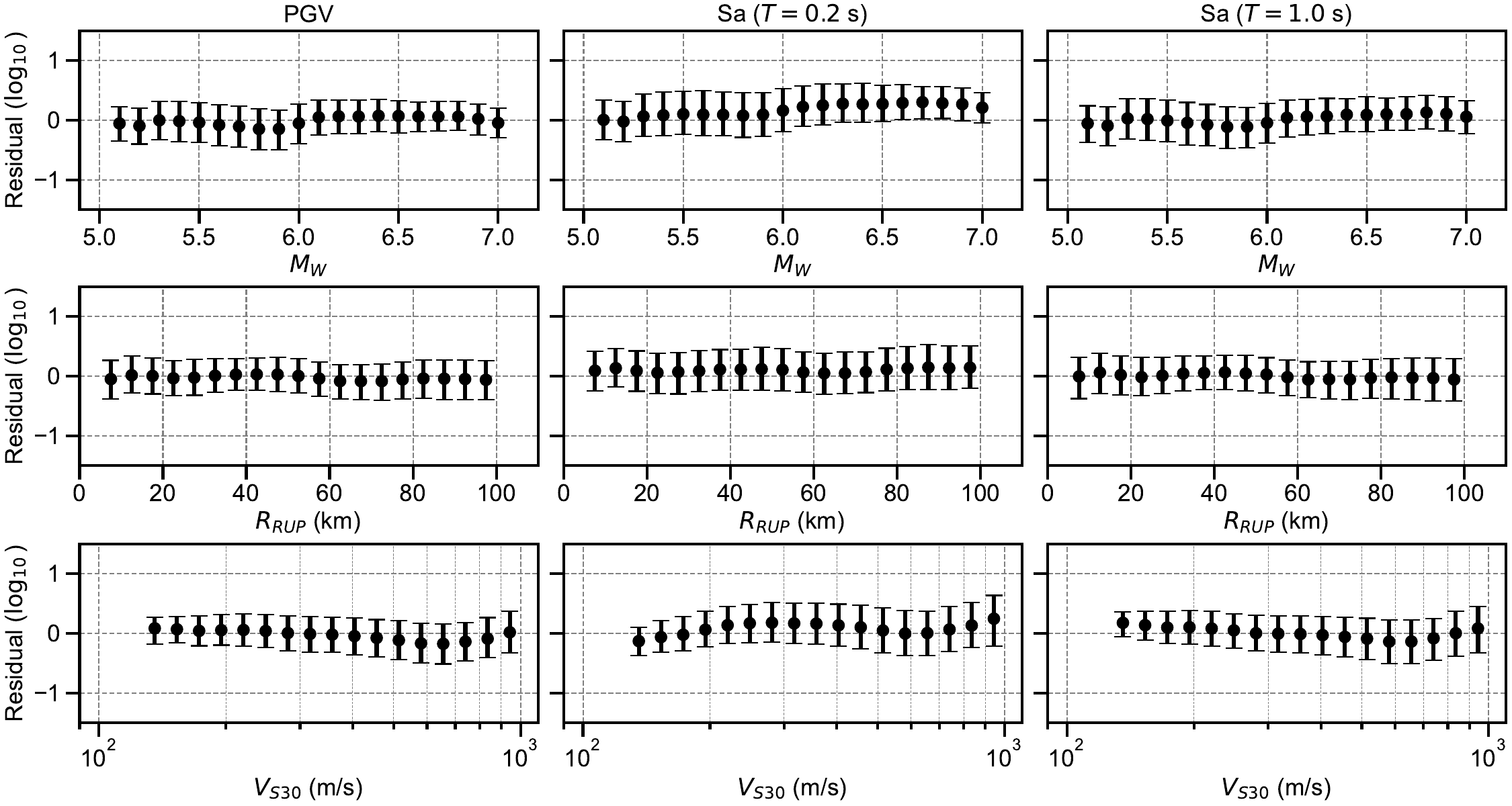}
    \caption{
      Residual plots comparing the CW-GMGM with the CY14 GMM.
      Each bar shows the median and the 16th and 84th percentiles of the residuals.
    }
    \label{fig:residual_cwgmgm_cy}
  \end{minipage}%
\end{figure}%
\clearpage

\begin{figure}[th]
  \centering
  \begin{minipage}{\columnwidth}
    \centering
    \includegraphics[width=0.4\columnwidth]{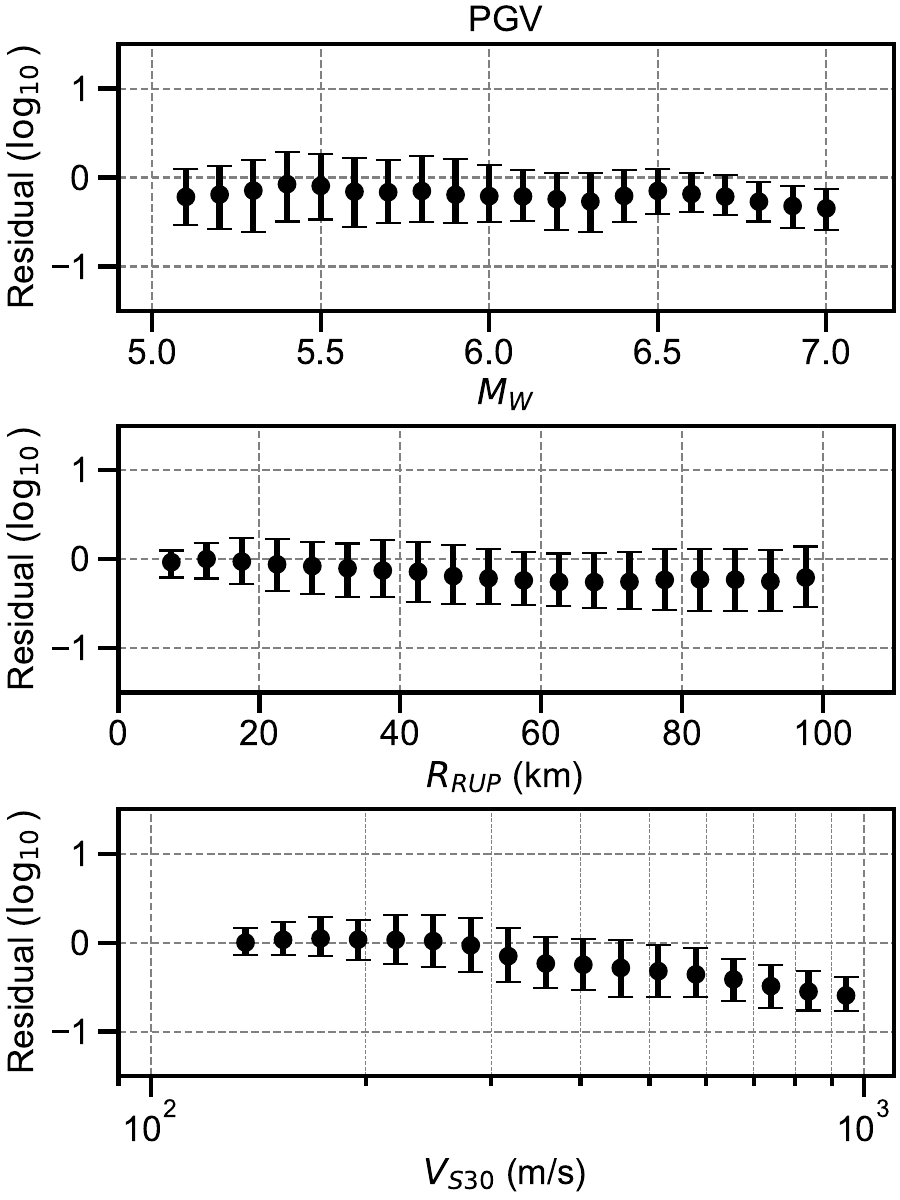}
    \caption{
      Residual plots comparing the CS-GMGM with the SM99 GMM.
      Each bar shows the median and the 16th and 84th percentiles of the residuals.
    }
    \label{fig:residual_csgmgm_sm}
  \end{minipage}%

  \vspace{2.0cm}

  \begin{minipage}{\columnwidth}
    \centering
    \includegraphics[width=\columnwidth]{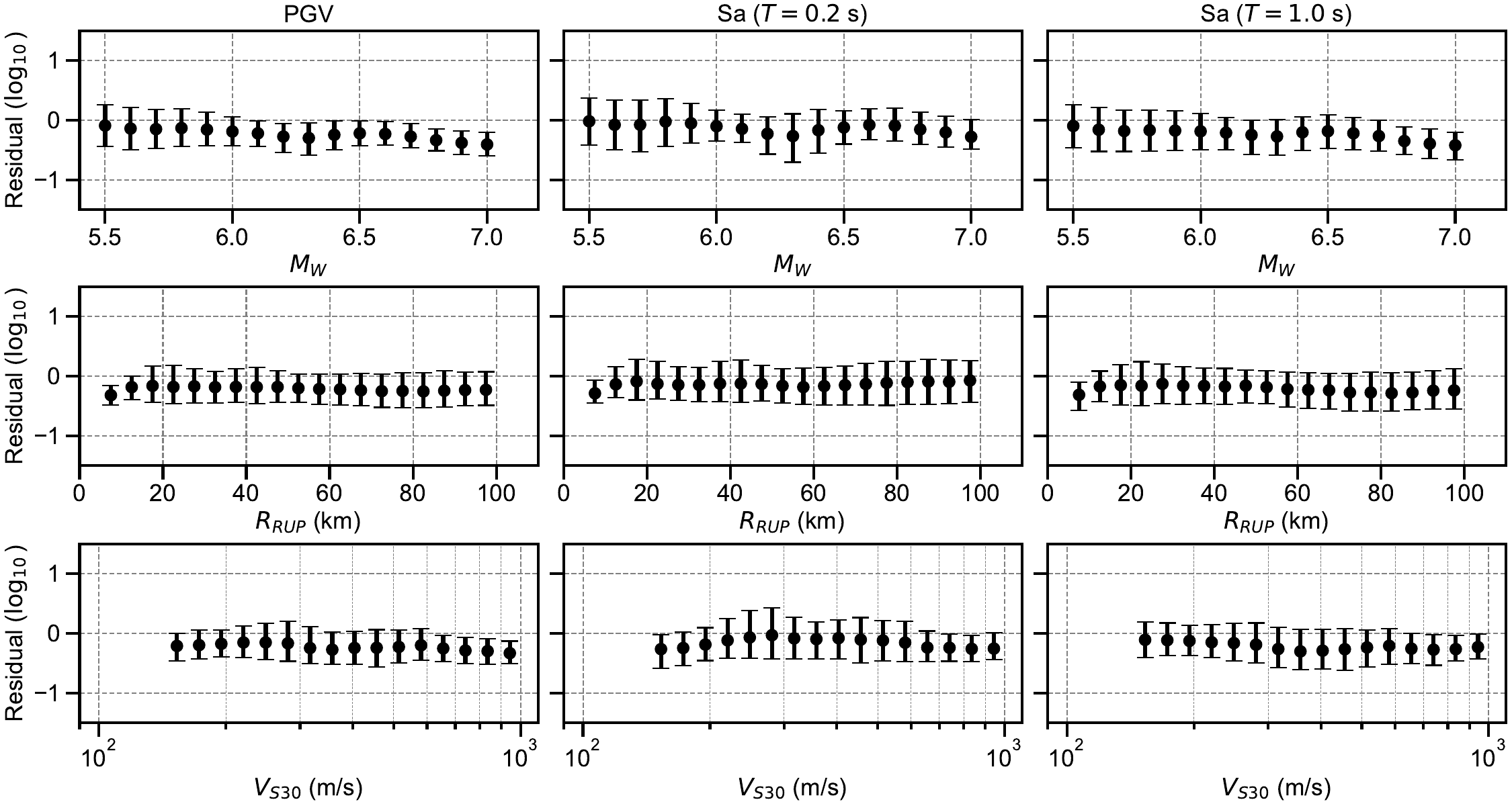}
    \caption{
      Residual plots comparing the CS-GMGM with the MF13 GMM.
      Each bar shows the median and the 16th and 84th percentiles of the residuals.
    }
    \label{fig:residual_csgmgm_mf}
  \end{minipage}%
\end{figure}%
\clearpage

\begin{figure}[th]
  \centering
  \begin{minipage}{\columnwidth}
    \centering
    \includegraphics[width=\columnwidth]{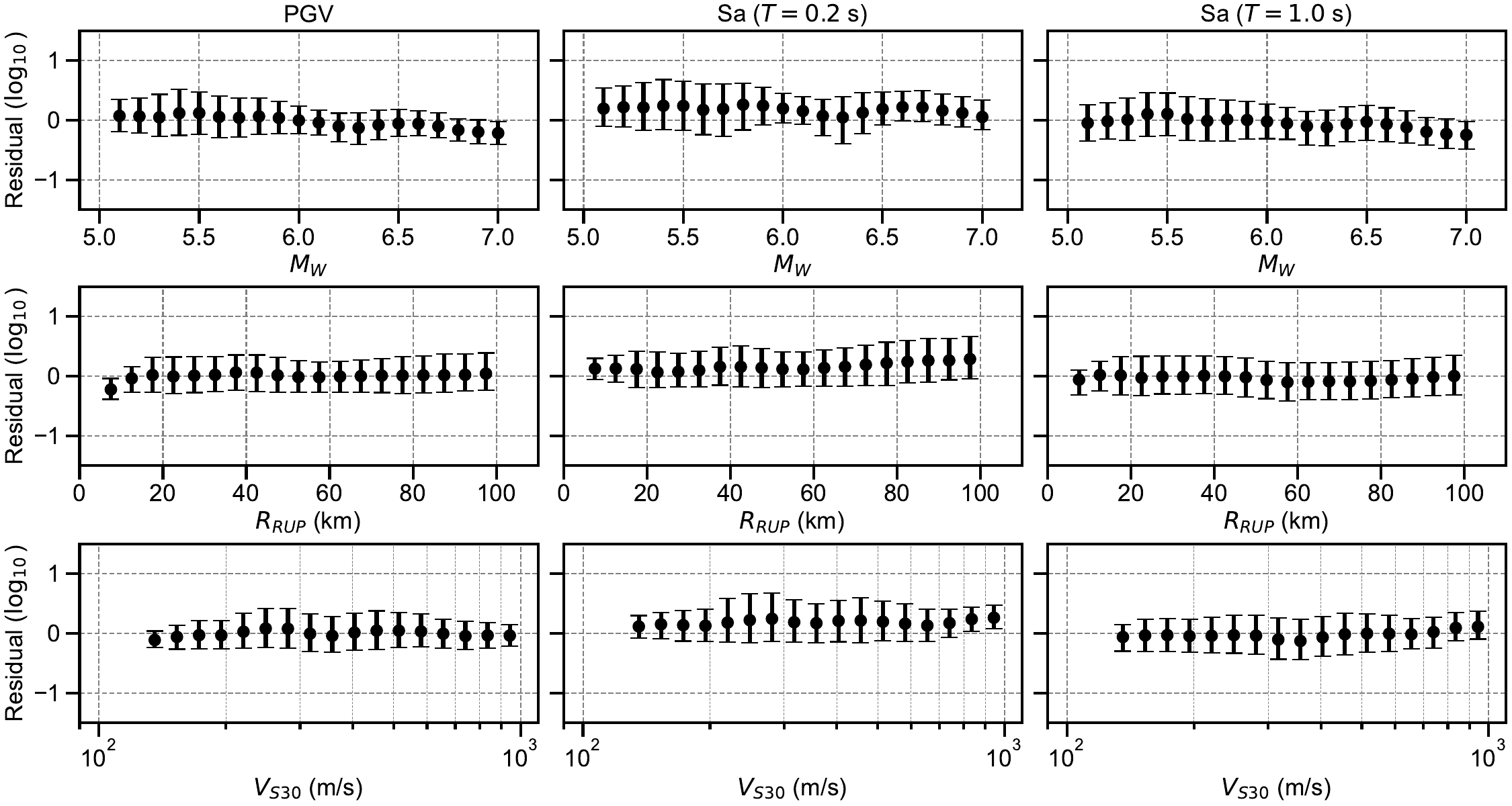}
    \caption{
      Residual plots comparing the CS-GMGM with the ASK14 GMM.
      Each bar shows the median and the 16th and 84th percentiles of the residuals.
    }
    \label{fig:residual_csgmgm_ask}
  \end{minipage}%

  \vspace{2.0cm}

  \begin{minipage}{\columnwidth}
    \centering
    \includegraphics[width=\columnwidth]{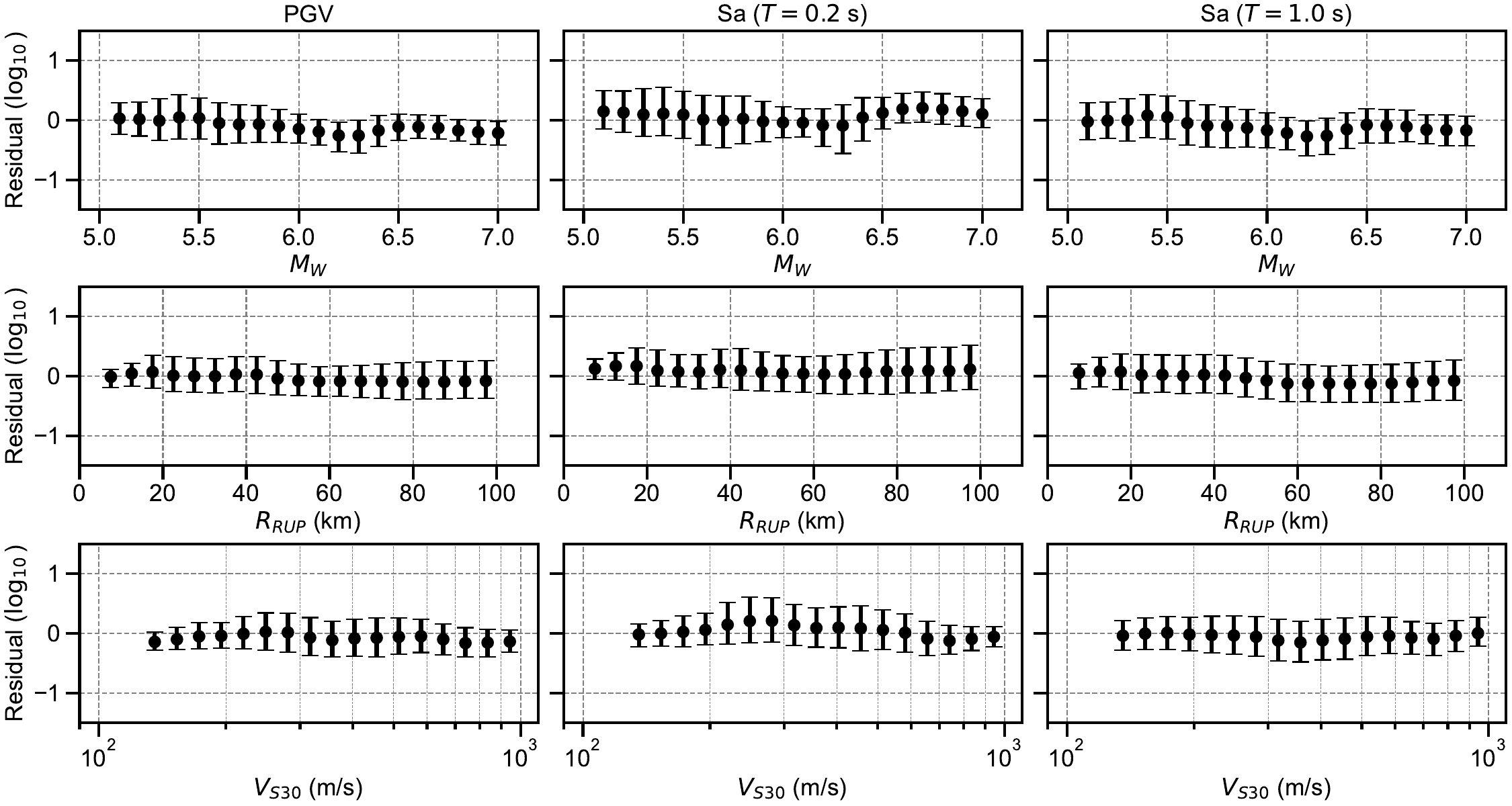}
    \caption{
      Residual plots comparing the CS-GMGM with the BSSA14 GMM.
      Each bar shows the median and the 16th and 84th percentiles of the residuals.
    }
    \label{fig:residual_csgmgm_bssa}
  \end{minipage}%
\end{figure}%
\clearpage

\begin{figure}[th]
  \centering
  \begin{minipage}{\columnwidth}
    \centering
    \includegraphics[width=\columnwidth]{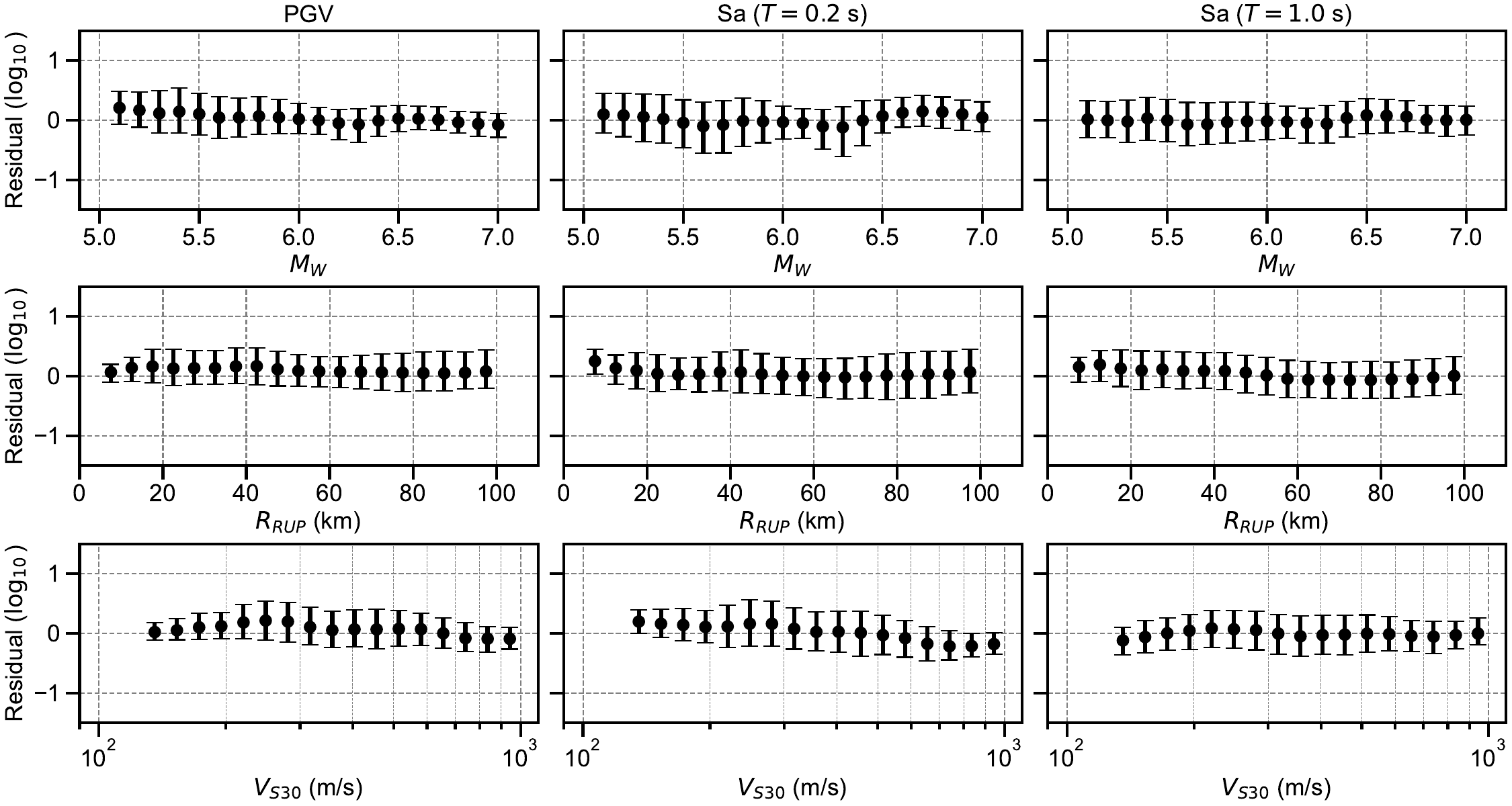}
    \caption{
      Residual plots comparing the CS-GMGM with the CB14 GMM.
      Each bar shows the median and the 16th and 84th percentiles of the residuals.
    }
    \label{fig:residual_csgmgm_cb}
  \end{minipage}%

  \vspace{2.0cm}

  \begin{minipage}{\columnwidth}
    \centering
    \includegraphics[width=\columnwidth]{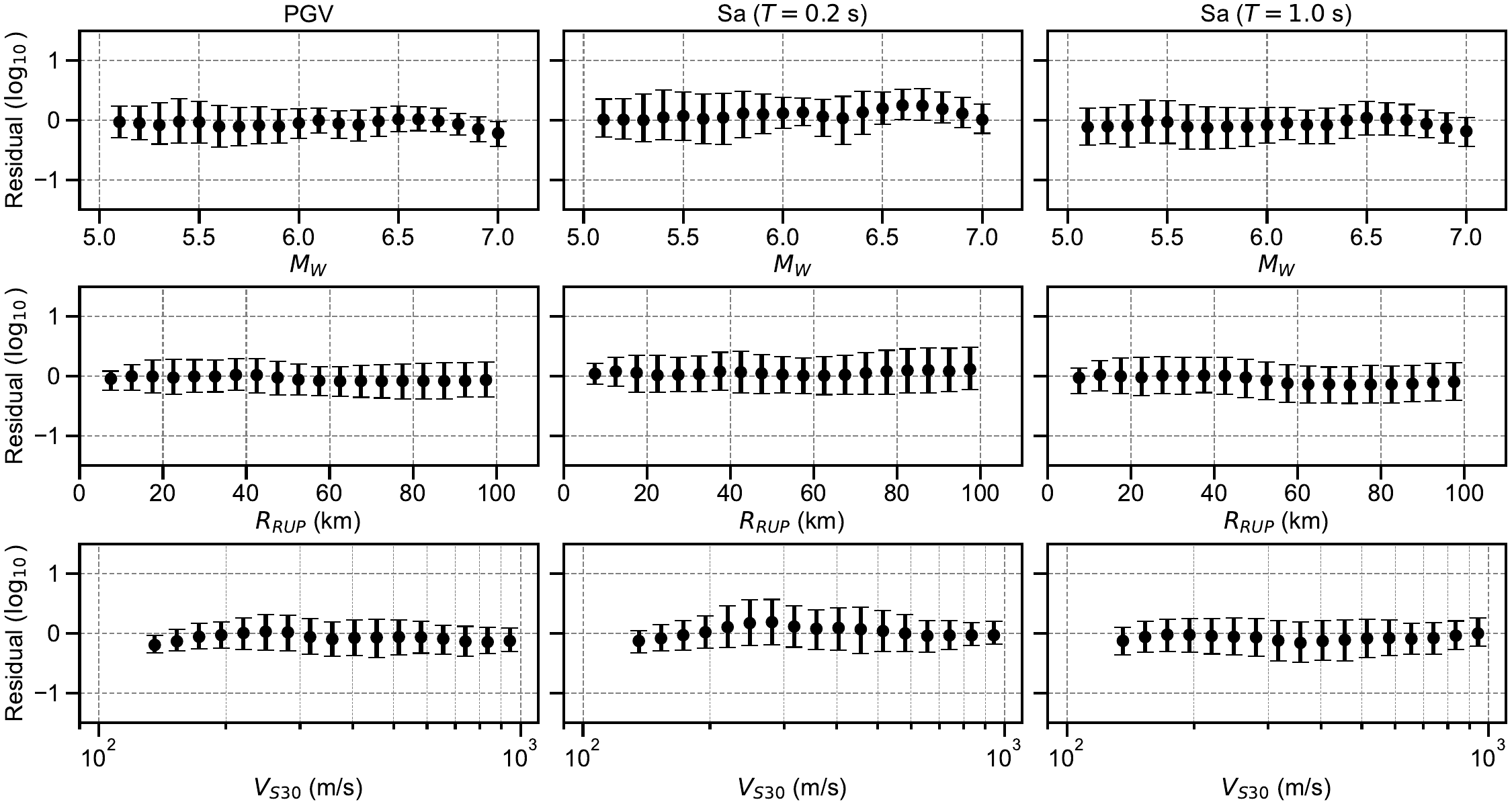}
    \caption{
      Residual plots comparing the CS-GMGM with the CY14 GMM.
      Each bar shows the median and the 16th and 84th percentiles of the residuals.
    }
    \label{fig:residual_csgmgm_cy}
  \end{minipage}%
\end{figure}%
\clearpage

\begin{figure}[ht]
  \centering
  \begin{minipage}{\columnwidth}
    \centering
  \includegraphics[width=\columnwidth]{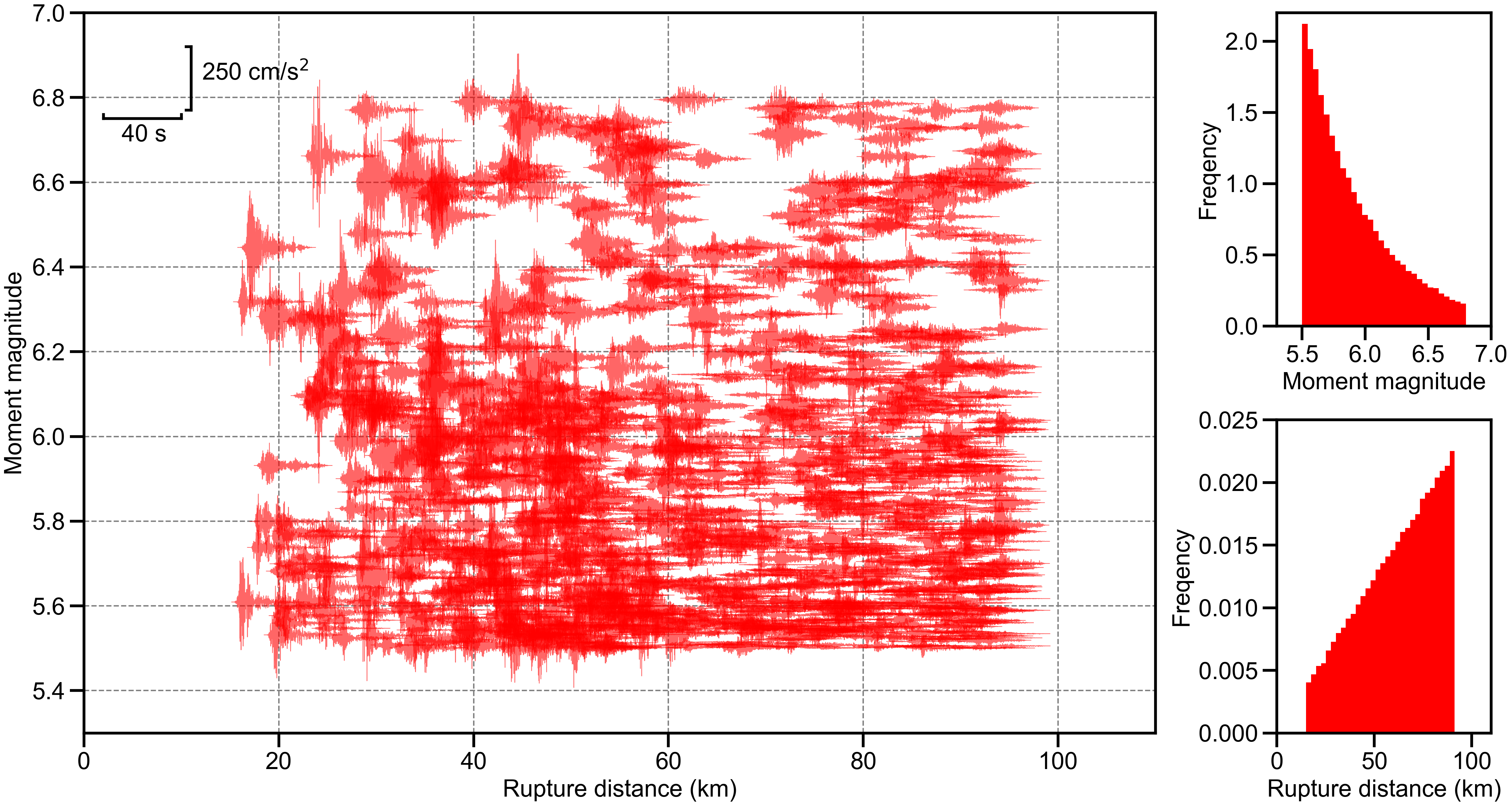}
  \caption{
    Hazard analysis results for numerical example 1 using the CW-GMGM.
    The left panel shows 1,000 acceleration waveforms selected from the sampled ground motions.
    The beginning of each waveform corresponds to its associated $M_W$ and $R_{\mathrm{RUP}}$ values.
    For clarity, only the first 40 s of each waveform are shown.
    On the right hand side, the upper histogram presents the distribution of $M_W$ for all samples,
    and the lower histogram shows the distribution of $R_{\mathrm{RUP}}$.
  }
  \label{fig:gm_scat_loc_2_site_cwgmgm}
  \end{minipage}%
\end{figure}%
\clearpage

\begin{figure}[ht]
  \centering
  \begin{minipage}{\columnwidth}
    \centering
  \includegraphics[width=\columnwidth]{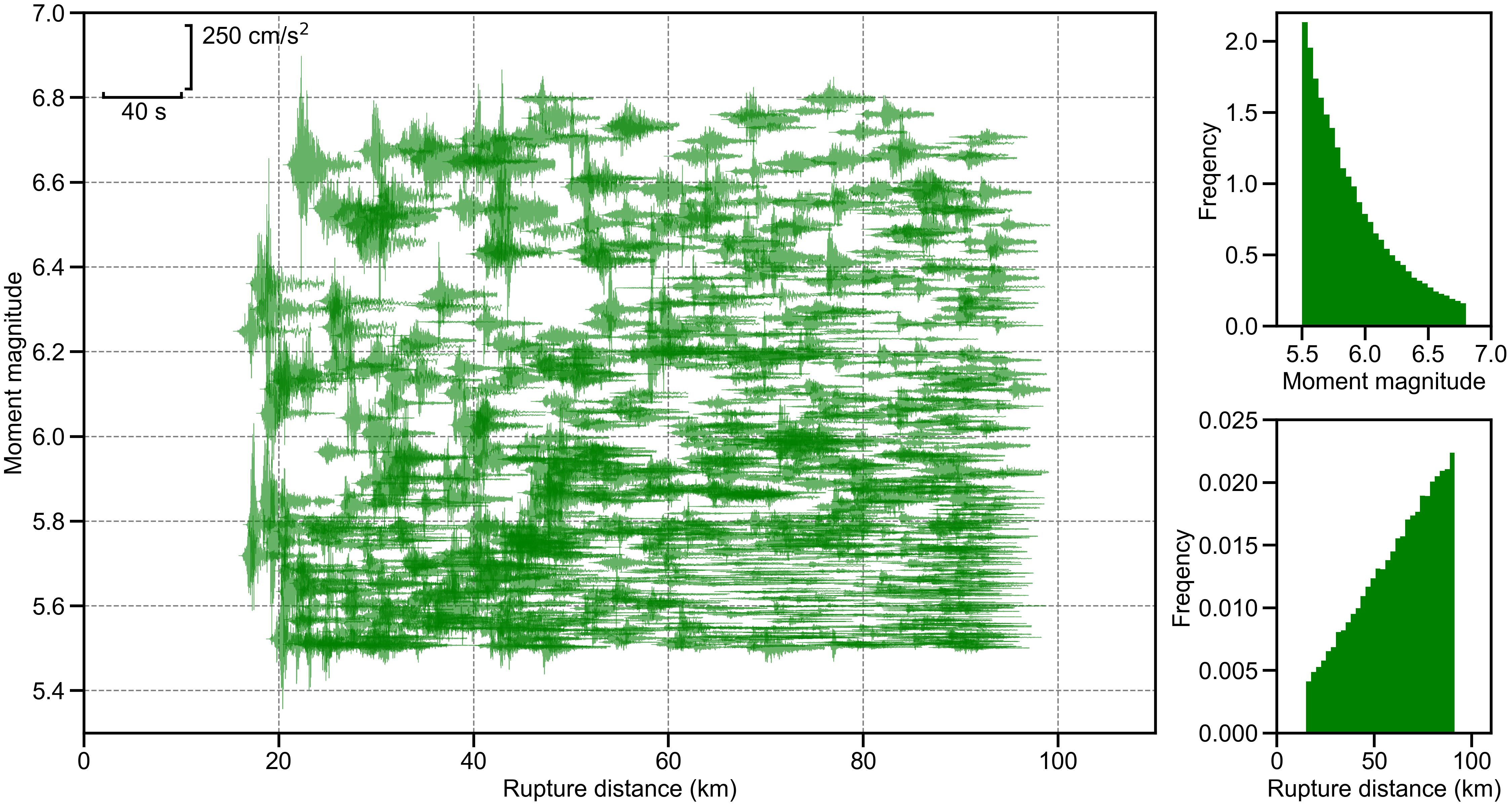}
  \caption{
    Hazard analysis results for numerical example 1 using the CS-GMGM.
    The left panel shows 1,000 acceleration waveforms selected from the sampled ground motions.
    The beginning of each waveform corresponds to its associated $M_W$ and $R_{\mathrm{RUP}}$ values.
    For clarity, only the first 40 s of each waveform are shown.
    On the right hand side, the upper histogram presents the distribution of $M_W$ for all samples,
    and the lower histogram shows the distribution of $R_{\mathrm{RUP}}$.
  }
  \label{fig:gm_scat_loc_4_site_csgmgm}
  \end{minipage}%
\end{figure}%
\clearpage

\begin{figure}[ht]
  \centering
  \begin{minipage}{\columnwidth}
    \centering
  \includegraphics[width=\columnwidth]{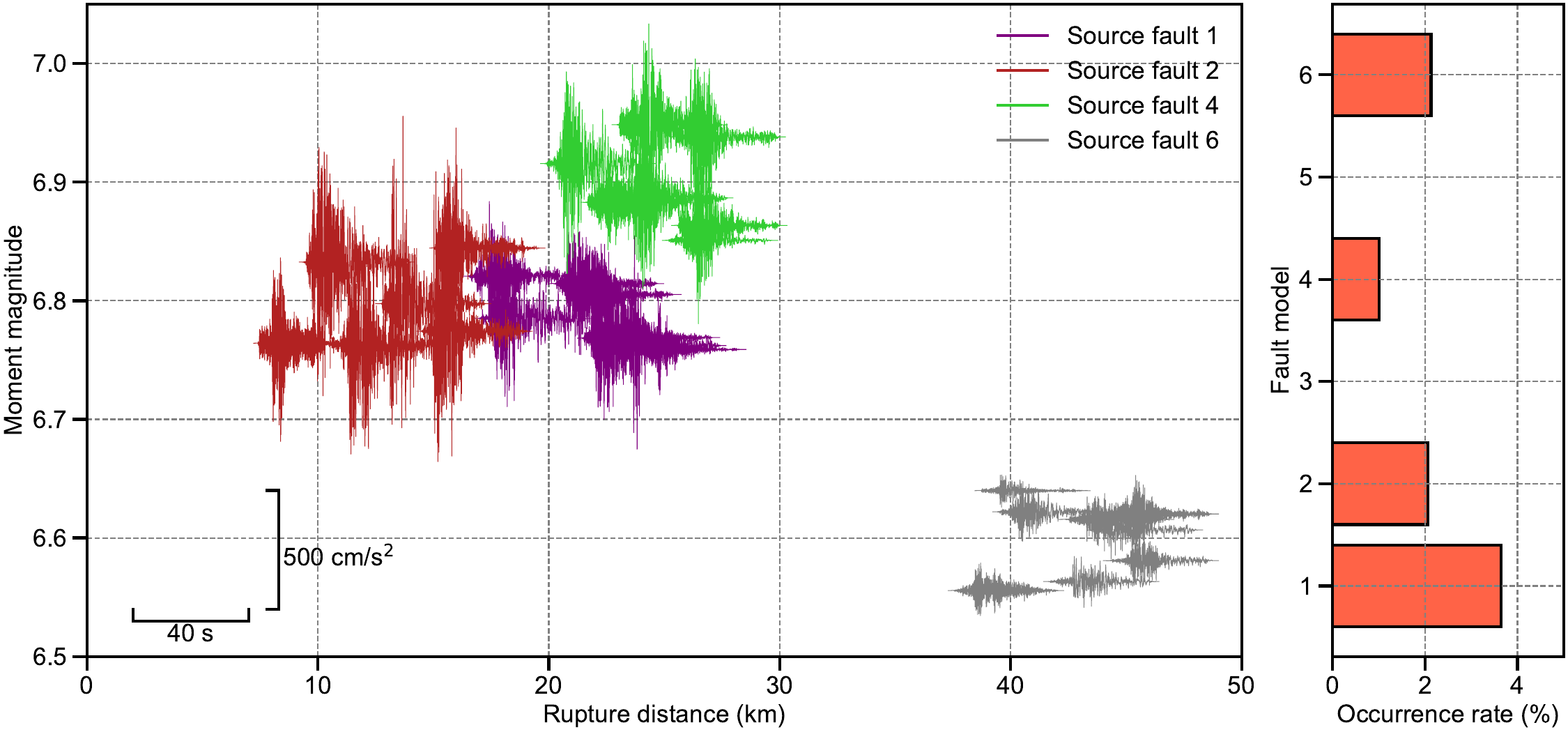}
  \caption{
    Hazard analysis results for numerical example 2 using the CW-GMGM.
    The left panel shows examples of ground-motion waveforms obtained through MCS,
    with colors indicating the corresponding source faults.
    The beginning of each waveform corresponds to its associated $M_W$ and $R_{\mathrm{RUP}}$ values,
    and its location is randomly offset to improve the visual clarity.
    The right panel shows the 50-year earthquake occurrence probabilities for each source fault, computed from the MCS results.
  }
  \end{minipage}%
\end{figure}%
\clearpage

\begin{figure}[ht]
  \centering
  \begin{minipage}{\columnwidth}
    \centering
  \includegraphics[width=\columnwidth]{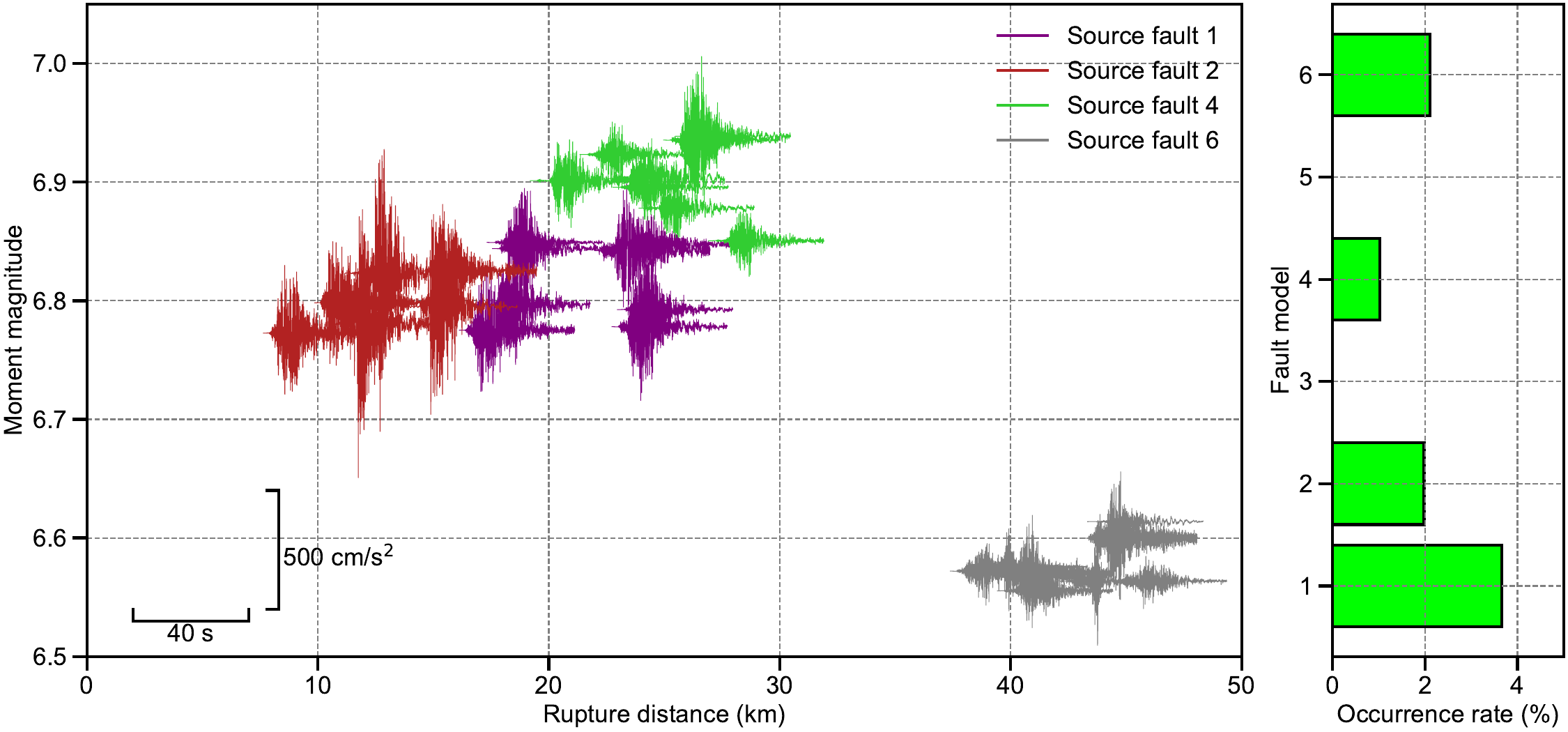}
  \caption{
    Hazard analysis results for numerical example 2 using the CS-GMGM.
    The left panel shows examples of ground-motion waveforms obtained through MCS,
    with colors indicating the corresponding source faults.
    The beginning of each waveform corresponds to its associated $M_W$ and $R_{\mathrm{RUP}}$ values,
    and its location is randomly offset to improve the visual clarity.
    The right panel shows the 50-year earthquake occurrence probabilities for each source fault, computed from the MCS results.
}
  \end{minipage}%
\end{figure}%
\clearpage
\end{document}